\newcommand\papertitle{Moment expansion of polarized dust SED: A new path towards capturing the CMB $B$-modes with \lb{}}

\documentclass[twocolumn]{aa} 

\makeatletter
\renewcommand*\aa@pageof{, page \thepage{} of \pageref*{LastPage}}
\makeatother

\usepackage{graphicx}
\usepackage{txfonts}
\usepackage{url}
\usepackage{natbib}
\bibpunct{(}{)}{;}{a}{}{,} 
\usepackage{color}
\usepackage{colortbl} 
\usepackage[dvipsnames]{xcolor}
\usepackage{enumitem}
\usepackage{caption}

\usepackage[switch]{lineno}

\usepackage[breaklinks, colorlinks, citecolor=blue]{hyperref}
\usepackage{subfigure}
\usepackage{epsfig}
\usepackage{multirow}
\usepackage{array}
\newcommand{\ie}{{i.e.}}

\usepackage{url}
\usepackage{psfrag}
\usepackage[T1]{fontenc}
\usepackage{rotating}
\usepackage{soul}
\usepackage{wrapfig}
\usepackage{tikz}
\def\checkmark{\tikz\fill[scale=0.3](0,.35) -- (.25,0) -- (1,.7) -- (.25,.15) -- cycle;}

\makeatletter
\def\env@cases{
  \let\@ifnextchar\new@ifnextchar
  \left\lbrace
  \def\arraystretch{1.2}
  \array{l@{}l@{}}
}
\makeatother

\definecolor{mygreen}{RGB}{104,198,107}
\definecolor{myred}{RGB}{252,137,125}
\definecolor{myyellow}{RGB}{252,225,126}
\definecolor{mygrey}{RGB}{215,215,215}

\newcommand{\hwline}{\noalign{\color{white}\hrule height 1pt}}

\def\lb{\textit{LiteBIRD}}
\def\wmap{\textit{WMAP}}
\def\planck{\textit{Planck}}

\begin{document}

\title{\papertitle}

\offprints{\url{leo.vacher@irap.omp.eu}}
\authorrunning{Vacher et al.}
\titlerunning{Moment expansion for \lb}

\author{L. Vacher \inst{\ref{inst1}} 
\and J. Aumont  \inst{\ref{inst1}}
\and L. Montier  \inst{\ref{inst1}}
\and S. Azzoni  \inst{\ref{inst2},\ref{inst3}}
\and F. Boulanger  \inst{\ref{inst4}}
\and M. Remazeilles  \inst{\ref{inst5},\ref{inst6}},\\ for the \lb\ Collaboration
} 

\institute{
IRAP, Universit\'e de Toulouse, CNRS, CNES, UPS, Toulouse, France\label{inst1}
\and
Department of Physics, University of Oxford, Denys Wilkinson Building, Keble Road, Oxford OX1 3RH, United Kingdom \label{inst2}
\and
Kavli Institute for the Physics and Mathematics of the Universe (Kavli IPMU, WPI),UTIAS, The University of Tokyo, Kashiwa, Chiba 277-8583, Japan \label{inst3}
\and
Laboratoire de Physique de l’Ecole normale supérieure, ENS, Université PSL, CNRS, Sorbonne Université,
Université Paris-Diderot, Sorbonne Paris Cité, Paris, France
\label{inst4}
\and
 Instituto de Fisica de Cantabria (CSIC-Universidad de Cantabria), Avda. de los Castros s/n, E-39005 Santander, Spain
 \label{inst5}
\and
Jodrell Bank Centre for Astrophysics, Department of Physics and Astronomy, The University of Manchester, Manchester M13 9PL, U.K.
\label{inst6}
}

\abstract{
Accurate characterization of the polarized dust emission from our Galaxy will be decisive in the quest for the cosmic microwave background (CMB) primordial $B$-modes. An incomplete modeling of its potentially complex spectral properties could lead to biases in the CMB polarization analyses and to a spurious measurement of the tensor-to-scalar ratio $r$. It is particularly crucial for future surveys like the \lb{} satellite, the goal of which is to constrain the faint primordial signal leftover by inflation with an accuracy on the tensor-to-scalar ratio $r$ of the order of $10^{-3}$. Variations of the dust properties along and between lines of sight lead to unavoidable distortions of the spectral energy distribution (SED) that cannot be easily anticipated by standard component-separation methods. This issue can be tackled using a moment expansion of the dust SED, an innovative parametrization method imposing minimal assumptions on the sky complexity.
In the present paper, we apply this formalism to the $B$-mode cross-angular power spectra computed from simulated \lb{} polarization data at frequencies between 100 and 402\,GHz that contain CMB, dust, and instrumental noise. The spatial variation of the dust spectral parameters (spectral index $\beta$ and temperature $T$) in our simulations lead to significant biases on $r$ ($\sim$21\,$\sigma_{r}$) if not properly taken into account. Performing the moment expansion in $\beta$, as in previous studies, reduces the bias but does not lead to sufficiently reliable estimates of $r$. We introduce, for the first time, the expansion of the cross-angular power spectra SED in both $\beta$ and $T$, showing that, at the sensitivity of \lb{},  the SED complexity due to temperature variations needs to be taken into account in order to prevent analysis biases on $r$. Thanks to this expansion, and despite the existing correlations between some of the dust moments and the CMB signal responsible for a rise in the error on $r$, we can measure an unbiased value of the tensor-to-scalar ratio with a dispersion as low as $\sigma_{r}=8.8\times10^{-4}$.}


\maketitle

\section{Introduction}
\label{sec:intro}

Our present understanding of the primordial Universe relies on the paradigm of inflation \citep{inflationhist1,inflationhist2,inflationhist3}, introducing a phase of accelerated expansion in the first fractions of a second after the primordial singularity. Such a phenomenon is expected to leave a background of gravitational waves propagating in the primordial plasma during recombination, leaving a permanent mark imprinted in the polarization anisotropies of the cosmic microwave background (CMB): the primordial $B$-modes \citep{InflationmodesB3,InflationsmodesB1,InflationsmodesB2}. The amplitude of the angular power spectrum of those primordial $B$-modes is characterized by the {tensor-to-scalar ratio} $r$, which is proportional to the energy scale at which inflation occurred \citep{renergyscale}. Hence, looking for this smoking gun of inflation allows us to test our best theories of fundamental physics in the primordial Universe at energy scales far beyond the reach of particle accelerators. In this scope, it is one of the biggest challenges of cosmology set out for the next decades. The best experimental upper limit on the $r$ parameter so far is $r<0.032$ \citep[95\,\% C.L.,][]{tristram,bicep2021,PlanckandBICEP}.

The JAXA Lite (Light) satellite, used for the $B$-mode polarization and Inflation from cosmic background Radiation Detection (\lb{}) mission, is designed to observe the sky at large angular scales in order to constrain this parameter $r$ down to $\delta r= 10^{-3}$, including all sources of uncertainty \citep{litebird,LiteBIRDUpdatedDesign}. Exploring this region of the parameter space is critical, because this order of magnitude for the tensor-to-scalar ratio is predicted by numerous physically motivated inflation models (for a review see e.g.,  \cite{EncyclopediaInflationaris})  

However, the success of this mission relies on our ability to treat polarized foreground signals. Indeed various diffuse astrophysical sources emit polarized $B$-mode signals above the primordial ones, the strongest being due to the diffuse polarized emission of our own Galaxy \citep{PlanckCompoSep}. Even in a diffuse region like the BICEP/Keck field, the Galactic $B$-modes are at least ten times stronger at 150\,GHz than the $r=0.01$ tensor $B$-modes targeted by the current CMB experiments \citep{BICEPKECKGW}.

The true complexity of polarized foreground emission that the next generation of CMB experiments will face is still mostly unknown today. Underestimation of this complexity can lead to the estimation of a spurious nonzero value of $r$ \citep[see e.g.,][]{PlanckL,Remazeilles_etal_2016}.

At high frequencies ($>100$ GHz), the thermal emission of interstellar dust grains is the main source of Galactic foreground contaminating the CMB \citep{dusthighfreq,PlanckDust2}. The canonical model of the spectral energy distribution (SED) of this thermal emission for intensity and polarization is given by the modified black body (MBB) law \citep{Desertdustmodel}. This model provides a good fit to the dust polarization SED at the sensitivity of the \planck{} satellite \citep{PlanckDust2} but it may not fully account for
it at the sensitivity of future experiments \citep{HensleyBull}. Furthermore, due to changes of physical conditions across the galaxy, spatial variations of the SEDs are present between and along the lines of sight. The former leads to what is known as \emph{frequency decorrelation} in the CMB community \citep[see e.g.][]{tassis,PlanckL,pelgrims2021}. Moreover, both effects lead to averaging MBBs when observing the sky (unavoidable line-of-sight or beam-integration effects). Because of the nonlinearity of the MBB law, those averaging effects will {distort} the SED, leading to deviations from this canonical model \citep{Chluba}. 

\cite{Chluba} proposed a general framework called {``moment expansion''} of the SED to take into account those distortions, using a Taylor expansion around the MBB with respect to its spectral parameters \citep[Taylor expansion of foreground SEDs was discussed in previous studies; see e.g.,][]{stolyarov2005}. This method is agnostic: it does not require any assumption on the real complexity of the polarized dust emission. The moment expansion approach thus provides a promising tool with which to model the unanticipatable complexity of the dust emission in real data.

\cite{Mangilli} generalized this formalism for the sake of CMB data analysis in harmonic space and for cross-angular power spectra and applied it successfully to complex simulations and \planck{} High-Frequency Instrument (HFI) intensity data. This latter work shows that the real complexity of Galactic foregrounds could be higher than expected, encouraging us to follow the path opened by the moment expansion formalism.

In the present work, we apply the moment expansion in harmonic space to characterize and treat the dust foreground polarized emission of \lb{} high-frequency simulations, using  dust-emission models of increasing complexity. We discuss the ability of this method to recover an unbiased value for the $r$ parameter, with enough accuracy to achievethe scientific objectives of the \textit{LiteBIRD} mission.

In Sect.~\ref{sec:formalism}, we first review the formalism of moment expansion in map and harmonic domains. We then describe in Sect.~\ref{sec:sims} how we realize several sets of simulations of the sky as seen by the \lb{} instrument with varying dust complexity and how we estimate the angular power spectra. In Sect.~\ref{sec:fit}, we describe how we estimate the moment parameters and the tensor-to-scalar ratio $r$ in those simulations. The results are then presented in Sect.~\ref{sec:results}. Finally, we discuss those results and the future work that has to be done in the direction opened by moment expansion in Sect.~\ref{sec:discussion}.

\section{\label{sec:formalism}Formalism}

\subsection{Characterizing the dust SED in real space}

\subsubsection{Modified black body model \label{sec:mbb}}

The canonical way to characterize astrophysical dust-grain emission in every volume element of the Galaxy is given by the modified black body (MBB) function, consisting of multiplying a standard black body SED $B_\nu(T)$ at a given temperature $T_0$ by a power-law of the frequency $\nu$ with a spectral index $\beta_0$. The dust intensity map $I_{\rm D}(\nu,\vec{n})$ observed at a frequency $\nu$ in every direction with respect to the unit vector $\vec{n}$, can then be written as:
\begin{equation}
\label{eq:MBB}
     I(\nu,\vec{n}) =  \left(\frac{\nu}{\nu_0}\right)^{\beta_0} \frac{B_{\nu}(T_0)}{B_{\nu_0}(T_0)} A(\vec{n}) 
     =\frac{I_{\nu}(\beta_0,T_0)}{I_{\nu_0}(\beta_0,T_0)} A(\vec{n}),
\end{equation}

\noindent where $A(\vec{n})$ is the dust intensity template at a reference frequency $\nu_0$\footnote{Throughout this work, we use $\nu_0 = 353$\,GHz.}.
We know that the physical conditions (thermodynamic and dust grain properties) change through the interstellar medium across the Galaxy, depending, in an intricate fashion, on the gas velocity and density, the interstellar radiation field, the distance to the Galactic center \citep[see e.g., ][]{dustacrossMW,dustvarMW,PlanckVardust,PlanckCompoSep,vardustdisk,dustradfield}. This change of physical conditions leads to variations in $\beta$ and $T$ depending on the direction of observation $\vec{n}$:

\begin{equation}
\label{eq:MBBn}
     I(\nu,\vec{n}) =  \frac{I_{\nu}(\beta(\vec{n}),T(\vec{n}))}{I_{\nu_0}(\beta(\vec{n}),T(\vec{n}))} A(\vec{n}).
\end{equation}

The SED amplitude and parameters (temperature and spectral index) are then different for every line of sight. It is therefore clear that, in order to provide a realistic model of the dust emission, the frequency and spatial dependencies may not be trivially separated. 

\subsubsection{\label{sect:limitsmbb}Limits of the modified black body}

The dust SED model given by the MBB has proven to be highly accurate \citep{Planck2014dust,Planck2015dust}. However, it must be kept in mind that this model is empirical and is therefore not expected to give a perfect description of the dust SED in the general case. Indeed, physically motivated dust grain emission models predict deviations from it \citep[e.g.,][]{modelbeyondmbb}. Surveys tend to show that the dust-emission properties vary across the observed 2D sky and the 3D Galaxy \citep{PlanckDust2}. Furthermore, in true experimental conditions, one can never directly access the pure SED of a single volume element with specific emission properties and unique spectral parameters. Averages are therefore made over different SEDs emitted from distinct regions with different physical emission properties, in a way that may not be avoided:  along the line of sight; between different lines of sight, inside the beam of the instrument or; when doing a spherical harmonic decomposition to calculate the angular power spectra over large regions of the sky.

The MBB function is nonlinear, and therefore summing MBBs with different spectral parameters does not return another MBB function and produces \emph{SED distortions}. For all these reasons, modeling the dust emission with a MBB is intrinsically limited, even when doing so with spatially varying spectral parameters. As a consequence, inaccuracies might appear when modeling the dust contribution to CMB data that will unavoidably impact the final estimation of the cosmological parameters. 

\subsubsection{Moment expansion in pixel space}
\label{sec:moment_pixel}

A way to address the limitation of the MBB model in accurately describing the dust emission is given by the {moment expansion} formalism proposed by \cite{Chluba}. This formalism is designed to take into account the SED distortions due to averaging effects by considering a multidimensional Taylor expansion of the distorted SED $I(\nu,\vec{p})$ around the mean values $\vec{p}_{0}$ of its spectral parameters $\vec{p} = \{p_i\}$. This is the so-called {moment expansion} of the SED, which can be written as
\begin{align}
    I(\nu,\vec{p}) = I(\nu,\vec{p}_{0}) &+ \sum_i \omega_1^{p_i}\langle\partial_{p_i}I(\nu,\vec{p})\rangle_{\vec{p}=\vec{p}_0}
    \nonumber \\
    &+ \frac{1}{2}\sum_{i,j} \omega_2^{p_ip_j}\langle\partial_{p_i}\partial_{p_j}I(\nu,\vec{p})\rangle_{\vec{p}=\vec{p}_0}
    \nonumber \\
    &+ \dots \nonumber\\
    &+\frac{1}{\alpha!} \sum_{i,\dots,k} \omega_\alpha^{p_i\dots p_k}\langle\partial_{p_i}\dots\partial_{p_k}I(\nu,\vec{p})\rangle_{\vec{p}=\vec{p}_0},
\label{eq:momentgeneral}
\end{align}

where the first term on the right-hand side is the SED without distortion $I(\nu,\vec{p}_{0})$ evaluated at $\vec{p}=\vec{p}_0$, and the other terms are the so-called {moments} of order $\alpha$, quantified by the {moment parameters} $\omega_\alpha^{p_i\dots p_k}$ for the expansion with respect to any parameter of $\vec{p}$. Performing the expansion to increasing order adds increasing complexity to the SED $I(\nu,\vec{p}_{0})$. 

For the MBB presented in Sect.~\ref{sec:mbb}, there are two parameters so that $\vec{p} = \{\beta,T\}$. Thus the dust moment expansion reads

\begin{align}
     I(\nu,\vec{n})  = \frac{I_{\nu}(\beta_0,T_0)}{I_{\nu_0}(\beta_0,T_0)} \bigg\{ & A(\vec{n}) + \omega^\beta_1(\vec{n}) \ln\left(\frac{\nu}{\nu_0}\right)+ \frac{1}{2}\omega^\beta_2(\vec{n}) \ln^2\left(\frac{\nu}{\nu_0}\right)\nonumber \\[2mm]
     &+ \omega^T_1(\vec{n})\Big( \Theta(\nu,T_0) - \Theta(\nu_0,T_0) \Big) + \dots \bigg\},
\label{eq:momentinttemp}
\end{align}

\noindent where the expansion has been written up to order two in $\beta$ (with moment expansion parameters $\omega^\beta_1$ at order one and $\omega^\beta_2$ at order two) and to order one in $T$ (with a moment expansion parameter $\omega^T_1$ at order one). The following expression has been introduced to simplify the black body derivative with respect to $T$:

\begin{align}
    \Theta(\nu,T) = \frac{x}{T}\frac{e^{x}}{e^{x}-1},\ {\rm with}\ x = \frac{h \nu}{k T}.
\end{align}

The moment expansion in pixel space can be used for component separation and possibly crossed with other methods \citep[see e.g.,][]{RemazeillesmomentsILC,Debabrata_2021}. However, in the present work, we are interested in the modeling of the dust at the $B$-mode angular power spectrum level. Performing the moment expansion at the angular power spectrum level adds some complexity to the SEDs due to the additional averaging occurring when dealing with spherical harmonic coefficients. Indeed, these coefficients are estimated on potentially large fractions of the sky and probe regions with various physical conditions. On the other hand, the expansion at the power spectrum level possibly drastically reduces the parameter space with respect to performing the expansion in every sky pixel.

\subsection{Characterizing the dust SED in harmonic space}

\subsubsection{Dust SED in spherical harmonic space \label{sec:formalismspectra}}

The expansion presented in Sect.~\ref{sec:moment_pixel} can be applied in spherical harmonic space using the same logic. The sky emission projection then reads

\begin{equation}
        I(\nu,\vec{n}) = \sum_{\ell = 0}^{\infty}\sum_{m=-\ell}^{\ell}I^\nu_{\ell m}Y_{\ell m} (\vec{n}).
\end{equation}

Applying the moment expansion to the spherical harmonics coefficients, with respect to $\beta$ and $T$, as in Eq.~\ref{eq:momentinttemp}, leads to

\begin{align}
     I^\nu_{\ell m} = & \frac{I_{\nu}(\beta_0(\ell),T_0(\ell))}{I_{\nu_0}(\beta_0(\ell),T_0(\ell))} \bigg\{ A_{\ell  m} + \omega^\beta_{1,\ell  m} \ln\left(\frac{\nu}{\nu_0}\right) + \frac{1}{2}\omega^\beta_{2,\ell  m} \ln^2\left(\frac{\nu}{\nu_0}\right) \nonumber \\[2mm]
     &+ \omega^T_{1,\ell  m} \Big( \Theta(\nu,T_0(\ell)) - \Theta(\nu_0,T_0(\ell)) \Big) + \dots \quad\bigg\},
\label{eq:momentint2temp}
\end{align}

\noindent where this time $\beta_0(\ell)$ and $T_0 (\ell)$ are the averages of $\beta$ and $T$ at a given multipole $\ell$ over the sky fraction we are looking at. We note that the moment parameters $\omega^{p_i}_{\alpha,\ell m}$ involved here are different from the $\omega^{p_i}_i(\vec{n})$ appearing in Eq.~\ref{eq:momentinttemp} in the map space because they involve different averaging. In principle, the moment expansion in harmonic space can take into account the three kinds of spatial averages presented in Sect.~\ref{sect:limitsmbb}.

As the dust spectral index and temperature are difficult to separate in the frequency range considered for CMB studies \citep[\ie, Rayleigh-Jeans domain, see e.g.][]{betatcorr}, the moment expansion in harmonic space has  only been applied in the past with respect to $\beta$, with the temperature being fixed to a reference value $T=T_0$ \citep{Mangilli,Azzoni}. In the present paper, for the first time, the moment expansion in harmonic space is instead performed with respect to both $\beta$ and $T$, as it was in real space in \citet{RemazeillesmomentsILC}.

\subsubsection{Cross-power spectra}

Relying on the derivation made by \cite{Mangilli} and Eq.~\ref{eq:momentint2temp}, we can explicitly write the cross-spectra between two maps $M_{\nu_i}$ and $M_{\nu_j}$ at frequencies $\nu_i$ and $\nu_j$, using the moment expansion in $\beta$ and $T$ as follows:

\begin{align}
    \mathcal{D}_\ell(\nu_i \times \nu_j) &= \frac{I_{\nu_i}(\beta_0(\ell),T_0(\ell))I_{\nu_j}(\beta_0(\ell),T_0(\ell))}{I_{\nu_0}(\beta_0(\ell),T_0(\ell))^2} \cdot \bigg\{ \nonumber \\[-0.5mm]
    0^{\rm th}\ \text{order}\;&
    \begin{cases}
    & \ \mathcal{D}_\ell^{A \times A}
    \end{cases}
    \nonumber \\[-0.5mm]
    1^{\rm st}\ \text{order}\ \beta\;&
    \begin{cases}
    &+\mathcal{D}_\ell^{A \times \omega^{\beta}_1}\left[ \ln\left(\frac{\nu_i}{\nu_0}\right) + \ln\left(\frac{\nu_j}{\nu_0}\right) \right] \nonumber \\
    &+ \mathcal{D}_\ell^{\omega^{\beta}_1 \times \omega^{\beta}_1} \left[\ln\left(\frac{\nu_i}{\nu_0}\right)\ln\left(\frac{\nu_j}{\nu_0}\right) \right]\nonumber \\  \end{cases}\\[-0.5mm]
    1^{\rm st}\ \text{order}\ T \;&
    \begin{cases}
    &+\mathcal{D}_\ell^{A \times \omega_1^T} \left( \Theta_i +  \Theta_j -  2\Theta_0\right) \\
    &+ \mathcal{D}_\ell^{\omega_1^T \times \omega_1^T}\Big(\Theta_i - \Theta_0\Big)\left(\Theta_j - \Theta_0\right)\nonumber
    \end{cases}\\[-0.5mm]
    1^{\rm st}\ \text{order}\ T\beta \;&
    \begin{cases}
    &+ \mathcal{D}_\ell^{\omega^{\beta}_1 \times \omega_1^T} \left[ \ln\left(\frac{\nu_j}{\nu_0} \right)\Big( \Theta_i - \Theta_0\Big) + \ln\left(\frac{\nu_i}{\nu_0} \right)\left( \Theta_j - \Theta_0\right)\right]  \nonumber \\
    \end{cases}\\[-0.5mm]
    2^{\rm nd}\ \text{order}\ \beta \;&
    \begin{cases}
    &+ \frac{1}{2} \mathcal{D}_{\ell}^{A \times\omega^{\beta}_{2}}  \left[ \ln^2\left(\frac{\nu_i}{\nu_0}\right)
    + \ln^2\left(\frac{\nu_j}{\nu_0}\right)
    \right]
    \\[-0.5mm]
    &+ \frac{1}{2} \mathcal{D}_{\ell}^{\omega^{\beta}_{1} \times \omega^{\beta}_{2}}  \Big[ \ln \left(\frac{\nu_i}{\nu_0}\right)  \ln^2\left(\frac{\nu_j}{\nu_0}\right)  
    +\ln\left(\frac{\nu_j}{\nu_0}\right)
    \ln^2 \left(\frac{\nu_i}{\nu_0}\right) \Big] 
    \\[-0.5mm]
    &+\frac{1}{4} \mathcal{D}_{\ell}^{\omega^{\beta}_{2} \times \omega^{\beta}_{2}}  \left[\ln^2 \left(\frac{\nu_i}{\nu_0}\right)  \ln^2 \left(\frac{\nu_j}{\nu_0}\right) \right]
    \,
    \end{cases}\nonumber\\[-0.5mm]
    &+ \dots \bigg\},
\label{eq:moments}
\end{align}

\noindent where we use the following abbreviation: $\Theta(\nu_k,T_0(\ell))\equiv \Theta_k$, so that $\Theta_0=\Theta(\nu_0,T_0(\ell))$, and we defined the moment expansion cross-power spectra between two moments $\mathcal{M}$ and $\mathcal{N}$ as

\begin{equation}
    \mathcal{C}_\ell^{\mathcal{M}\times\mathcal{N}} = \sum_{m, m'=-\ell}^{\ell} \mathcal{M}_{\ell m} \mathcal{N}_{\ell m'},\ {\rm with}\ (\mathcal{M},\mathcal{N})\in\left\{A,\omega^{\beta}_1,\omega^{T}_1, \omega^{\beta}_2,\ \dots\right\}.
\label{eq:moments_spectra}
\end{equation}

In the remainder of this article, we use the $\mathcal{D}_\ell$ quantity, which is a scaling of the angular power spectra, and is defined as 
\begin{equation}
    \mathcal{D}_\ell \equiv \frac{\ell(\ell +1)}{2 \pi} \mathcal{C}_\ell.
\label{eq:dl}
\end{equation}

Equation~\ref{eq:moments} has been written using the expansion with respect to $\beta$ at order two and $T$ at order one, as in Eq.~\ref{eq:momentint2temp}. Nevertheless, the terms involving power spectra between order two in $\beta$ and order one in $T$ have been neglected so as to match the needs of the implementation of our method in the following. 

Hereafter, when we refer to "order $k$" at the angular power spectrum level, we are referring to moment expansion terms involving the pixel space moment up to order $k$. For example, $\mathcal{D}_\ell^{A\times\omega_1^T}$ and $\mathcal{D}_\ell^{\omega^\beta_1\times\omega_1^T}$ are order one, while $\mathcal{D}_\ell^{A\times\omega^\beta_2}$, $\mathcal{D}_\ell^{\omega^\beta_1\times\omega^\beta_2}$ and $\mathcal{D}_\ell^{\omega^\beta_2\times\omega^\beta_2}$ are order two. At order zero, one retrieves the MBB description of the cross-angular power spectra SED $\mathcal{D}_\ell(\nu_i\times\nu_j)$ as a function of the frequencies $\nu_i$ and $\nu_j$.

This formalism was originally introduced to analyze the complexity of intensity data in \cite{Mangilli}. In the present work, we focus on $B$-mode polarization power spectra. This was put forward after analyzing the \planck{} and balloon-borne Large Aperture Submillimeter Telescope for Polarimetry (BLASTPol) data and finding that the polarization fraction appears to be constant in the far-infrared-to-millimetre wavelengths {\citep{blastpol1,blastpol2}}. This allows us to assume that the same grain population is responsible for the total and polarized foreground emission \citep{dustmodels}. As a result, intensity and polarization SED complexity may be similar.
Nevertheless, $Q$ and $U$ can have a different SED because of the polarization angle frequency dependence \citep[see e.g.,][]{tassis,moment_polar} and so can $E$ and $B$. This could be a limitation when analyzing the dust $E$ and $B$ with a single moment expansion, especially when SED variations occur along the line of sight.
Even when trying to model a single polarization component ---as we do in the present work, dealing only with $B$ modes--- it is not clear whether the distorted SED can be modeled in terms of $\beta$ and $T$ moments only. Further work needs to be done to assess this question. However, they should not impact the present study in which variations along the line of sight are not simulated.

Modeling the complexity of the foreground signals by means of the moment expansion of the $B$-mode angular power spectrum has already been successfully applied to Simons Observatory \citep{SimonsObservatory} simulated data  \citep{Azzoni}. However, the approach taken by these latter authors is different from the one presented above. They apply a \emph{minimal} moment expansion: assumptions are made to keep only the $\mathcal{D}_\ell^{\omega^{\beta}_1\times\omega^{\beta}_1}$ and $\mathcal{D}_\ell^{A\times\omega^{\beta}_2}$ parameters, which are modeled with a power-law scale dependence. These assumptions may not hold for experiments with higher sensitivity and observing wider sky patches. Furthermore, they assume a scale-invariant dust spectral index. In this work, on the other hand, we relax these assumptions in order to characterize the required spectral complexity of the dust emission for \lb{}.

\section{\label{sec:sims}Simulations and cross-spectra estimation}

\subsection{\label{sec:LiteBIRD}\lb{}}

\lb{} is an international project proposed by the Japanese spatial agency (JAXA), which selected it in May 2019 as a strategic large class mission. The launch is planned for 2029 for a minimal mission duration of 3 years. \citep{2020SPIE,Ptep}

\lb{} is designed to realize a full sky survey of the CMB at large angular scales in order to look for the reionization bump of primordial $B$-modes and explore the scalar-to-tensor ratio ($r$) parameter space with a total uncertainty $\delta r$ below $10^{-3}$, including foreground cleaning and systematic errors.
\lb{} is composed of three telescopes observing in different frequency intervals: the Low-, Medium- and High-Frequency Telescopes (LFT, MFT and HFT).
Each of the telescopes illuminates a focal plane composed of hundreds of polarimetric detectors. The whole instrument will be cooled down to 5\,K \citep{LiteBIRDUpdatedDesign} while the focal plane will be cooled down to 100\,mK \citep{LBsubK}. In order to mitigate the instrumental systematic effects, the polarization is modulated by a continuously rotating half-wave plate. \lb{} will observe the sky in 15 frequency bands from 40 to 402\,GHz. Table~\ref{tab:litetab} gives the details of the frequency bands and their sensitivities in polarization \citep[adapted from][see Sect.~\ref{sec:Instrsim}]{2020SPIE}.

\begin{table}
\begin{center}
\begin{tabular}{cccc}
        \multirow{2}{*}{Telescope} & Frequency & Sensitivity  $\sigma^{\rm noise}_{Q,U}(\nu)$ & $\boldsymbol \theta_{\rm FWHM}$  \\
         & [GHz]&  [$\mu$K$\cdot$arcmin] & {arcmin}\\
        \hline
         LFT & 40.0 & 37.42 & {70.5}     \\
        LFT & 50.0 & 33.46 & {58.5} \\
        LFT & 60.0 & 21.31 & {51.1} \\
        LFT & 68.0 & {19.91/31.77} & {41.6/47.1} \\
        LFT & 78.0 & {15.55/19.13} & {36.9/43.8} \\
        LFT & 89.0 & {12.28/28.77} & {33.0/41.5} \\
     LFT/MFT & 100.0 & {10.34/8.48} & {30.2/37.8} \\
     LFT/MFT & 119.0 & {7.69/5.70} & {26.3/33.6} \\
     LFT/MFT & 140.0 & {7.25/6.38} & {23.7/30.8}  \\
         MFT & 166.0 & 5.57 & {28.9} \\
     MFT/HFT & 195.0 & {7.05/10.50} & {28.0/28.6} \\
         HFT & 235.0 & 10.79 & {24.7}\\
        HFT & 280.0 & 13.8 & {22.5} \\
        HFT & 337.0 & 21.95 & {20.9} \\
        HFT & 402.0 & 47.45 & {17.9} \\
\end{tabular}
\caption{\footnotesize Instrumental characteristics of \lb{} used in this study \citep[adapted from][see Sect.~\ref{sec:Instrsim}]{2020SPIE}. Some frequency bands are shared by two different telescopes or detector arrays. If so, the two values of polarization sensitivities $\sigma^{\rm noise}_{Q,U}(\nu)$ and instrumental beam full width at half maximum $\theta_{\rm FWHM}$ are displayed on the same line.}
\label{tab:litetab}
\end{center}
\end{table}

\subsection{Components of the simulations}
\label{sec:ingredients}

We build several sets of \lb{} sky simulations. These multi-frequency sets of polarized sky maps are a mixture of CMB, dust, and instrumental noise. The simulations are made at the nine highest frequencies accessible by the instrument ($\geq 100$ GHz), where dust is the predominant source of foreground contamination.
For every studied scenario, we built $N_{\rm sim} = 500$ simulations, each composed of a set of $N_{\rm freq}=9$ pairs of sky maps $(Q,U)$ built using the {\sc HEALPix} package, with $N_{\rm side} = 256$ \citep{healpix}. All the signals will be expressed in $\mu{\rm K}_{\rm CMB}$ units.

\subsubsection{Cosmic microwave background signal}

To generate the CMB signal, we use the {\it Code for Anisotropies in the Microwave Background } \citep[CAMB,][]{CAMB} to create a fiducial angular power spectrum from the best-fit values of cosmological parameters estimated by the recent \planck{} data analysis \citep{PlanckOverview}.

For the $B$-modes, we consider the two different components of the spectrum: lensing-induced and primordial (tensor), so that $\mathcal{D}_{\ell}^{BB} =\mathcal{D}_{\ell}^{{\rm lensing}} + r_{\rm sim} \cdot \mathcal{D}_{\ell}^{{\rm tensor}}$, where $\mathcal{D}_{\ell}^{{\rm tensor}}$ refers to the tensor $B$-modes for $r=1$ and $r_{\rm sim}$ labels the input values of the tensor-to-scalar ratio $r$ contained in the simulation. We use two different values throughout this work: $r_{\rm sim}=0$, which is used in the present work as the reference simulations and $r_{\rm sim}=10^{-2}$ used for consistency checks when the CMB primordial signal is present.

For all simulations, we then generate the Stokes $Q$ and $U$ CMB polarization Gaussian realization maps $S^{\rm CMB}_{\nu,r_{\rm sim}}$ from the angular power spectra using the {\tt synfast} function of {\sc HEALPix}.

\subsubsection{Foregrounds: dust}

Our study focuses on high frequencies ($\geq 100$\,GHz) only, where thermal dust emission is the main source of polarized foreground as mentioned in Sect.~\ref{sec:intro}. We make use of two different scenarios of increasing complexity included in the {\sc PySM} \citep{Pysm} and one of intermediate complexity not included in the {\sc PySM}:

\begin{itemize}
    \item {\tt d0}, included in the {\sc PySM}: the dust polarization $Q$ and $U$ maps are taken from $S^{Planck}_{\nu=353}$, the \planck{} 2015 data at 353\,GHz \citep{planck_2015_overview}, extrapolated to a frequency $\nu$ using the MBB given in Eq.~\ref{eq:MBB} with a temperature $T_0=T_{\tt d0}=20$\,K and spectral index $\beta_0=\beta_{\tt d0}=1.54$ constant over the sky:
    
    \begin{equation}S^{\rm dust}_\nu=S_\nu^{\tt d0}=\frac{I_\nu(\beta_{\tt d0},T_{\tt d0})}{I_{\nu_0}(\beta_{\tt d0},T_{\tt d0})}\cdot S^{Planck}_{353},
    \end{equation}
    
    \item {\tt d1T}, introduced here: the dust polarization $Q$ and $U$ maps are also taken from \citet{planck_2015_overview} but they are extrapolated to a frequency $\nu$ using the MBB given in Eq.~\ref{eq:MBBn}, with spatially varying  spectral index $\beta(\vec{n})$, as in {\tt d1} and a fixed temperature $T_0=T_{\tt d1T}=21.9$\,K, obtained as the mean of the \planck{} {\sc Commander} dust temperature map  \citep{planck_2015_commander} on our $f_{\rm sky}=0.7$ sky mask: 
    
    \begin{equation}
        S^{\rm dust}_\nu=S_\nu^{\tt d1T}=\frac{I_\nu(\beta(\vec{n}),T_{\tt d1T})}{I_{\nu_0}(\beta(\vec{n}),T_{\tt d1T})}\cdot S^{Planck}_{353}.
    \end{equation}
    
    \item {\tt d1}, included in the {\tt PySM}: similar to { \tt d1T} with both a spatially varying temperature $T(\vec{n})$ and spectral index $\beta(\vec{n})$ obtained from the \planck{} data using the {\sc Commander} code \citep{planck_2015_commander}: 
    
    \begin{equation}
        S^{\rm dust}_\nu=S_\nu^{\tt d1}=\frac{I_\nu(\beta(\vec{n}),T(\vec{n}))}{I_{\nu_0}(\beta(\vec{n}),T(\vec{n}))}\cdot S^{Planck}_{353}.
    \end{equation}

\end{itemize}

\subsubsection{ \label{sec:Instrsim}Instrumental noise}

The band polarization sensitivities  $\sigma^{\rm noise}_{Q,U}(\nu)$ are derived from the noise equivalent temperature (NET) values converted into $\mu$K$\cdot$arcmin for each telescope (LFT, MFT and HFT). As seen in Table~\ref{tab:litetab}, some frequency bands are overlapping between two telescopes. In this situation, we take the mean value of the two NETs, weighted by the beam full width at half maximum (FWHM) $\theta$ as:

\begin{equation}
    \sigma^{\rm noise}_{Q,U}(\nu_{\rm overlapping}) = \sqrt{\frac{1}{ \left(\frac{\theta_{\rm min}}{{\theta_{\rm max}}}\sigma^{\rm noise}_{Q,U}(\nu_{\theta_{\rm min}})\right)^{-2} + \left(\sigma^{\rm noise}_{Q,U}(\nu_{\theta_{\rm max}})\right)^{-2} }},
\end{equation}

{\noindent where $\theta_{\rm min}$ is the smallest FWHM among the two and $\theta_{\rm max}$ the largest. The band polarization sensitivities} are displayed in Table~\ref{tab:litetab}. For every simulation, the noise component $N_\nu$ is generated in every pixel of the maps with a Gaussian distribution centered on zero, with standard deviation $\sigma^{\rm noise}_{Q,U}(\nu)$ weighted by the pixel size (and $\sqrt{2}\cdot\sigma^{\rm noise}_{Q,U}(\nu)$ for the maps used to compute the auto-power spectra, see Sect.~\ref{sec:spectra_estimation}).

For simplicity, we choose to ignore beam effects in our simulations, assuming they can be taken into account  perfectly. Simulations are thus produced at infinite (0\,arcmin) resolution and no beam effect is corrected for when estimating the angular power spectrum. This is equivalent to convolving the maps by Gaussian beams of finite resolution and correcting the power spectra for the associated Gaussian beam window functions.

\subsection{Combining signals and building the simulated maps}

\label{sec:simulations}

The simulated $(Q,U)$ maps $M_\nu$, for a given simulation, can be expressed as the sum:

\begin{equation}
M_\nu=S^{\rm CMB}_{\nu,r_{\rm sim}}+S^{\rm dust}_\nu+N_\nu.
\label{eq:map}
\end{equation}
\begin{figure*}
    \centering
    \includegraphics[scale = 0.2]{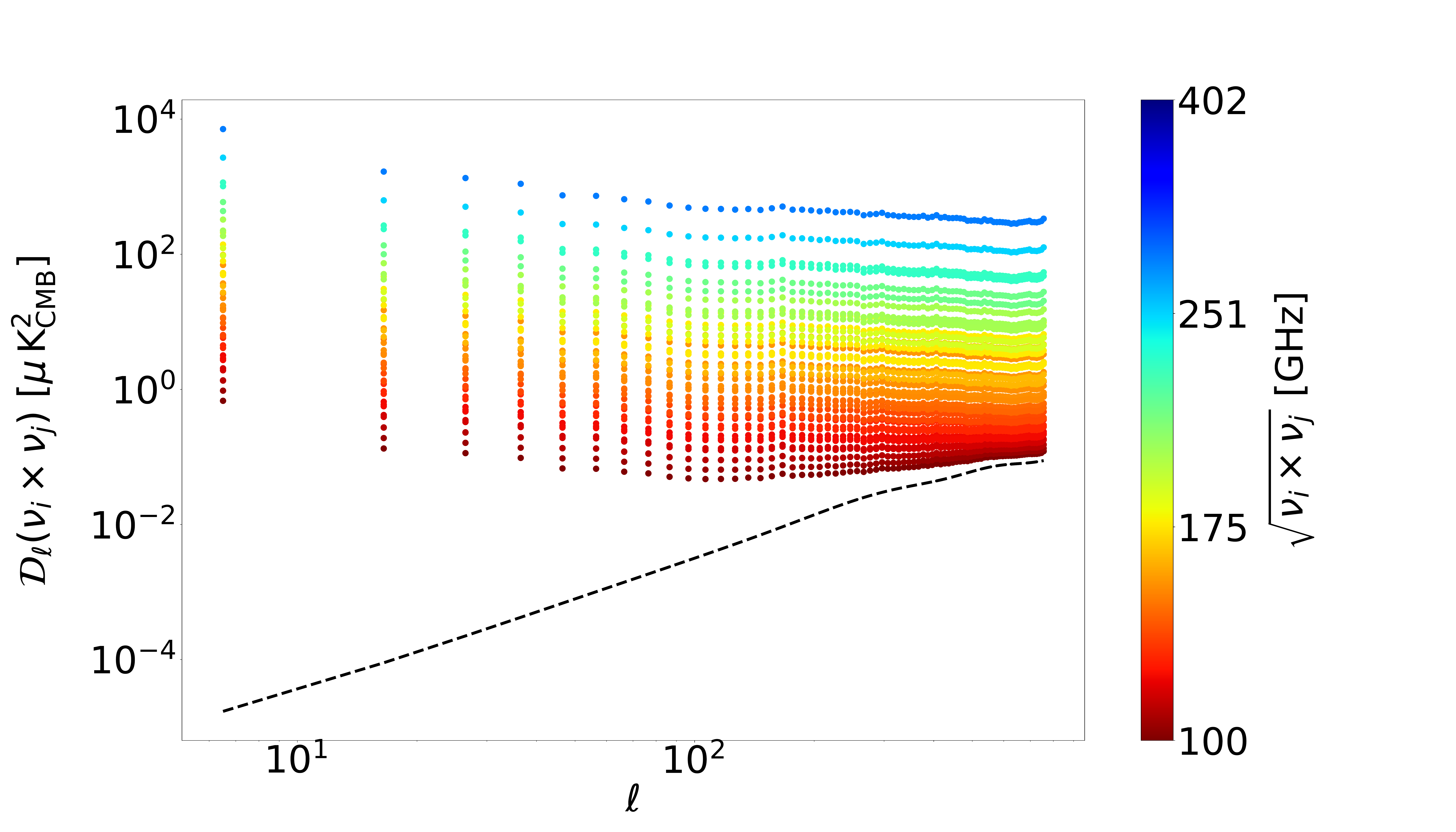}
    \caption{\footnotesize Mean value over the $N_{\rm sim}$ simulations of the $B$-mode angular power spectra $\mathcal{D}_{\ell}(\nu_i \times \nu_j)$ for the {\tt d1c} simulation type, with $r_{\rm sim}=0$. The color bar spans all the $N_{\rm cross}$ spectra $\mathcal{D}_{\ell}(\nu_i \times \nu_j)$, associated to their reduced cross-frequency $\nu_{\rm red.}=\sqrt{\nu_i \nu_j}$, from 100\,GHz (dark red) to 402\,GHz (dark blue). The input CMB lensing power spectrum is shown as a black dashed line.}
    \label{fig:simulations}
\end{figure*}

Cosmic microwave background and noise are simulated stochastically: for each simulation, we generate a new realization of the CMB maps $S^{\rm CMB}_{\nu,r_{\rm sim}}$ and the noise maps $N_\nu$. The dust map $S^{\rm dust}_\nu$ is the same for each simulation, at a given frequency.

Hereafter, we use the notation {\tt d0}, {\tt d1T,} and {\tt d1} to refer to simulations containing only dust and \lb{} noise, {\tt d0c}, {\tt d1Tc,} and {\tt d1c} for simulations including CMB, dust, and \lb{} noise and, finally, and {\tt c} for the simulation containing only CMB and \lb{} noise. The different components present in these different {simulation types} are summarized in Table~\ref{tab:sims}.

\begin{table}[t!]
\centering
        \begin{tabular}{c !{\color{white}\vrule width 1pt} c!{\color{white}\vrule width 1pt}c!{\color{white}\vrule width 1pt}c!{\color{white}\vrule width 1pt}c!{\color{white}\vrule width 1pt}c}
        & $S^{\rm CMB}_{\nu,r_{\rm sim}}$ & $S^{\tt d0}_\nu$ & $S^{\tt d1T}_\nu$ & $S^{\tt d1}_\nu$ & $N_\nu$ \\[0.5ex]\hwline
        {\tt c} & \cellcolor{mygreen}\checkmark & \cellcolor{myred}$\times$  & \cellcolor{myred}$\times$ & \cellcolor{myred}$\times$ & \cellcolor{mygreen}\checkmark\\\hwline
        {\tt d0} & \cellcolor{myred}$\times$ & \cellcolor{mygreen}\checkmark  & \cellcolor{myred}$\times$ & \cellcolor{myred}$\times$ & \cellcolor{mygreen}\checkmark\\
        \hwline
        {\tt d1T} & \cellcolor{myred}$\times$ & \cellcolor{myred}$\times$  & \cellcolor{mygreen}\checkmark & \cellcolor{myred}$\times$ & \cellcolor{mygreen}\checkmark\\
        \hwline
        {\tt d1} & \cellcolor{myred}$\times$ & \cellcolor{myred}$\times$  & \cellcolor{myred}$\times$ & \cellcolor{mygreen}\checkmark & \cellcolor{mygreen}\checkmark\\
        \hwline
        {\tt d0c} & \cellcolor{mygreen}\checkmark & \cellcolor{mygreen}\checkmark  & \cellcolor{myred}$\times$ & \cellcolor{myred}$\times$ & \cellcolor{mygreen}\checkmark\\
        \hwline
        {\tt d1Tc} & \cellcolor{mygreen}\checkmark & \cellcolor{myred}$\times$  & \cellcolor{mygreen}\checkmark & \cellcolor{myred}$\times$ & \cellcolor{mygreen}\checkmark\\
        \hwline
        {\tt d1c} & \cellcolor{mygreen}\checkmark & \cellcolor{myred}$\times$  & \cellcolor{myred}$\times$ & \cellcolor{mygreen}\checkmark & \cellcolor{mygreen}\checkmark\\
    \end{tabular}

\caption{\footnotesize Summary of the different components present in the simulated maps $M_\nu$ in Eq.~\ref{eq:map}, for every \emph{simulation type}. A tick on a green background signifies that the component is present in the simulations, red with a cross symbol shows that it is absent.}
\label{tab:sims}
\end{table}

\subsection{ \label{sect:spectra} Angular power spectra of the simulations}

\subsubsection{\label{sect:mask}Mask}

A mask is applied on the simulated maps presented in Sect.~\ref{sec:simulations} in order to exclude the Galactic plane from the power-spectrum estimation. The mask is created by setting a threshold on the polarized intensity ($P=\sqrt{Q^2+U^2}$) of the \planck{} 353\,GHz map \citep{PlanckOverview} \footnote{\url{http://pla.esac.esa.int/pla/}}, smoothed with a $10^{\circ}$ beam. In order to keep $f_{\rm sky}= 0.7$,$f_{\rm sky}= 0.6,$ and $f_{\rm sky}= 0.5$, the cut is applied at 121\,$\mu$K, 80\,$\mu$K,  and 53\,$\mu$K, respectively. We then realize a C2 apodization of the binary mask with a scale of $5^{\circ}$ using {\sc Namaster} \citep{namaster}. The resulting Galactic masks are displayed in Fig.~\ref{fig:mask}. These masks are similar to those used in \citet{PlanckDust2}.

\subsubsection{Estimation of the angular power spectra}
\label{sec:spectra_estimation}

We use the {\sc Namaster}\footnote{\url{https://github.com/LSSTDESC/NaMaster}} software \citep{namaster} to compute the angular power spectra of each simulation. {\sc Namaster} allows us to correct for the $E$ to $B$ leakage bias due to the incomplete sky coverage. Therein we use a \emph{purification} process to suppress the effect of the $E$ to $B$ leakage in the variance. For every simulation, from the set of maps $M_{\nu_i}$,  we compute all the possible auto-frequency and cross-frequency spectra $\mathcal{D}_\ell({\nu_i\times\nu_j})\equiv\mathcal{D}_\ell({M_{\nu_i}\times M_{\nu_j}})$ with

\begin{align}
\nu_i\times \nu_j \in \left\{\right.&100\times 100, 100 \times 119, 100 \times 140,\   \dots, 100\times402,\nonumber\\
&119\times140,\ \dots, 119\times402,\nonumber\\[-2mm]
&\qquad\qquad\vdots\nonumber\\
&337\times337, 337\times402,\nonumber\\
&\left.\!\!402 \times 402 \right\},
\label{eq:all_cross}
\end{align}

\noindent leading to $N_{\rm cross}=N_{\rm freq}\cdot(N_{\rm freq}+1)/2=45$ cross-frequency spectra. These spectra are displayed in Fig.~\ref{fig:simulations} for the case of the {\tt d1c} simulation
type.\\

In order to avoid noise auto-correlation in the auto-spectra (\ie, $\mathcal{D}_\ell(\nu_i\times\nu_j)$ when $i=j$), the latter are estimated in a way that differs slightly from what is presented in Sect.~\ref{sec:Instrsim}. We simulate two noise-independent data subsets at an observing frequency $\nu_i$, with a noise amplitude $\sqrt{2}$ higher than that of the frequency band, and compute the cross-angular power spectrum between those. Thus, $\mathcal{D}_\ell(\nu_i\times\nu_i)$ is free from noise auto-correlation bias at the expense of multiplying the noise amplitude in the spectrum by a factor of two. This approach is similar to that commonly used by the \planck{} Collaboration \citep[see e.g.,][]{PlanckDust,PlanckDust2,tristram}.

The spectra are evaluated in the multipole interval $\ell \in [1,200]$ in order to be able to focus on the reionization and recombination bumps of the primordial $B$-modes spectra. 
The spectra are binned in $N_{\ell}=20$ bins of size $\Delta \ell = 10$ using {\sc Namaster}. The same binning is applied throughout this article such that, in the following, the multipole $\ell$ denotes the multipole bin of size $\Delta\ell=10$ centered on $\ell$\footnote{ The $N_\ell$ multipole bins are centered on the following $\ell$ values: $[6.5,  16.5,  26.5,  36.5,  46.5,  56.5,  66.5,  76.5,  86.5,
96.5, 106.5,\\ 116.5, 126.5, 136.5, 146.5, 156.5, 166.5, 176.5,
186.5, 196.5]$}.

From the sets of $(Q,U)$ maps, {\sc Namaster} computes the $\mathcal{D}_\ell^{EE}$, $\mathcal{D}_\ell^{BB}$, and $\mathcal{D}_\ell^{EB}$ angular power spectra; for the sake of the present analysis, we keep only $\mathcal{D}_\ell^{BB}$. Hence, when we discuss or analyze power spectra, we are referring to the $B$-mode power spectra  $\mathcal{D}_\ell^{BB}$. All spectra are expressed in $(\mu{\rm K}_{\rm CMB})^2$.

\section{\label{sec:fit}Best-fit implementation}

In order to characterize the complexity of the dust SED that will be measured by \lb{}, we modeled the angular power spectra of our simulations described in Sect.~\ref{sec:sims} over the whole frequency and multipole ranges with the moment expansion formalism introduced in Sect.~\ref{sec:formalism}.

\subsection{General implementation}
\label{sec:fit_dust}

For each multipole $\ell$, we ordered the angular power spectra $\mathcal{D}_\ell^{BB}(\nu_i\times\nu_j$) as in Eq.~\ref{eq:all_cross} in order to build a SED that is a function of both $\nu_i$ and $\nu_j$. We fit this SED with models, as in Eq.~\ref{eq:moments} for example, using a Levenberg-Marquardt $\chi^2$ minimization with {\tt mpfit} \citep{mpfit}\footnote{\url{https://github.com/segasai/astrolibpy/tree/master/mpfit}}. All the fits performed with {\tt mpfit} were also realized with more computationally heavy Monte Carlo Markov Chains (MCMC) with {\tt emcee} \citep{emcee}, giving compatible results, well within the error bars.

The reduced $\chi^2$ minimization is given by

\begin{equation}
    \chi^2 = \frac{1}{N_{\rm d.o.f.}}\vec{R}^T\mathbb{C}^{-1}\vec{R},
\label{eq:chi2}
\end{equation}

\begin{figure}[t]
    \centering
    \includegraphics[scale = 0.3]{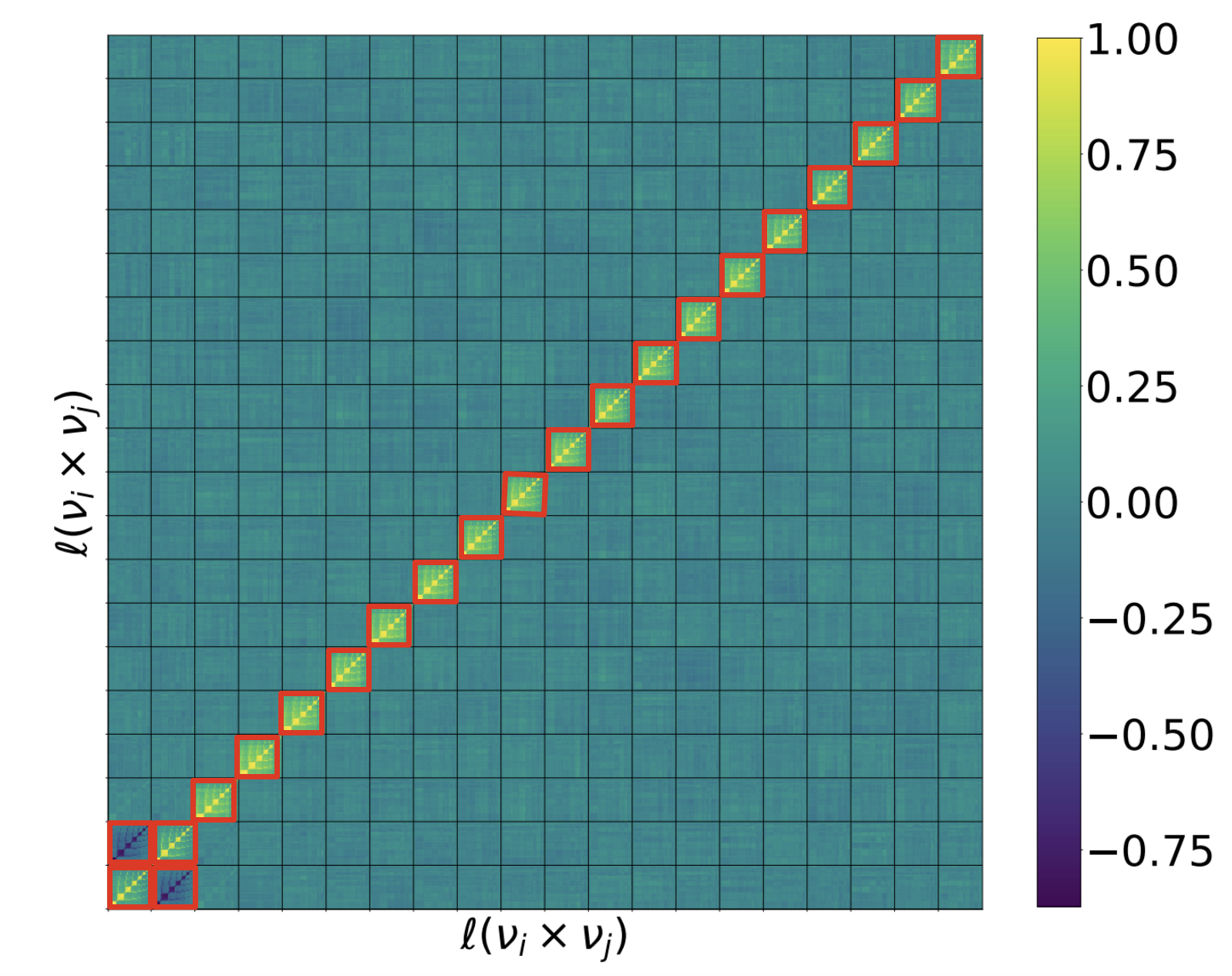}
    \caption{\footnotesize Correlation matrix (${\rm Corr}_{\ell\ell'} \equiv \mathbb{C}_{\ell\ell'}/\sqrt{\mathbb{C}_{\ell\ell}\mathbb{C}_{\ell'\ell'}}$) for the $N_{\rm sim}$ simulations in {\tt d1c}. Every block represents a value of $\ell$ and contains the ordered $N_{\rm cross} = 45$ cross-spectra. The red squares represent the truncation of the full covariance matrix applied in the analysis (kept entries in red, other entries set to zero).}
    \label{fig:cov}
\end{figure}

\noindent where $N_{\rm d.o.f.}$ is the number of degrees of freedom and $\mathbb{C}$ is the covariance matrix of our $N_{\rm sim}$ simulations, represented in Fig.~\ref{fig:cov}, of dimension $(N_{\ell}\cdot N_{\rm cross})^2$:
\begin{equation}
    \mathbb{C}_{\ell,\ell'}^{i\times j,k\times l} = {\rm cov}\left(\mathcal{D}^{\rm sim}_\ell (\nu_i \times \nu_j),\mathcal{D}^{\rm sim}_{\ell'} (\nu_k \times \nu_l)\right).
\label{eq:cov}
\end{equation}

The entire covariance matrix $\mathbb{C}$ is, in general, not invertible. To avoid this, we kept only the $\ell=\ell'$ block-diagonal of $\mathbb{C}$ with the strongest correlation values\footnote{${\rm Corr}_{\ell\ell'} \equiv \mathbb{C}_{\ell\ell'}/\sqrt{\mathbb{C}_{\ell\ell}\mathbb{C}_{\ell'\ell'}}$}, as well as the $(\ell=6.5,\ell'=16.5)$ off-diagonal blocks showing a significant anti-correlation, as illustrated in Fig.~\ref{fig:cov}. It was then possible to invert the thus-defined truncated correlation matrix  with the required precision most of the time. 

In the case of the {\tt d1} simulation type, we experienced a fit convergence issue for $\sim20$\,\% of the simulations, leading to a very large $\chi^2$. In order to overcome this problem, two options lead to identical results: throwing away the outliers from the analysis or fitting using only the block-diagonal matrix (i.e., the $\ell=6.5$, $\ell'=16.5$ block is set to zero). This last option solves the conversion issue while providing sufficient precision. The results presented in the following are using the block-diagonal matrix when the simulation type is {\tt d1}.

Finally, in Eq.~\ref{eq:chi2}, $\vec{R}$ is the residual vector associated with every simulation of size $N_{\ell}\times N_{\rm cross}$:

\begin{equation}
\vec{R} =
\begin{pmatrix}
\mathcal{R}_{\ell = 6.5}(100 \times 100) \\
\mathcal{R}_{\ell = 6.5}(100 \times 119) \\
\vdots \\
\mathcal{R}_{\ell = 16.5}(100 \times 100)\\
\vdots \\
\mathcal{R}_{\ell = 196.5}(402 \times 402)\\
\end{pmatrix}, 
\end{equation}

\noindent with $\mathcal{R}_\ell (\nu_i \times \nu_j) = \mathcal{D}^{\rm sim}_\ell (\nu_i \times \nu_j) - \mathcal{D}^{\rm model}_\ell(\nu_i \times \nu_j)$.

The expression used for the model to fit is given by:

\begin{equation}
\begin{split}
  \mathcal{D}_{\ell}^{\rm model}(\nu_i \times \nu_j) &= \mathcal{D}_{\ell}^{{\rm dust}}\left(\beta_0(\ell), T_0(\ell), \mathcal{D}^{\mathcal{M}\times\mathcal{N}}_{\ell}(\nu_i\times\nu_j)\right) \\ & + A_{\rm lens} \cdot \mathcal{D}_{\ell}^{{\rm lensing}} + r \cdot  \mathcal{D}_{\ell}^{{\rm tensor}}, 
\label{eq:model}
\end{split}
\end{equation}

\noindent  where $A_{\rm lens}$ is not a free parameter and will remain fixed to zero (when there is no CMB, simulation types {\tt d0}, {\tt d1T,} and {\tt d1}) or one (when the CMB is included, simulation types {\tt d0c}, {\tt d1Tc} and {\tt d1c}). We leave the question of the impact of dust modeling with moments on the lensing measurement for future work. In Eq.~\ref{eq:model}, the free parameters can thus be $\beta_{0}(\ell)$, $T_{0}(\ell)$, and $\mathcal{D}^{\mathcal{M}\times\mathcal{N}}_{\ell}(\nu_i\times\nu_j)$ and the tensor-to-scalar ratio $r$. The estimated value of $r$ is referred to as $\hat{r}$

No priors on the parameters are used in order to explore the parameter space with minimal assumptions. Finally, a frequency-dependent conversion factor is included in $\mathcal{D}_{\ell}^{{\rm dust}}$ -- from (MJy$\cdot$sr$^{-1})^2$ to $(\mu$K$_{\rm CMB})^2$ -- to express the dust spectra in ($\mu$K$_{\rm CMB})^2$ units. In those units, $\mathcal{D}_{\ell}^{{\rm lensing}}$ and $\mathcal{D}_{\ell}^{\rm tensor}$ are frequency-independent.

To mitigate the impact of outliers in our simulations, all the final values of the best-fit parameters and $\chi^2$ distributions are represented by their median and median absolute deviations over $N_{\rm sim}$ values. For the tensor-to-scalar ratio $\hat{r}$, we chose to represent all the best-fit values from the $N_{\rm sim}$ simulations in a histogram and we assume its distribution is normal. Fitting a Gaussian curve on this histogram and getting the mean and standard deviation gives us the final values of $\hat{r}$ and $\sigma_{\hat{r}}$ presented in the paper.

\subsection{Implementation for the dust component}
\label{sec:fit_general}

For the dust component, we consider four different \emph{fitting schemes}, corresponding to four expressions for the dust model $\mathcal{D}^{\rm dust}_\ell$ in Eq.~\ref{eq:model}, which are referred to as "MBB", "$\beta$-1", "$\beta$-$T$", and "$\beta$-2". Each of them corresponds to a truncation of Eq.~\ref{eq:moments}, keeping only some selected terms of the moment expansion: MBB stands for those of the modified black body, $\beta$-1 for those of the expansion in $\beta$ at first order, $\beta$-2 for the expansion in $\beta$ at second order, and $\beta$-$T$ for the expansion in both $\beta$ and $T$ at first order. We chose the $\beta$-1 and $\beta$-2 truncations based on the studies of \citet{Mangilli} and \citet{Azzoni}, where the dust SED moment expansion is performed only with respect to $\beta$. The $\beta$-$T$ fitting scheme is instead the first-order truncation in both $\beta$ and $T$, introduced here for the first time at the power spectrum level. The parameters fitted in each of these fitting schemes are summarized in Table~\ref{tab:parameter}. We note that the $\beta$-2 and $\beta$-$T$ fitting schemes share the same number of free parameters. Finally, when we fit $\hat{r}$ at the same time as the dust parameters, the fitting schemes will be referred to as $r$MBB, $r\beta$-1, $r\beta$-$T,$ and $r\beta$-2.

Different physical processes are expected to occur at different angular scales, leading to different SED properties. Thus, we estimate the dust-related parameters with one parameter per multipole bin. As an example, we estimate $\beta_0 = \beta_0(\ell)$ and $T_0 = T_0(\ell)$ to be able to take into account their scale dependence, at the cost of increasing the number of free parameters in our model. This is also true for the higher order moments. On the other hand, $\hat{r}$ is not scale dependent and, when it is fitted, we add one single parameter over the whole multipole range.

In \cite{Mangilli}, the first-order moment expansion parameter $\mathcal{D}_\ell^{A\times\omega^\beta_1}$ is considered to be  the leading order correction to the MBB spectral index. We applied a similar approach in the present work, extending it to the dust temperature when it is fitted. In our pipeline, we proceed iteratively:

\begin{enumerate}
\item(i)  we fit $\beta_0(\ell)$ and $T_0(\ell)$ at order zero (MBB), for each $\ell$, 

\item (ii) we fix $\beta_0(\ell)$ and $T_0(\ell)$ and fit the higher order parameters, as in Eq.~\ref{eq:moments}, (iii) we update the $\beta_0(\ell)$ to $\beta_{\rm corr}(\ell)$ {(and $T_0(\ell)$ to $T_{\rm corr}(\ell)$ in the case of $\beta$-$T$)} as:

\begin{equation}
    \beta_{\rm corr}(\ell) = \beta_0(\ell) + \frac{\mathcal{D}^{A \times \omega^{\beta}_1 }_{\ell}}{\mathcal{D}^{A \times A}_{\ell}},\quad T_{\rm corr}(\ell) = T_0(\ell) + \frac{\mathcal{D}^{A \times \omega^{T}_1 }_{\ell}}{\mathcal{D}^{A \times A}_{\ell}},
\label{eq:iteration}
\end{equation}
\item[iv)] and we iterate from (ii) fixing $\beta_0(\ell)=\beta_{\rm corr}(\ell)$, until $\mathcal{D}_\ell^{A \times \omega^{\beta}_1}$ converges to be compatible with zero (and $T_0(\ell)=T_{\rm corr}(\ell)$, until $\mathcal{D}_\ell^{A \times \omega^{T}_1}$ converges to zero in the case of $\beta$-$T$).
\end{enumerate}
We used three such iterations, which we found to be sufficient to guarantee the convergence. As the moment expansion is a nonorthogonal and incomplete basis \citep{Chluba}, this iterative process is performed to ensure that the expansions up to different orders share the same $\beta_0(\ell)$ and $T_0(\ell)$ with $\mathcal{D}^{A \times \omega^{\beta}_1 }_\ell=0$ and $\mathcal{D}^{A \times \omega^{T}_1 }_\ell=0$.

\begin{table}[t!]
\centering
        \normalsize
        \begin{tabular}{c!{\color{white}\vrule width 1pt}c!{\color{white}\vrule width 1pt}c!{\color{white}\vrule width 1pt}c!{\color{white}\vrule width 1pt}c}
        & MBB & $\beta$-1 & $\beta$-$T$ & $\beta$-2\\
        \hwline
        $N_{\rm param.}$ & \cellcolor{mygrey} $3N_\ell$& \cellcolor{mygrey} $2N_\ell$ & \cellcolor{mygrey} $5N_\ell$ & \cellcolor{mygrey} $5N_\ell$ \\\hwline
        $\beta_0(\ell)$ & \cellcolor{mygreen}\checkmark& \cellcolor{myyellow}$\circ$ & \cellcolor{myyellow}$\circ$ & \cellcolor{myyellow}$\circ$\\\hwline
        $T_0(\ell)$ & \cellcolor{mygreen}\checkmark& \cellcolor{myred}$\times$ & \cellcolor{myyellow}$\circ$ & \cellcolor{myred}$\times$\\\hwline
        $\mathcal{D}_\ell^{A\times A}$ & \cellcolor{mygreen}\checkmark& \cellcolor{myred}$\times$ & \cellcolor{myred}$\times$ & \cellcolor{myred}$\times$\\\hwline
        $\mathcal{D}_\ell^{A\times\omega_1^\beta}$ & \cellcolor{myred}$\times$ & \cellcolor{mygreen}\checkmark & \cellcolor{mygreen}\checkmark & \cellcolor{mygreen}\checkmark\\\hwline
        $\mathcal{D}_\ell^{\omega_1^\beta\times\omega_1^\beta}$ & \cellcolor{myred}$\times$& \cellcolor{mygreen}\checkmark & \cellcolor{mygreen}\checkmark & \cellcolor{mygreen}\checkmark\\\hwline
        $\mathcal{D}_\ell^{A\times\omega_1^T}$ & \cellcolor{myred}$\times$& \cellcolor{myred}$\times$ & \cellcolor{mygreen}\checkmark &\cellcolor{myred}$\times$\\\hwline
        $\mathcal{D}_\ell^{\omega_1^T\times\omega_1^T}$ & \cellcolor{myred}$\times$& \cellcolor{myred}$\times$ & \cellcolor{mygreen}\checkmark & \cellcolor{myred}$\times$\\\hwline
        $\mathcal{D}_\ell^{\omega_1^\beta\times\omega_1^T}$ & \cellcolor{myred}$\times$ & \cellcolor{myred}$\times$ & \cellcolor{mygreen}\checkmark & \cellcolor{myred}$\times$\\\hwline
        $\mathcal{D}_\ell^{A\times\omega_2^\beta}$ & \cellcolor{myred}$\times$ & \cellcolor{myred}$\times$ & \cellcolor{myred}$\times$ & \cellcolor{mygreen}\checkmark\\\hwline
        $\mathcal{D}_\ell^{\omega_1^\beta\times\omega_2^\beta}$ & \cellcolor{myred}$\times$ & \cellcolor{myred}$\times$ & \cellcolor{myred}$\times$ & \cellcolor{mygreen}\checkmark\\\hwline
        $\mathcal{D}_\ell^{\omega_2^\beta\times\omega_2^\beta}$ & \cellcolor{myred}$\times$ & \cellcolor{myred}$\times$ & \cellcolor{myred}$\times$ & \cellcolor{mygreen}\checkmark\\\hwline
    \end{tabular}
    \normalsize
\caption{\footnotesize Summary of the fitted parameters in the four dust moment expansion \emph{fitting schemes} we consider (MBB, $\beta$-1, $\beta$-$T,$ and $\beta$-2), in Eq.~\ref{eq:moments}. A tick on a green background signifies that the parameter is fitted, red with a cross symbol shows that the parameter is not fitted, and a circle symbol on yellow means that the parameter is fixed and corrected through an iterative process as presented in Sect.~\ref{sec:fit_general}. $\mathcal{D}_\ell^{A\times A}$ is fixed to the MBB best-fit value in the case of $\beta$-1, $\beta$-$T,$ and $\beta$-2 and all the other moments are set to zero when they are not fitted. When $\hat{r}$ is fitted at the same time, the fitting schemes are denoted $r$MBB, $r\beta$-1, $r\beta$-$T,$ and $r\beta$-2, and they have one more parameter than the number of parameters reported in the first line.}
\label{tab:parameter}
\end{table}

\section{\label{sec:results} Results}

\begin{figure}[t!]
    \centering
    \includegraphics[width=\columnwidth]{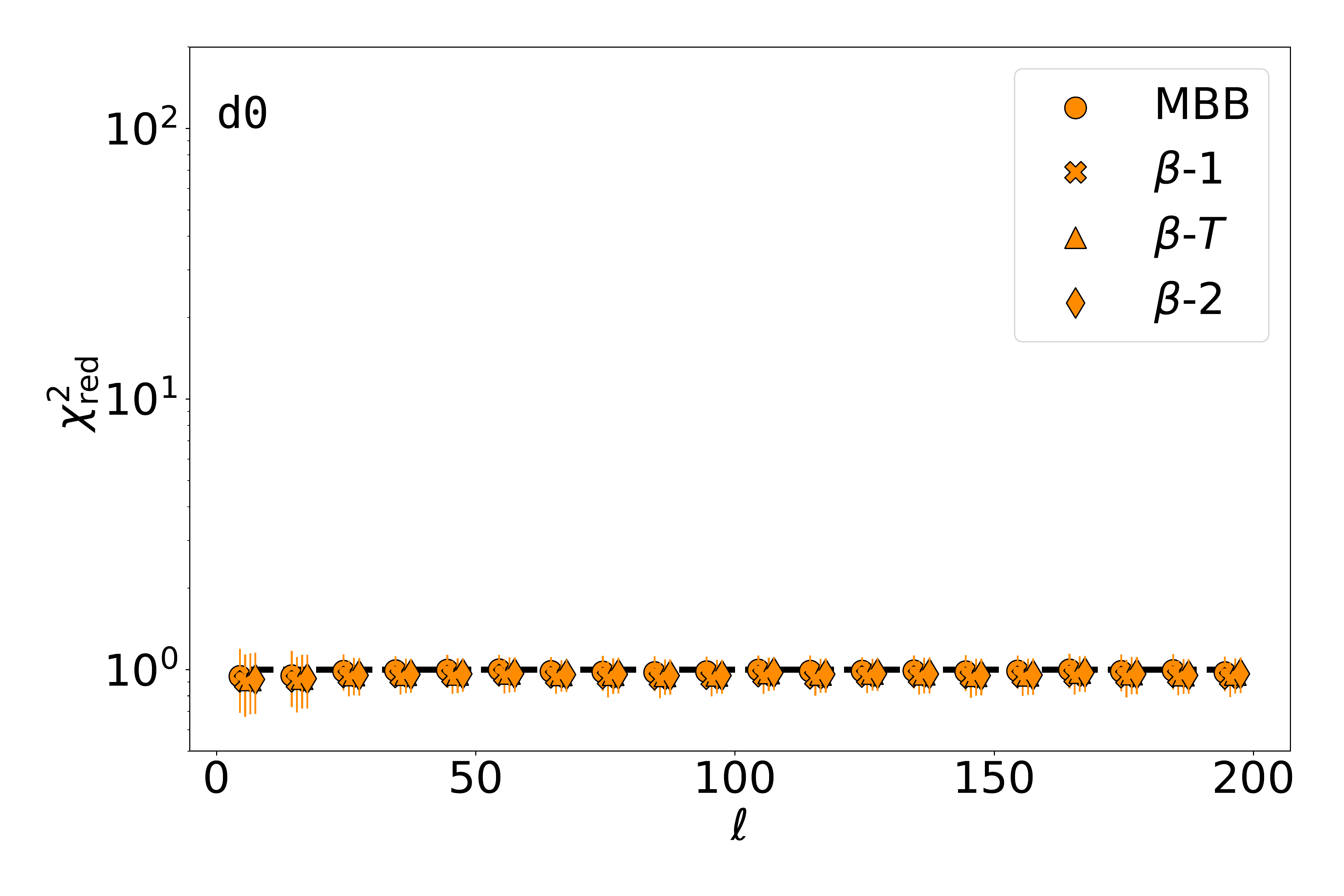}
    \includegraphics[width=\columnwidth]{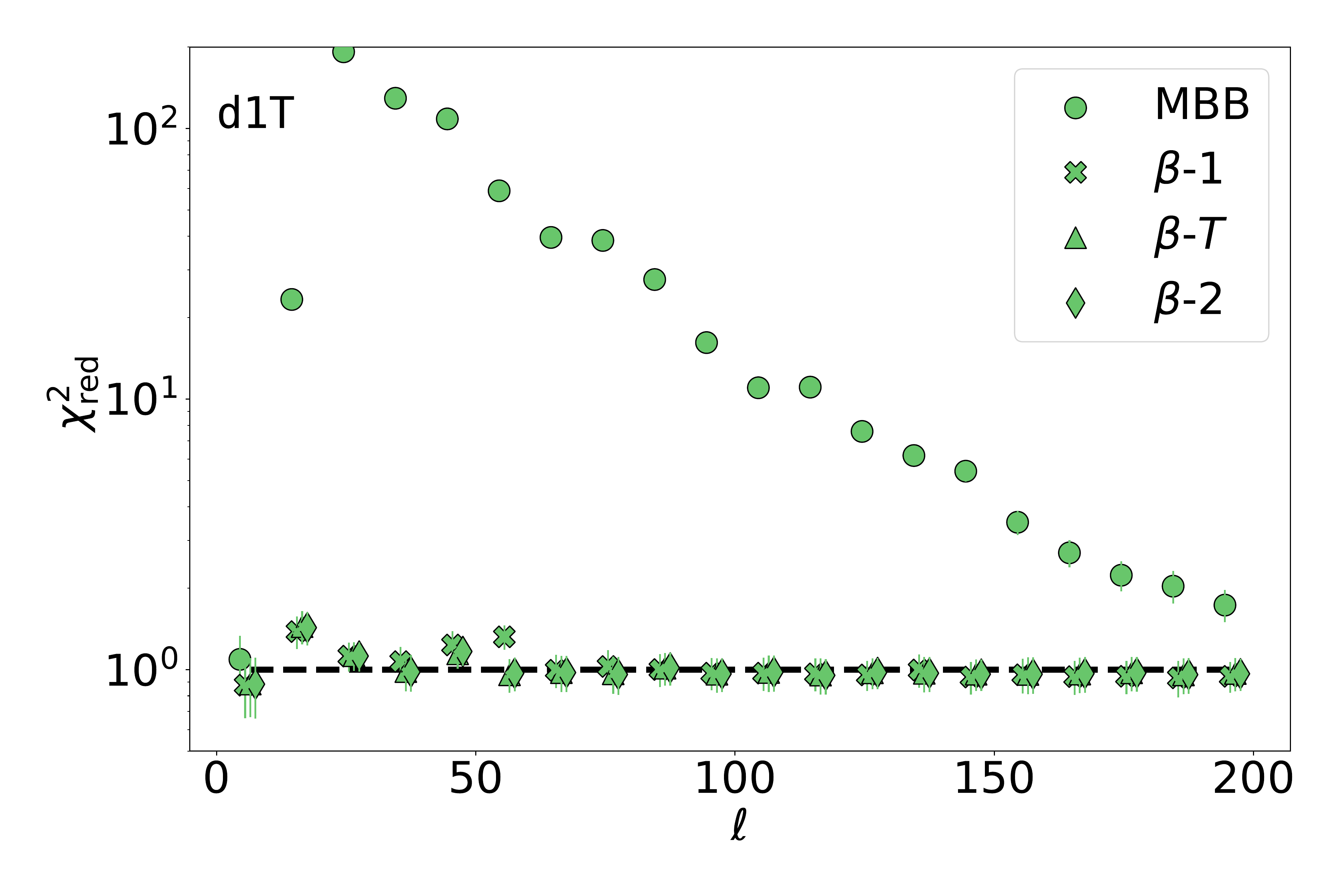}
    \includegraphics[width=\columnwidth]{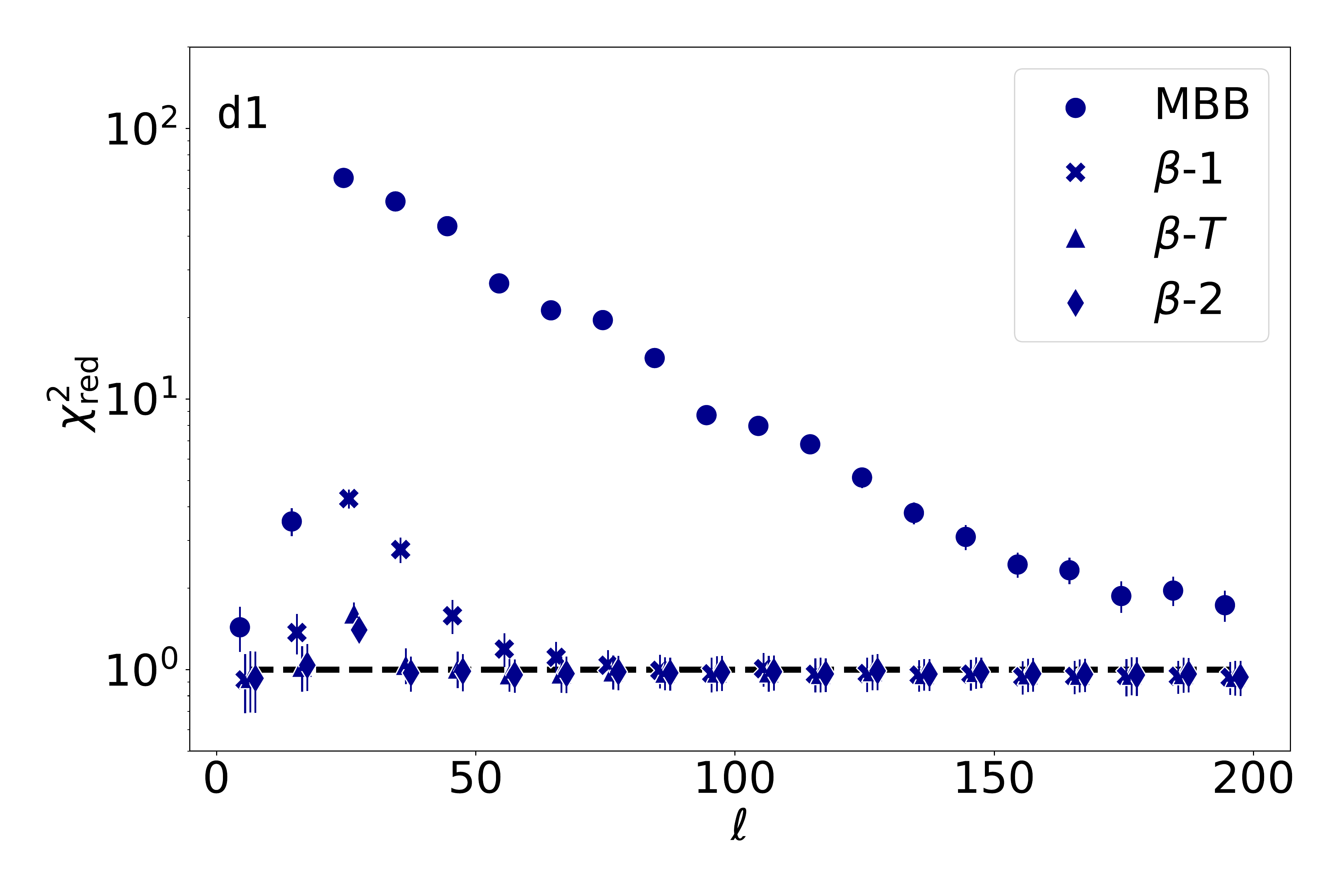}
    \caption{\footnotesize Median of the reduced $\chi^2$ in every multipole bin $\ell$, for all the $N_{\rm sim}$ simulations of {\tt d0} (top, orange), {\tt d1T} (middle, green) and {\tt d1} (bottom, blue), on $f_{\rm sky}=0.7$. The reduced $\chi^2$ values are reported for the four different fitting schemes: MBB (circles), $\beta$-1 (crosses), $\beta$-$T$ (diamonds) and $\beta$-2 (triangles). The values for the four fitting schemes are shifted from each others by $\ell=2$, in order to distinguish them. The black dashed line represents $\chi^2_{\rm red}=1$.}
    \label{fig:chi2dust}
\end{figure}

In this section, we present our evaluation of the best-fit parameters for the different {fitting schemes} presented in Sect.~\ref{sec:fit_dust} on the $B$-mode cross-angular power spectra computed from the different {simulation types} presented in Sect.~\ref{sec:simulations} and on the Galactic mask keeping $f_{\rm sky}=0.7$, which is defined in Sect.~\ref{sect:mask}.
We first tested the simulation types containing only dust and noise in order to calibrate the dust complexity of our data sets in Sect.~\ref{sec:results_dust_only}. We then used CMB only plus noise simulations to assess the minimal error on $\hat{r}$ in Sect.~\ref{sec:c} and, finally, we explored the dust, CMB, and noise simulation types to assess the impact of the dust complexity on $\hat{r}$ in Sect.~\ref{sec:results_dust_CMB}.

\subsection{Dust only}
\label{sec:results_dust_only}

To evaluate the amplitude of the dust moment parameters contained in the dust simulations in the absence of CMB, we ran the fitting schemes presented in Sect.~\ref{sec:fit_dust} in the three simulation types {\tt d0}, ${\tt d1T,}$ and ${\tt d1}$ presented in Sect.~\ref{sec:simulations}. In these cases, $A_{\rm lens}$ and $r$ in Eq.~\ref{eq:model} are both fixed to zero and the fitted parameters are given  in Table~\ref{tab:parameter} for every fitting scheme.

\subsubsection{{\tt d0}}
\label{sec:d0}

The {\tt d0} dust maps presented in Sect.~\ref{sec:ingredients} extrapolate between frequency bands with a MBB SED with constant parameters over the sky: $\beta_{\tt d0} = 1.54$ and $T_{\tt d0} = 20$\,K. We performed the fit with the four fitting schemes presented in Sect.~\ref{sec:fit_dust}.

\begin{figure}[t!]
    \centering
    \includegraphics[width=\columnwidth]{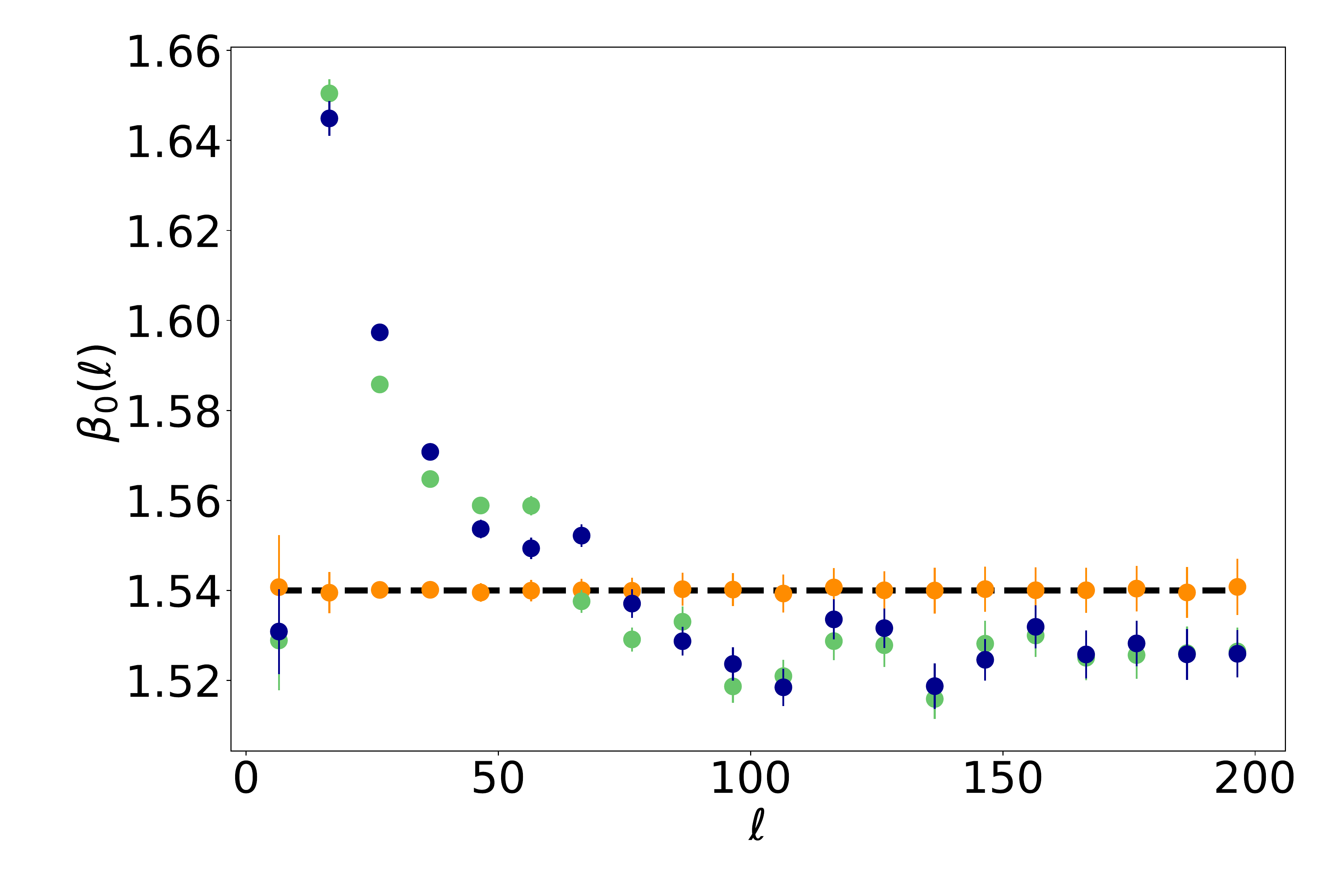}
    \includegraphics[width=\columnwidth]{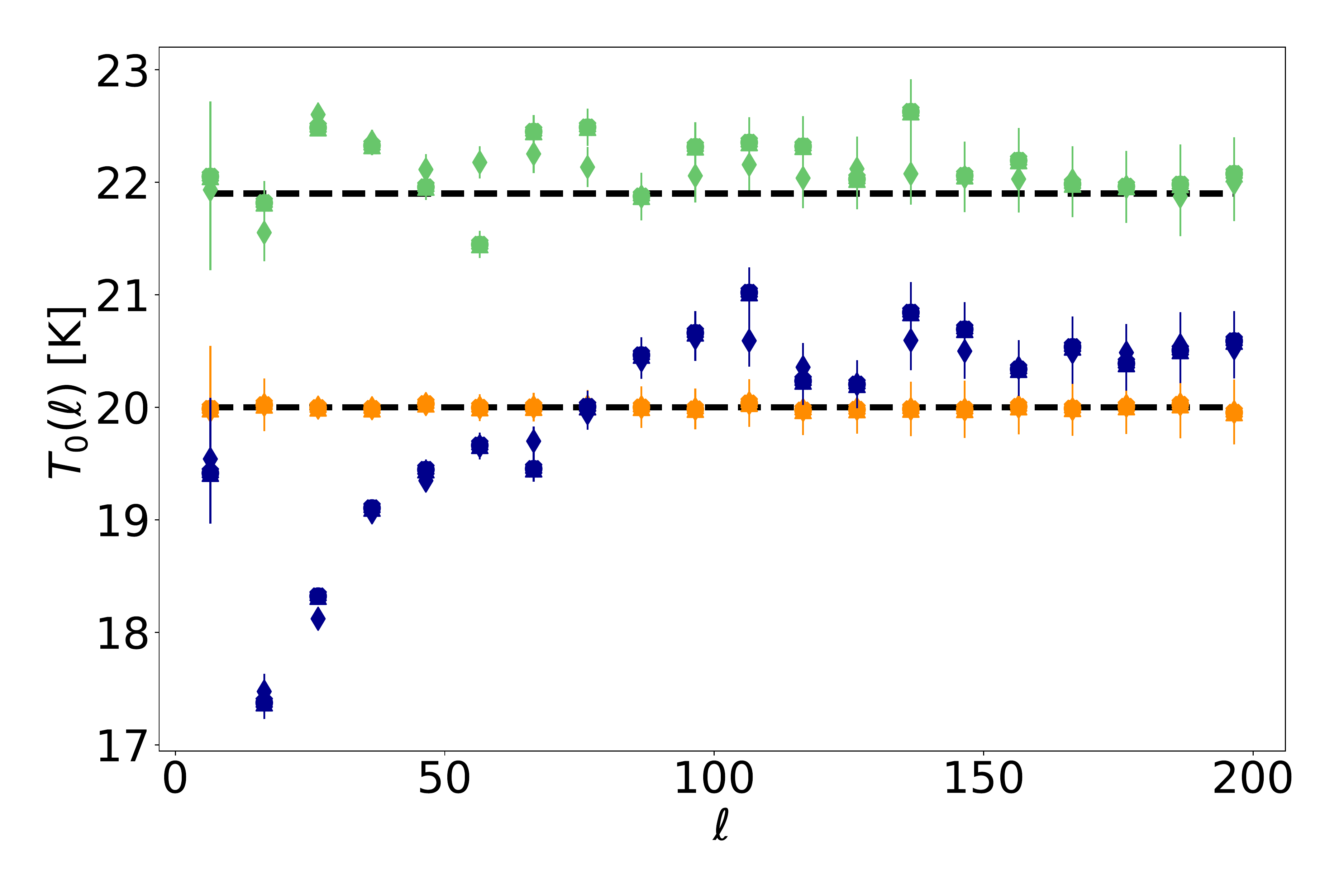}
    \caption{\footnotesize (Top): Median of the best fit values of $\beta_0(\ell)$ in {\tt d0} (orange), {\tt d1T} (green), and {\tt d1} (blue) for the MBB (circles). $\beta_{\tt d0}$ is marked by the dashed  black line. (Bottom): Same as above but with $T_0(\ell)$, the black dashed-lines being $T_{\tt d0}=20$\,K and $T_{\tt d1T}=21.9$\,K.}
    \label{fig:betaTdust}
\end{figure}

In Fig.~\ref{fig:chi2dust} the values of the reduced $\chi^2(\ell)$ for each fitting scheme are displayed. For every fitting scheme (MBB, $\beta$-1, $\beta$-$T$ and $\beta$-2), the reduced $\chi^2$ are close to 1 over the whole multipole range (slightly below 1 for the $\beta$-1, $\beta$-$T$ and $\beta$-2 fitting scheme). This indicates that the MBB is a good fit to the cross-angular power spectra computed from the  {\tt d0} maps with a spatially invariant MBB SED, as expected. Adding additional (higher order) parameters, such as with $\beta$-1, $\beta$-$T$ and $\beta$-2, has no significant effect on the $\chi^2$.

In Fig.~\ref{fig:betaTdust} we can see that the best-fit values of $\beta_0(\ell)$ and $T_0(\ell)$ are compatible with constant values $\beta_0(\ell)=\beta_{\tt d0}$ and $T_0(\ell)=T_{\tt d0}$, as expected for this simulated data set.

\begin{figure}[t!]
    \centering
    \includegraphics[width=\columnwidth]{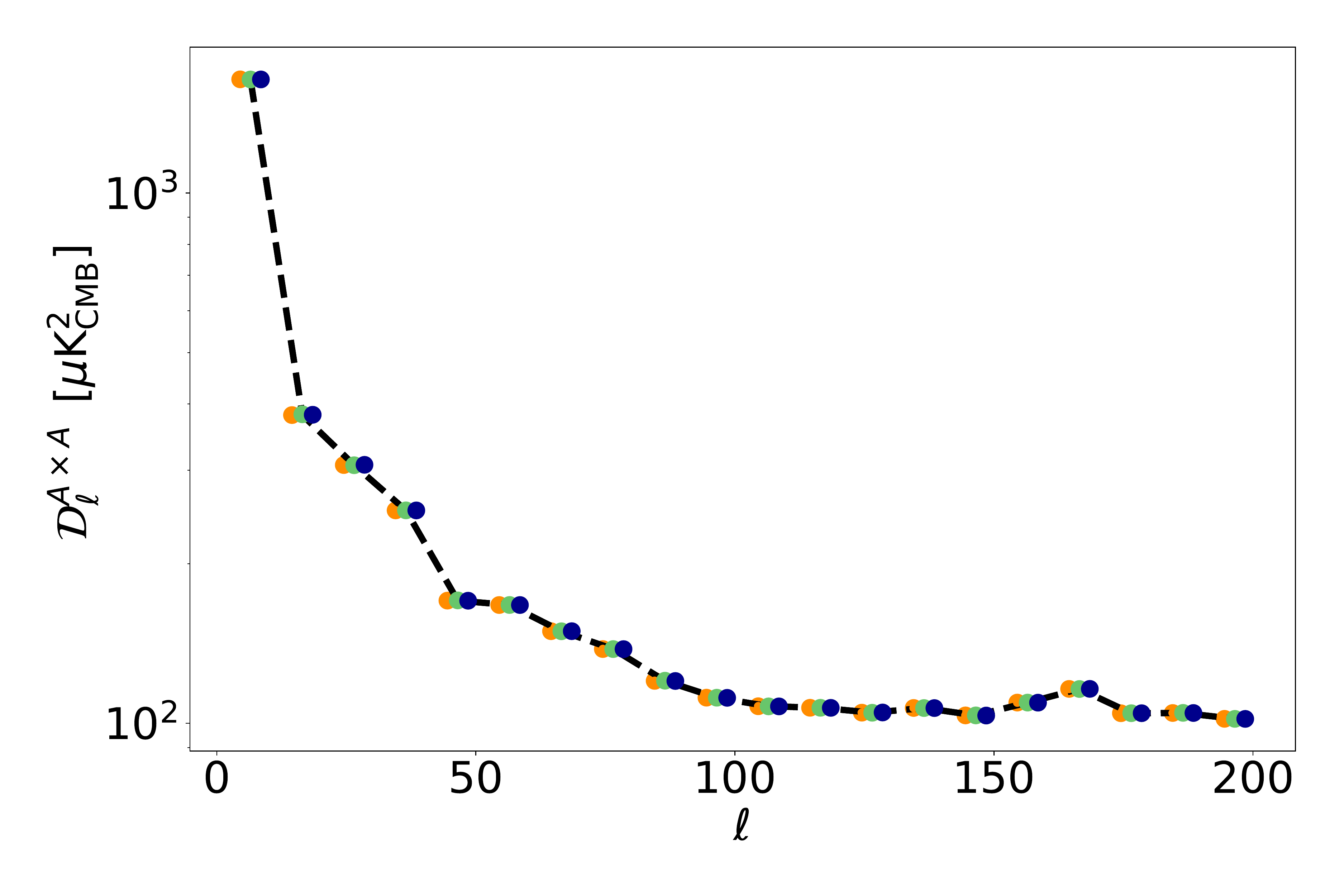}
    \caption{\footnotesize Median of the best-fit values of $\mathcal{D}_\ell^{A \times A}$ for {\tt d0} (orange), {\tt d1T} (green), and {\tt d1} (blue) using the MBB fitting scheme. The values for the three simulation types are shifted with respect to one another by $\ell=2$ in order to distinguish them. The black dashed line is the amplitude power spectrum of the dust template map used to build the three simulation sets {\tt d0}, {\tt d1T,} and  {\tt d1}.}
    \label{fig:Adust}
\end{figure}

\begin{figure}[t!]
    \centering
    \includegraphics[width=\columnwidth]{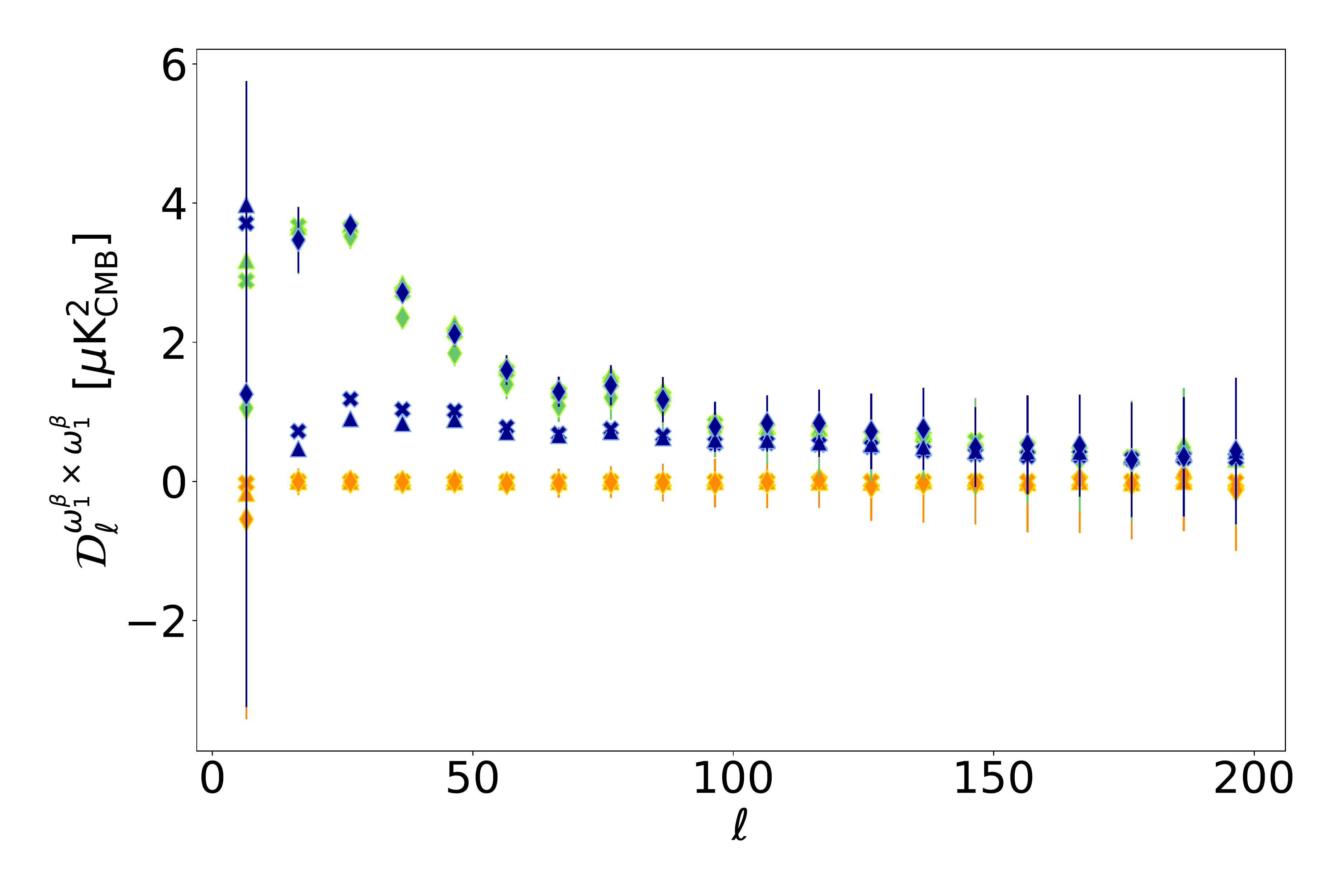}
    \caption{\footnotesize Best-fit values of the first-order moment $\mathcal{D}_\ell^{\omega^{\beta}_1\times\omega^{\beta}_1}$ for {\tt d0} (orange), {\tt d1T} (green), and {\tt d1} (blue), fitting with $\beta$-1 (crosses), $\beta$-2 (triangles), and $\beta$-$T$ (diamonds).}
    \label{fig:moments1}
\end{figure}

\begin{figure}[t!]
    \centering
    \includegraphics[width=\columnwidth]{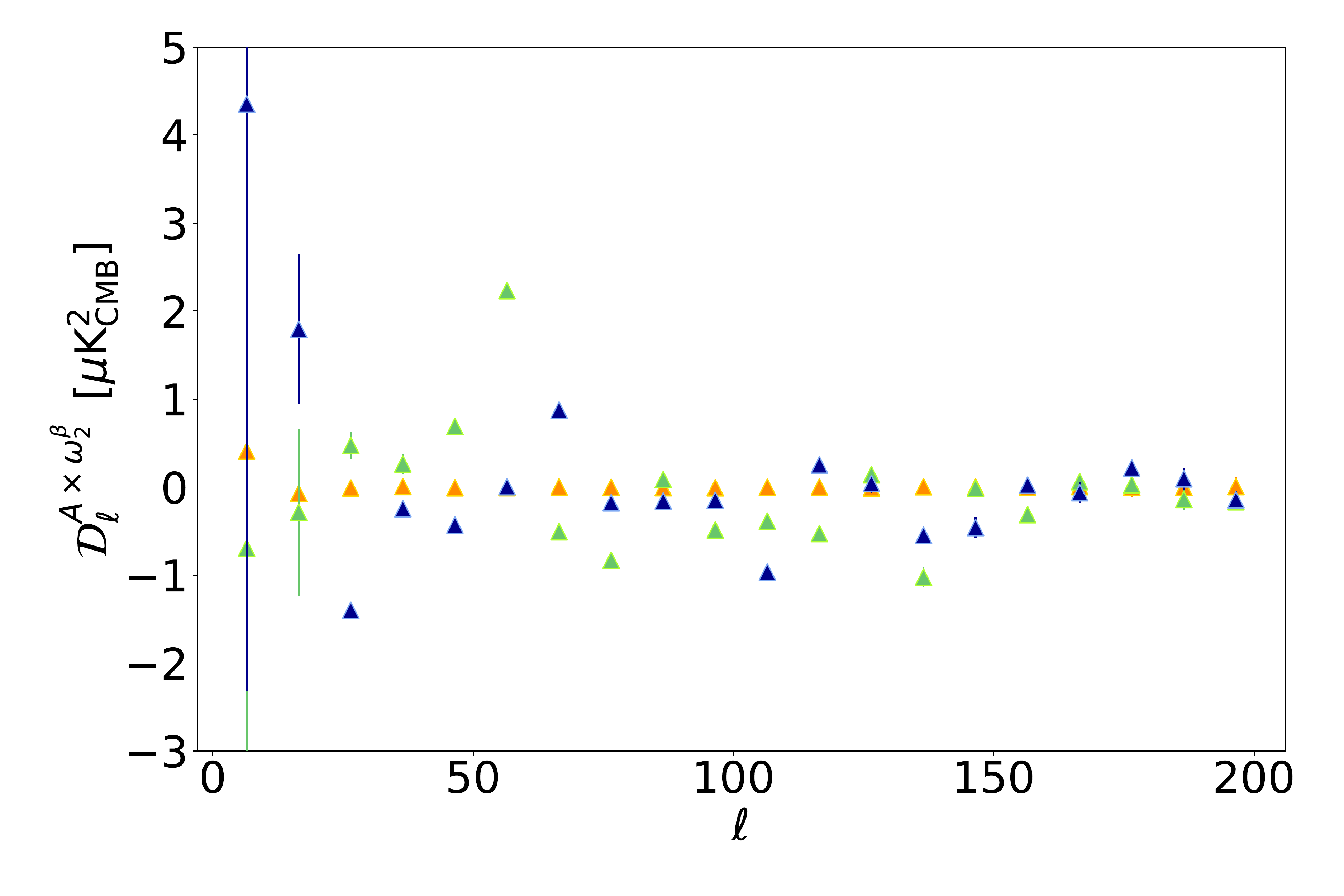}
    \includegraphics[width=\columnwidth]{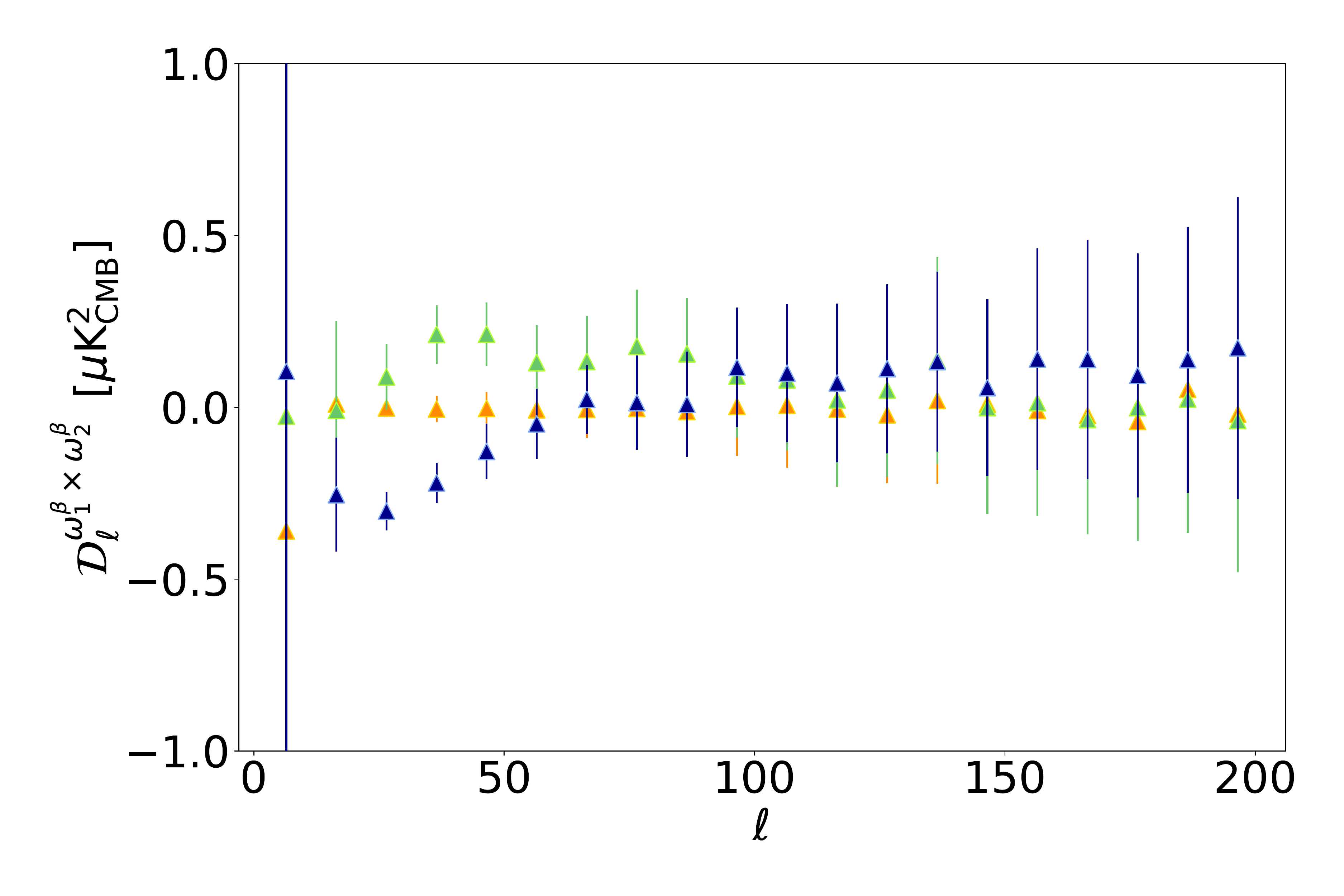}
    \includegraphics[width=\columnwidth]{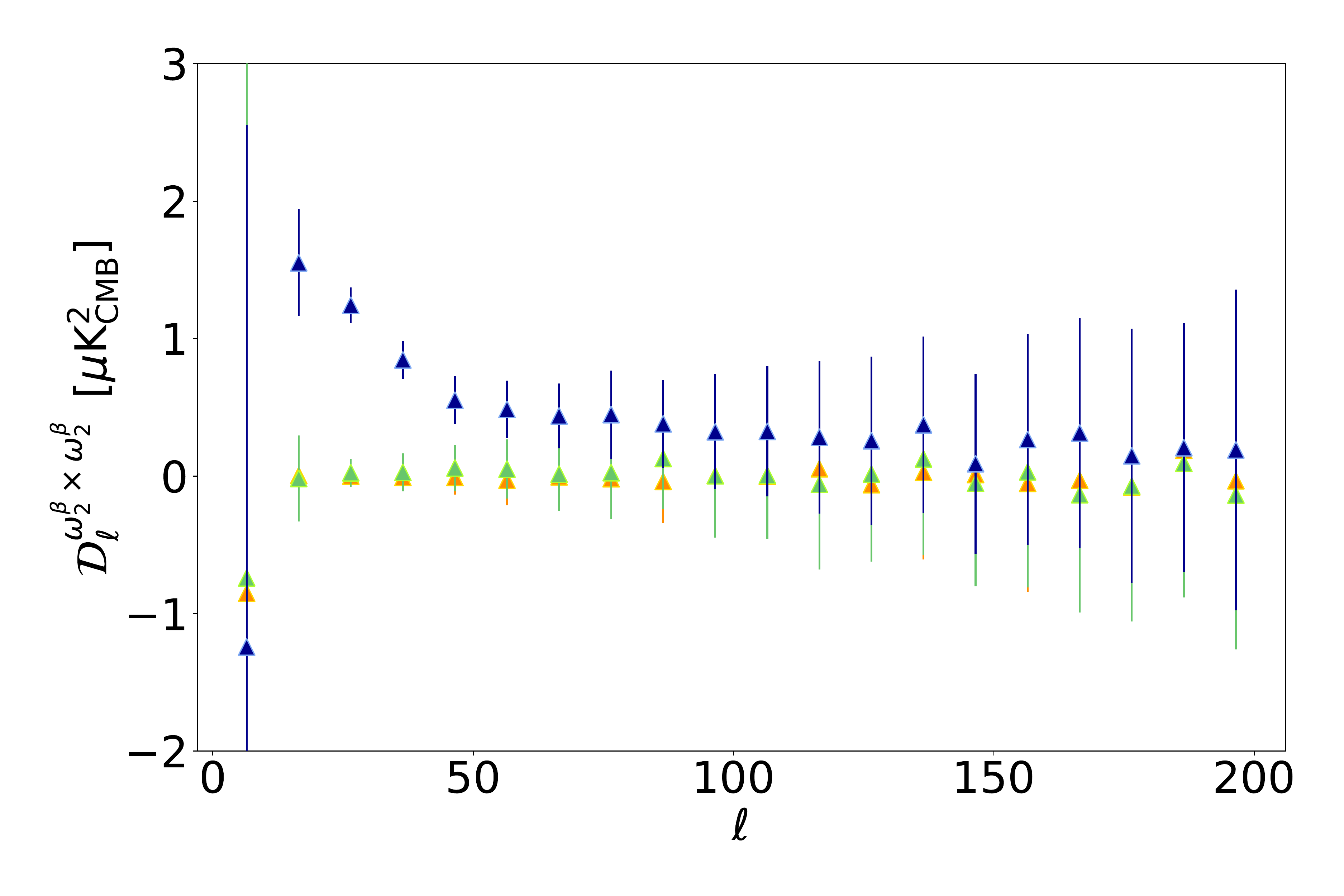}
    \caption{\footnotesize Best-fit values of the second-order $\mathcal{D}_\ell^{A\times\omega^{\beta}_2}$, $\mathcal{D}_\ell^{\omega^{\beta}_1\times\omega^{\beta}_2}$ and $\mathcal{D}_\ell^{\omega^{\beta}_2\times\omega^{\beta}_2}$ moment parameters in {\tt d0} (orange), {\tt d1T} (green), and {\tt d1} (blue) for $\beta$-2 (triangles). }
    \label{fig:moments2}
\end{figure}

\begin{figure}[t!]
    \centering
    \includegraphics[width=\columnwidth]{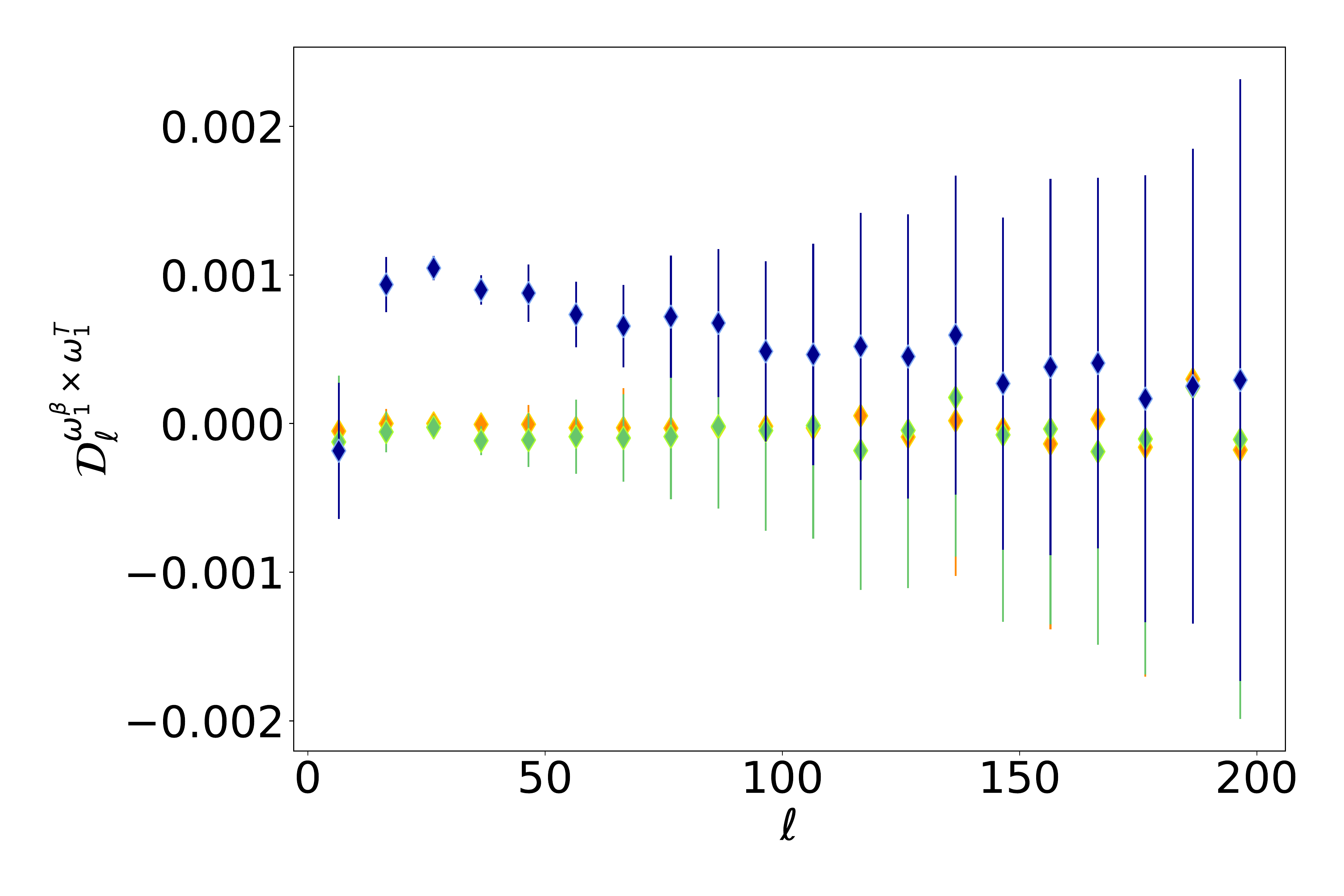}
    \includegraphics[width=\columnwidth]{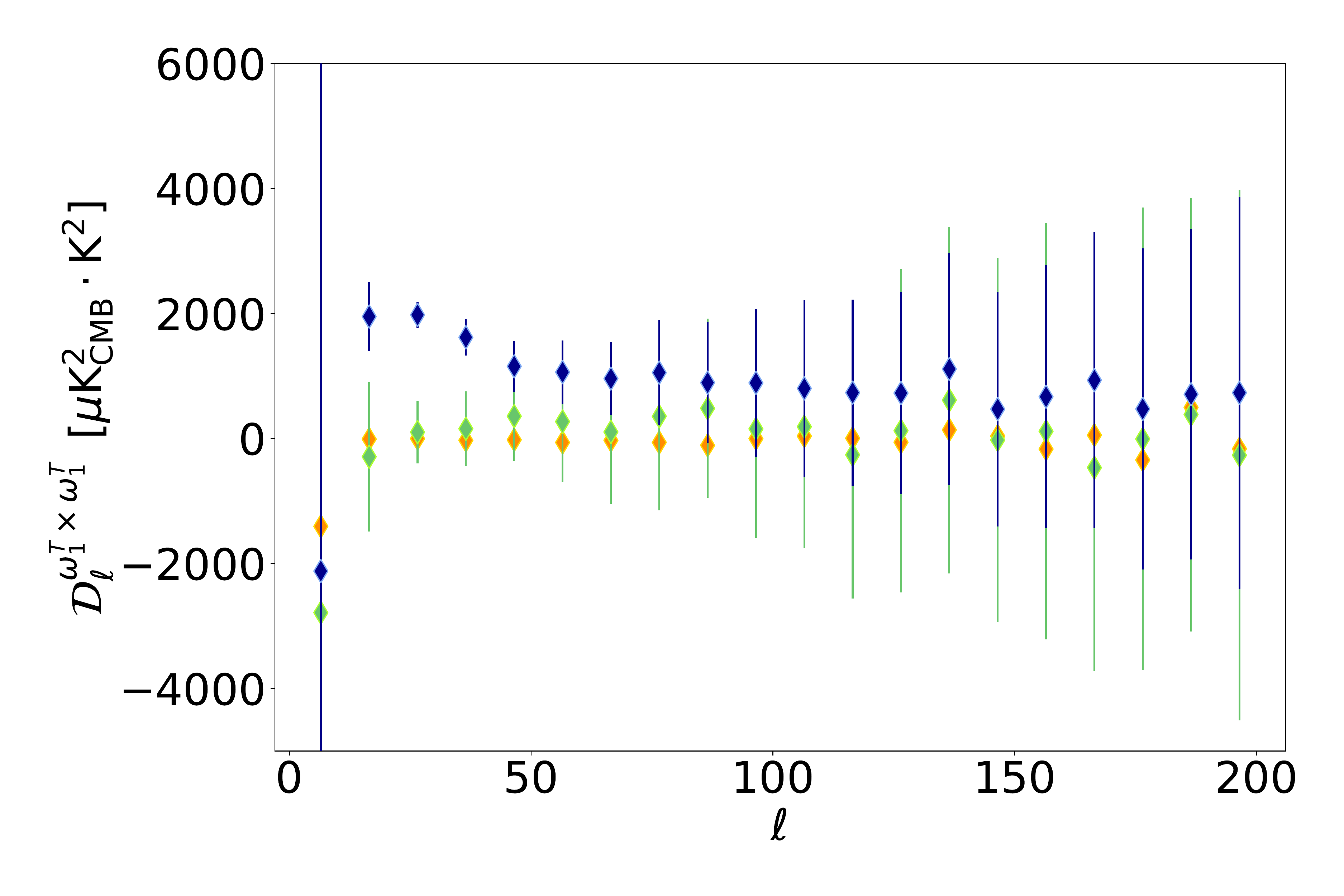}
    \caption{\footnotesize Best-fit values of the first-order  $\mathcal{D}_\ell^{\omega^{\beta}_1\times\omega^{T}_1}$ and $\mathcal{D}_\ell^{\omega^{T}_1\times\omega^{T}_1}$ moment parameters in {\tt d0} (orange), {\tt d1T} (green), and {\tt d1} (blue) for $\beta$-$T$ (diamonds). }
    \label{fig:moments3}
\end{figure}

The best-fit values of the dust amplitude and the moment-expansion parameters are presented in Figs.~\ref{fig:Adust}, \ref{fig:moments1}, \ref{fig:moments2}, and \ref{fig:moments3}, respectively. The amplitude power spectrum is compatible with that of the dust template map used to build {\tt d0} and the moment-xpansion parameters are compatible with zero for every fitting scheme, as expected with no spatial variation of the SED. 
Therefore, the moment expansion method presented in Sect.~\ref{sec:formalism} passes the \emph{null test} in the absence of SED distortions, with the {\tt d0} simulated data set.  


\subsubsection{{\tt d1T}}
\label{sec:d1T}

We now introduce, as a first layer of complexity, the spatial variations of the spectral index associated to a fixed temperature over the sky with the {\tt d1T} simulation type. The dust temperature was fixed to $T_{\rm d1T}=21.9$\,K while the spectral index $\beta(\vec{n})$ was allowed to vary between lines of sight. The four different fitting schemes presented in Sect.~\ref{sec:fit_dust} are fitted over the cross-spectra of our simulations as in Sect.~\ref{sec:d0}.

The reduced $\chi^2(\ell)$ values for each fitting scheme can be found in Fig.~\ref{fig:chi2dust}. It can be seen that the MBB no longer provides a good fit for the dust SED, especially at low multipoles. Averaging effects of spatially varying SEDs are more important over large angular scales and thus SED distortions and moments are expected to be more significant at low multipoles. Indeed, the moments added to the fit in $\beta$-1 are enough to lower the reduced $\chi^2$ such that it becomes compatible with 1 over almost all of the multipole range. The fitting schemes $\beta$-$T$ and $\beta$-2, including more parameters than $\beta$-1, provide a fit of similar goodness, except in the multipole bin $\ell=66.5$ where they are closer to 1.

Figure~\ref{fig:betaTdust} presents the best-fit values of $\beta_0(\ell)$ in the case of the MBB fit. For the sake of clarity, the values after iteration (see Sect.~\ref{sec:fit_dust}) for $\beta$-1, $\beta$-$T,$ and $\beta$-2 are not shown, but they present comparable trends. We can see that the best-fit values of $\beta_0(\ell)$ for this {\tt d1T} simulation type are no longer compatible with a constant. $\beta_0(\ell)$ fitted values show a significant increase at low (<100) multipoles, up to $\beta_0(\ell=16.5)=1.65$. For $\ell>100$, $\beta_0(\ell)$ is close to a constant of value $\sim1.53$. This increase towards the low $\ell$ is correlated to the increase of the MBB $\chi^2$ discussed in the previous paragraph. However, we note that in the lowest $\ell$-bin, the $\beta_0(\ell)$ value is close to 1.53 and that the $\chi^2$ of the MBB fit is close to unity.

The best-fit values of $T_0(\ell)$ are also presented in Fig.~\ref{fig:betaTdust} in the case of the MBB fit. Here again, the values after iteration for the other fitting schemes are not presented, but are similar. The {\tt d1T} $T_0(\ell)$ best-fit values oscillate around $T_{\tt d1T}=21.9$\,K, without being strictly compatible with a constant value, as would be expected for this simulation type. This tends to indicate that the SED distortions due to the spectral index spatial variations are affecting the accuracy at which we can recover the correct angular dependence of the sky temperature. 

The amplitude power spectrum is displayed in Fig.~\ref{fig:Adust} for the MBB fitting scheme. The other fitting scheme results are not presented for clarity and would not be distinguishable from those of the MBB. The fitted $\mathcal{D}_\ell^{A\times A}$ is compatible with the one of the dust template map used to build the simulations. 

All the parameters of the moment expansion with respect to $\beta$ can be found in Figs.~\ref{fig:moments1} and \ref{fig:moments2}, and are now significantly detected, except for $\mathcal{D}_\ell^{\omega^\beta_2 \times \omega^\beta_2}$. In Fig.~\ref{fig:moments3}, we can observe that the parameters of the moment expansion with respect to the temperature (only present in the $\beta$-$T$ fit) remain undetected. The SED distortions due to the spatial variations of $\beta$ are well detected, while no SED distortion linked to the temperature is seen, as expected for the ${\tt d1T}$ simulation type.

\subsubsection{{\tt d1}}
\label{sec:d1}

We now discuss the {\tt d1} simulations, with the highest complexity in the polarized dust SED. In this more physically relevant simulation type, the dust emission is given by a MBB with variable index $\beta(\vec{n})$ and temperature $T(\vec{n})$ over the sky.
We ran the four different fitting schemes on the {\tt d1} simulation type, as we did in Sect.~\ref{sec:d0} and \ref{sec:d1T}.

The values of the reduced $\chi^2(\ell)$ are displayed in Fig.~\ref{fig:chi2dust}. For the MBB and $\beta$-1, the reduced $\chi^2$ are not compatible with unity, especially at low multipole. This indicates that none of them are a good fit anymore for the spatially varying SED with $\beta(\vec{n})$ and $T(\vec{n})$. With $\beta$-2 and $\beta$-$T$, the $\chi^2(\ell)$ values become compatible with unity, except for the $\ell=26.5$ bin. We note that $\beta$-$T$ provides a slightly better fit than $\beta$-2 in this bin.

Looking at the medians of the best-fit values of $\beta_0(\ell)$ for {\tt d1} in Fig.~\ref{fig:betaTdust}, we can see that the spectral index is changing with respect to $\ell$, as discussed in Sect.~\ref{sec:d1T}, in a similar manner as for the {\tt d1T} simulation type. The fitted temperature $T_0(\ell)$ values for {\tt d1} show an increasing trend from $\sim17$ to $\sim20.5$\,K and from $\ell=16.5$ to $\ell\sim100$. At higher multipoles, $T_0(\ell)$ is close to a constant temperature of 20.5\,K. In {\tt d1}, as for {\tt d1T}, the angular scales at which we observe strong variations of $\beta_0(\ell)$ and $T_0(\ell)$ are the ones for which we observe a poor $\chi^2$ for some fitting schemes. Also, as for {\tt d1T}, the largest angular scale $\ell$-bin, at $\ell=6.5,$ shows $\beta$ and $T$ values close to the constant value at high $\ell$, which are associated with $\chi^2$ values closer to unity.
The best-fit values of the amplitude $\mathcal{D}_\ell^{A\times A}$ are shown in Fig.~\ref{fig:Adust}. These are similar to those of the other simulation types. 

The moment-expansion parameters fitted on {\tt d1} are shown in Figs.~\ref{fig:moments1}, \ref{fig:moments2}, and \ref{fig:moments3}. For this simulation type, the moment parameters are all significantly detected with respect to both $\beta$ and $T$. This was already the case with the \planck{} intensity simulations, produced in a similar way, as discussed in \cite{Mangilli}. Their detections quantify the complexity of dust emission and SED distortions from the MBB present in the {\tt d1} simulation type, due to the spatial variations of $\beta(\vec{n})$ and $T(\vec{n})$.

\subsection{\label{sec:nodust} CMB only}
\label{sec:c}

In order to calibrate the accuracy at which the $r$ parameter can be constrained with the \lb{} simulated data sets presented in Sect.~\ref{sec:simulations}, we tested the simulation type with no dust component, $M_\nu^{\tt c}$, and with no tensor modes ($r_{\rm sim}=0$, only CMB lensing and noise). We fit the expression in Eq.~\ref{eq:model} with $\mathcal{D}_\ell^{\rm dust}$ fixed to zero and $A_{\rm lens}$ fixed to one (\ie, $r$ is the only parameter we fit in this case).
Doing so over the $N_{\rm sim}$ simulations, we obtain $\hat{r}= (0.7 \pm 3.5) \times 10^{-4}$.
This sets the minimal value we can expect to retrieve for $\hat{r}$ with our assumptions if the dust component is perfectly taken into account.

\subsection{Dust and CMB}
\label{sec:results_dust_CMB}

We now present our analysis of the simulations including dust, CMB (lensing), and noise ({\tt d0c}, {\tt d1Tc} and {\tt d1c}) with no primordial tensor modes ($r_{\rm sim}=0$). As described above, we applied the four fitting schemes for the dust on the three simulation types, fitting $\hat{r}$ and fixing $A_{\rm lens}$ to~one    (namely $r$MBB, $r\beta$-1, $r\beta$-$T$ and $r\beta$-2) simultaneously. 

The best-fit values of $\beta_0(\ell)$, $T_0(\ell)$ and the moment expansion parameters $\mathcal{D}_\ell^{\mathcal{M}\times\mathcal{N}}$ derived with the simulation types {\tt d0c}, {\tt d1Tc,} and {\tt d1c} are not discussed further when they are compatible with the ones obtained for the {\tt d0}, {\tt d1T,} and {\tt d1} simulation types and presented in Sect.~\ref{sec:results_dust_only}.

\begin{figure}[t!]
    \centering
    \includegraphics[width=\columnwidth]{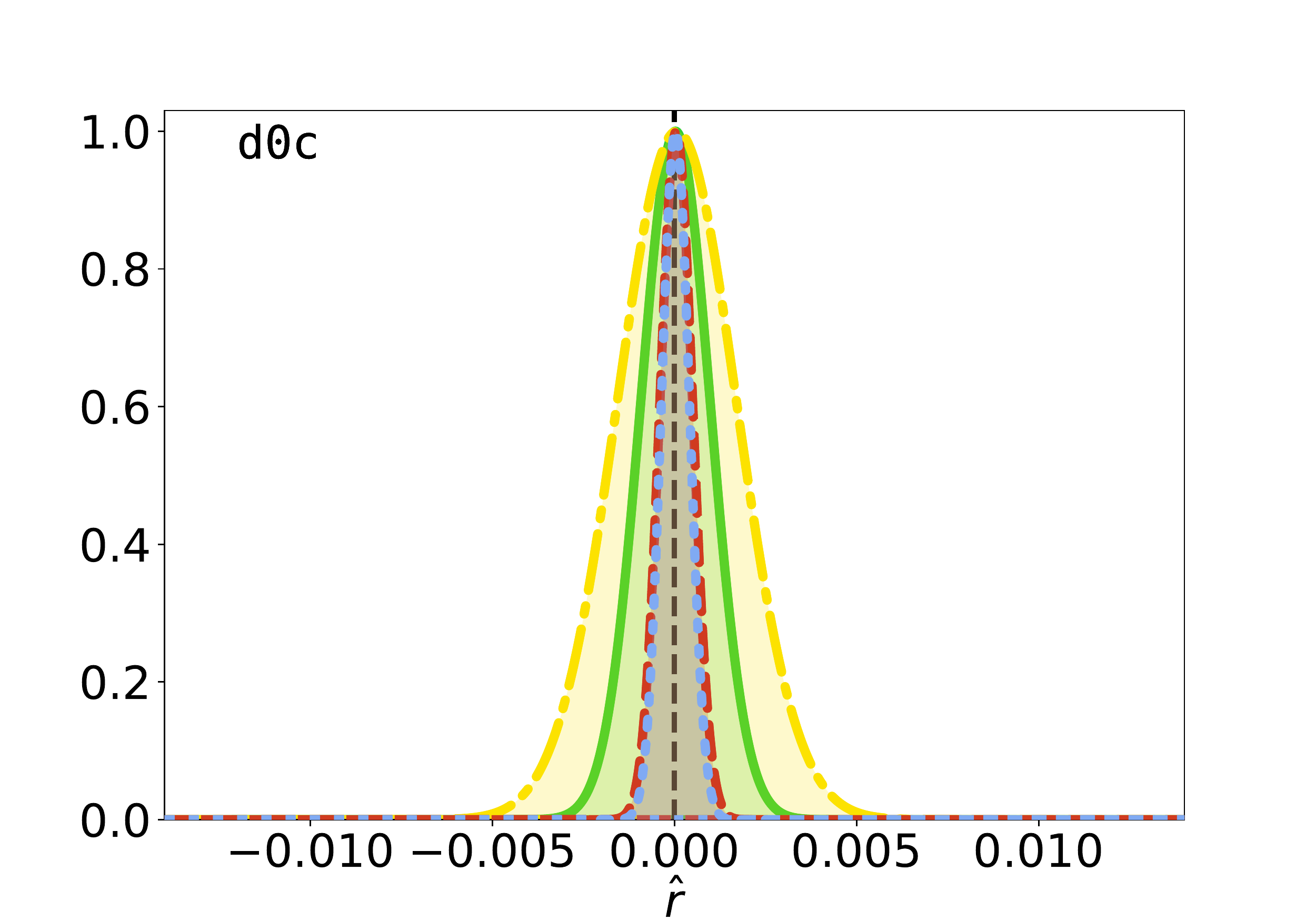}
    \includegraphics[width=\columnwidth]{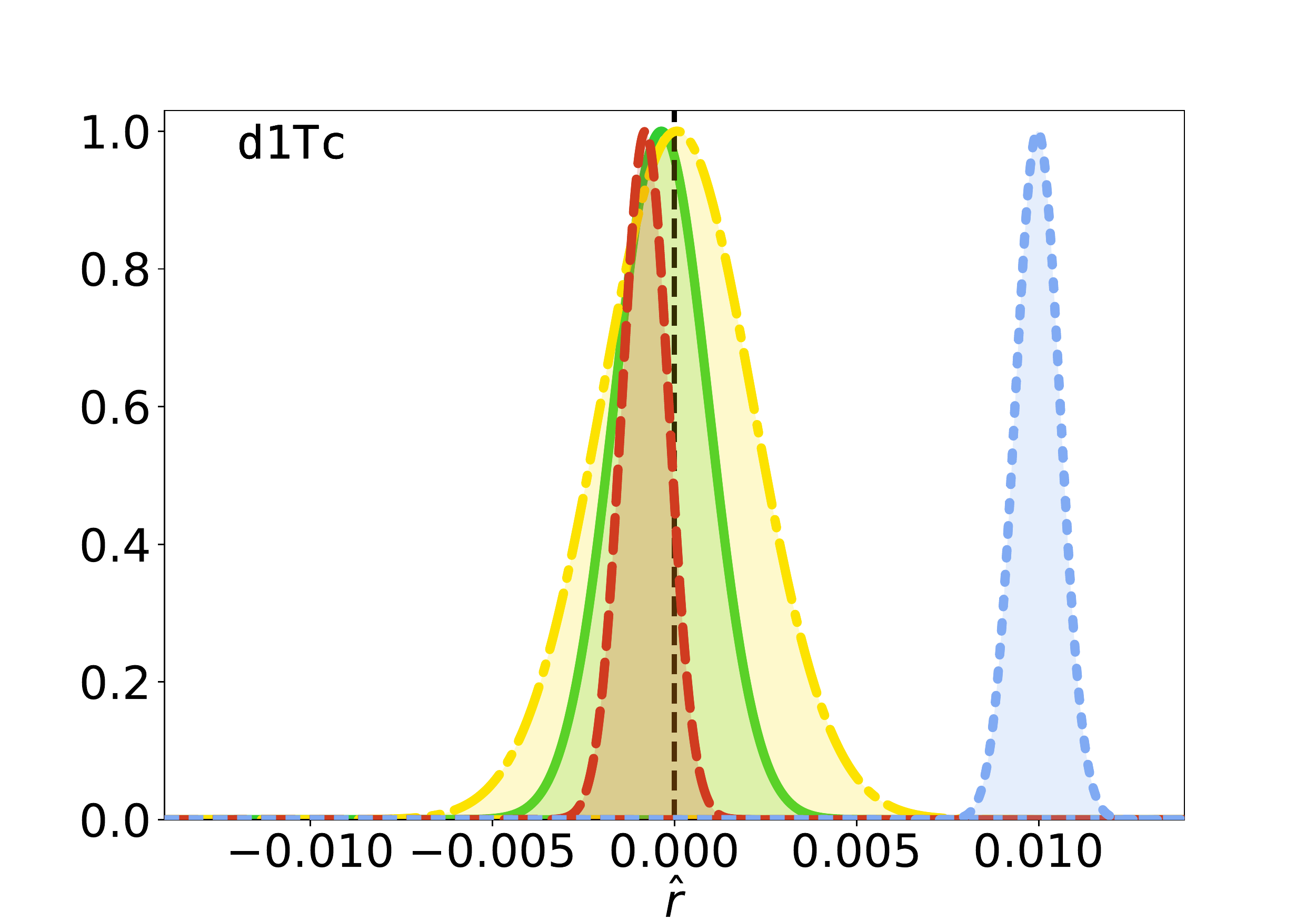}
    \includegraphics[width=\columnwidth]{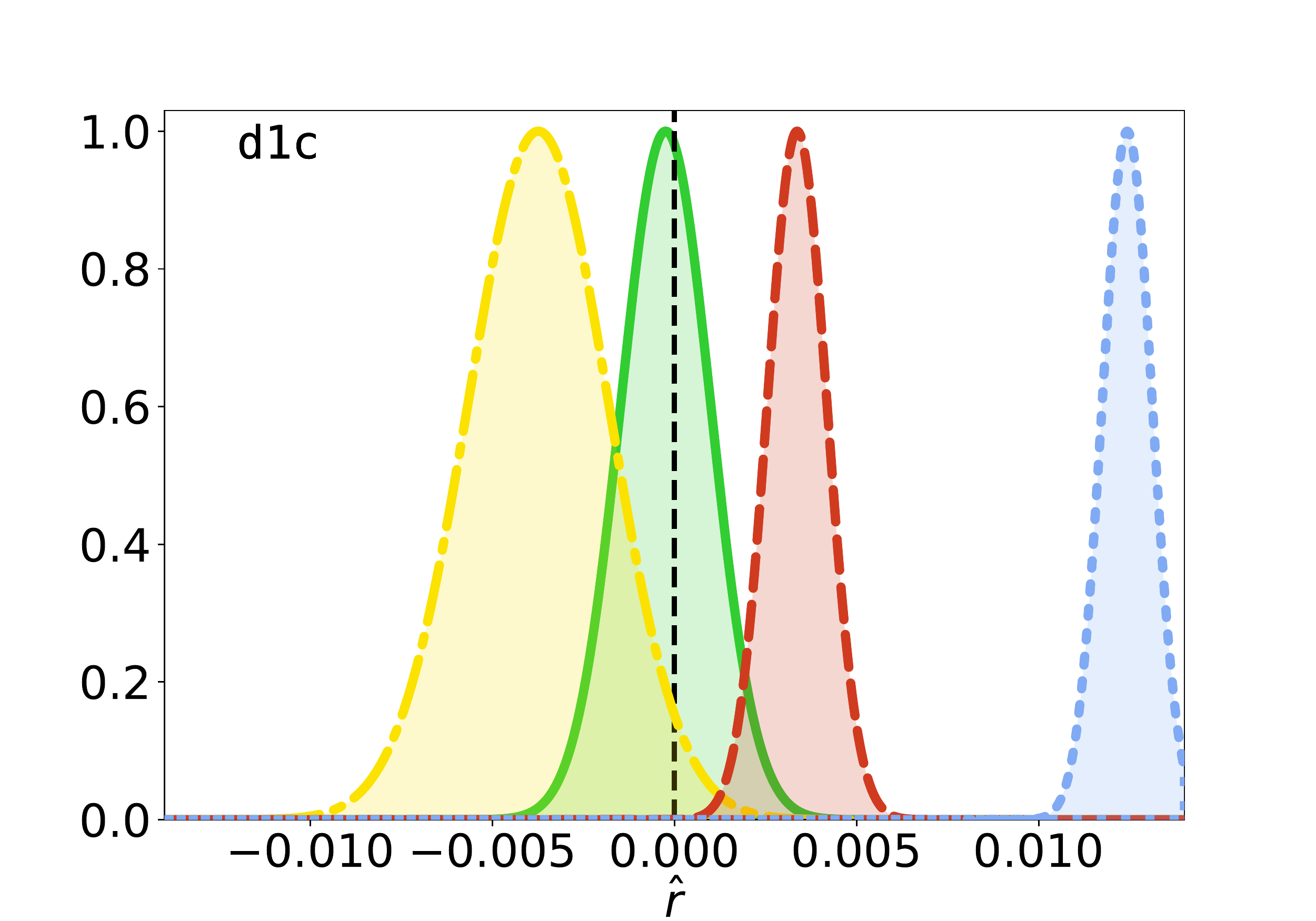}
    \caption{\footnotesize \emph{(Top panel)}: Posterior on $\hat{r}$ in the {\tt d0c} simulation type for the different fitting schemes: $r$MBB (blue, dotted line), $r\beta$-1 (red, dashed line), $r\beta$-$T$ (green, solid line), and $r\beta$-2 (yellow, dash-dotted line). The vertical black dashed line marks the value of $r_{\rm sim}=0$. \emph{(Central panel)}: Same, but in the case of the {\tt d1Tc} simulation type. \emph{(Bottom panel)}: Same, but in the case of the {\tt d1c} simulation type.}
    \label{fig:rgauss1}
\end{figure}

\subsubsection{{\label{sec:d0c}}{\tt d0c}}

For {\tt d0c}, as for {\tt d0}, we recover the input constant spectral index and temperature $\beta_{\tt d0}$ and $T_{\tt d0}$ at all angular scales for every fitting scheme. Furthermore, we do not detect any moment, when fitting $r\beta$-1, $r\beta$-$T,$ and $r\beta$-2. This simulation type therefore constitutes our {null-test} when $\hat{r}$ and the dust parameters are fitted at the same time. The addition of the CMB lensing in the simulations and the addition of $r$ to the fits thus does not lead to the detection of the moment parameters nor biases the recovery of the spectral index and the temperature.

The posterior distributions of the estimated tensor-to-scalar ratio $\hat{r}$ are displayed in Fig.~\ref{fig:rgauss1} and their mean and standard deviations are summarized in Table~\ref{tab:rvalues}. We note that $\hat{r}$ is compatible with the input value ($r_{\rm sim}=0)$ for all the fitting schemes. For $r$MBB and $r\beta$-1, the dispersion $\sigma_{\hat{r}}$ is comparable with the CMB-only scenario discussed in Sect.~\ref{sec:nodust}. For $r\beta$-$T$ and $r\beta$-2, the width of the distribution increases by a factor of $\sim 2$ and $\sim 4$,  respectively.

\subsubsection{{\tt d1Tc}}
\label{sec:d1Tc}

The posterior distribution of $\hat{r}$ in the case of the {\tt d1Tc} simulation type is displayed in Fig.~\ref{fig:rgauss1} for the four fitting schemes and the mean value and standard deviation of these distributions are summarized in Table~\ref{tab:rvalues}. We can see that in the case of $r$MBB, we fit $\hat{r}\pm\sigma_{\hat{r}}=(99.7\pm6.2)\times10^{-4}$. In that case, the input tensor-to-scalar ratio $r_{\rm sim}=0$ is not recovered and we obtain a bias on the central value of $\hat{r}$ of $\sim 16\,\sigma_{\hat{r}}$. As discussed in Sect.~\ref{sect:limitsmbb}, this is expected because we know that the MBB is not a good dust model for a SED with spatially varying spectral index, as we also verify in Sect.~\ref{sec:d1T} looking at the $\chi^2$ values.

Using the $r\beta$-1 fitting scheme allows us to recover $\hat{r}=(-8.0\pm6.4)\times10^{-4}$, where $r_{\rm sim}$ is recovered within $\sim\,2\sigma_{\hat{r}}$, while $r\beta$-2 and $r\beta$-$T$ recover the input value within $1\,\sigma_{\hat{r}}$ (with $\hat{r}=(-3.6\pm 13.0)\times10^{-4}$ and $\hat{r}=(0.7\pm20.9)\times10^{-4}$, respectively). As in Sect~\ref{sec:d0c}, the deviation remains similar between $r$MBB and $r\beta$-1 and increases by a factor of $\sim 2$ and $\sim 4$ from $r\beta$-1 to $r\beta$-$T$ and $r\beta$-2,  respectively.

\begin{table}[t!]
\centering\scalebox{1}{
    \centering
    \begin{tabular}{c | ccc}
        ($\hat{r} \pm \sigma_{\hat{r}})\times10^{4}$ & {\tt d0c} & {\tt d1Tc} & {\tt d1c}\\
        \hline
        $r$MBB     &  \cellcolor{mygreen} $0.3 \pm 3.9$ & \cellcolor{myred}$99.7 \pm 6.2$  & \cellcolor{myred} $125.1 \pm 5.9$\\
        $r\beta$-1   & \cellcolor{mygreen} $0.5 \pm 4.5$ & \cellcolor{myyellow} $-8.0 \pm 6.4$& \cellcolor{myred} $32.9 \pm 6.5$\\
        $r\beta$-$T$   & \cellcolor{mygreen} $0.3 \pm 9.5$ & \cellcolor{mygreen} $-3.6 \pm 13.0$  & \cellcolor{mygreen} $-3.3 \pm 11.7$\\
        $r\beta$-2   & \cellcolor{mygreen} $0.7 \pm 16.4$ & \cellcolor{mygreen} $0.7 \pm 20.9$ & \cellcolor{myyellow} $-37.4 \pm 19.4$\\ 
        \hline 
    \end{tabular}}
\caption{\footnotesize Best-fit values of $\hat{r}$ in units of $10^{-4}$ on $f_{\rm sky}=0.7$. The green values are compatible with $r_{\rm sim}=0$ at $1\,\sigma_{\hat{r}}$, the yellow values are compatible with $r_{\rm sim}=0$ at $2\,\sigma_{\hat{r}}$ and the red values are incompatible with $r_{\rm sim}=0$ at more than $2\,\sigma_{\hat{r}}$.}
\label{tab:rvalues}
\end{table}

\subsubsection{{\tt d1c}}
\label{sec:d1c}

In the case of the {\tt d1c} simulation type, as in {\tt d0c} and {\tt d1Tc}, we fit $\hat{r}$ in addition to the dust-related parameters. In that case, dust moment parameters are recovered as for {\tt d1} (see Sect.~\ref{sec:d1}), except for the $r\beta$-2 fitting scheme.

Figure~\ref{fig:momentsd1d1c} compares the moment parameters between $\beta$-2 on the {\tt d1c} simulations type, fitting only the dust-related parameters and $r\beta$-2 on {\tt d1c} when jointly fitting the dust parameters and $\hat{r}$. We observe that $\mathcal{D}_\ell^{\omega^{\beta}_2 \times \omega^{\beta}_2}$ is not consistently recovered when fitting $\hat{r}$ in addition to the dust parameters.

A similar comparison can be found in Fig.~\ref{fig:moments2d1d1c} for the moment parameters between $\beta$-$T$ and $r\beta$-$T$ on the {\tt d1c} simulation type. Using this fitting scheme, we can see that all the moments are correctly recovered when adding $\hat{r}$ to the fit.

The $\hat{r}$ posterior distributions in the case of {\tt d1c} are displayed in  Fig.~\ref{fig:rgauss1} and summarized in Table~\ref{tab:rvalues}. As discussed in Sect.~\ref{sect:limitsmbb} and observed in Sect.~\ref{sec:d1c}, the $r$MBB fit is highly biased, with $\hat{r}=(125.1 \pm 5.9)\times10^{-4}$ (by more than 21\,$\sigma_{\hat{r}}$). When fitting the $r\beta$-1, this bias is significantly reduced ($\hat{r}=(32.9 \pm 6.5)\times10^{-4}$, $~5\,\sigma_{\hat{r}}$ away from $r_{\rm sim}=0$), illustrating the ability of the first-moment parameters to correctly capture part of the SED complexity. However, performing the expansion in both $\beta$ and $T$ with $r\beta$-$T$ allows us to recover $r_{\rm sim}$ without bias ($\hat{r}=(-3.3\pm 11.7)\times10^{-4}$), highlighting the need for the description of the SED complexity in terms of dust temperature for this simulated data set where both $\beta$ and $T$ vary spatially. On the other hand, for $r\beta$-2, a negative tension ($1.9\,\sigma_{\hat{r}}$) can be observed: $\hat{r}=(-37.4\pm 19.4)\times10^{-4}$. This tension is discussed in Sect.~\ref{sec:bias}.  

For {\tt d1c}, the $\hat{r}$ distribution widths roughly meet the foreground cleaning requirements of \lb{} presented in Sect.~\ref{sec:LiteBIRD} for $r$MBB and $r\beta$-1 but are higher for $r\beta$-$T$ and $r\beta$-2. We also note that, with the same number of free parameters, all the standard deviations $\sigma_{\hat{r}}$ slightly increase compared to the {\tt d0c} simulation type. This is expected due to the increasing dust complexity. 

\section{\label{sec:discussion} Discussion}

\subsection{Lessons learnt}

In Sect.~\ref{sec:results}, we apply the fitting pipeline introduced in Sect.~\ref{sec:fit} on \lb{} simulated data sets on $f_{\rm sky}=0.7$ and for $r_{\rm sim}=0$, including the various dust simulation types defined in Sect.~\ref{sec:ingredients}. We fitted the estimated $B$-mode power-spectra with the four different fitting schemes summarized in Table~\ref{tab:parameter}. Our main results can be summarized as follows: 

\begin{itemize}
\item The MBB fitting scheme provides a good fit for the dust component in the {\tt d0} and {\tt d0c} simulation types. 
    However, when the spectral index changes with the angular scale, such as in the {\tt d1T}, {d1Tc}, {\tt d1,} and {\tt d1c} simulations, this approach no longer provides a good fit because of the complexity of the dust SED. As a consequence, in the $r$MBB case, $r_{\rm sim}$ cannot be recovered without a significant bias. 

\item The $\beta$-1 fitting scheme  allows us to perform a good fit for the dust complexity using the {\tt d0} and {\tt d1T} simulations but not for {\tt d1}, while the $r\beta$-1 fitting  scheme yields estimates of $\hat{r}$ close to $r_{\rm sim}$ within $1\,\sigma_{\hat{r}}$ for {\tt d0c}, and within $2\,\sigma_{\hat{r}}$ for {\tt d1Tc}, but presenting a bias of $\sim 6\,\sigma_{\hat{r}}$ for {\tt d1c}. 

\item The $\beta$-$T$ fitting scheme provides a good fit for every dust model, while using the $r\beta$-$T$ fitting  scheme allows us to recover $\hat{r}$  values consistent with $r_{\rm sim}$ within $1\,\sigma_{\hat{r}}$ for all the simulation types, but is associated with an increase of $\sigma_{\hat{r}}$ by a factor $\sim 2$ compared to the $r\beta$-1 case.

\item The $\beta$-2 fitting scheme also provides a good fit for each dust model, and the $r\beta$-2 fitting scheme leads to values of $\hat{r}$ compatible with $r_{\rm sim}$ within $1\,\sigma_{\hat{r}}$ for all the simulation types but {\tt d1c}. In this last case, there is a negative tension of $\sim2\,\sigma_{\hat{r}}$.  
    For all the simulation types, there is an increase of $\sigma_{\hat{r}}$ by a factor of $\sim 4$ compared to the $r\beta$-1 case.

\end{itemize}

The present analysis shows that the temperature could be a critical parameter for the moment expansion in the context of \lb{}.

Indeed, for simulations including a dust component with a spectral index and a temperature that both vary spatially, as in {\tt d1}, the only fitting scheme allowing us to recover $r_{\rm sim}$ within $1\,\sigma_{\hat{r}}$ is $r\beta$-$T$, the expansion to first order in both $\beta$ and $T$. This shows that expanding in $\beta$ only, without treating $T$, is not satisfactory when looking at such large fractions of the sky. Indeed, when applying the $\beta$-2 fitting scheme, the $\mathcal{D}_\ell^{\omega^{\beta}_2 \times \omega^{\beta}_2}$ parameter remains undetected for the {\tt d1T} simulation type (Sect~\ref{sec:d1T}), while it is significantly detected using the {\tt d1} simulation type (Sect~\ref{sec:d1}). Nevertheless, {\tt d1T} and {\tt d1}share the same template of $\beta(\vec{n})$ (Sect.~\ref{sec:ingredients}) and they only differ by the sky temperature (constant for {\tt d1T} and varying for {\tt d1}). This suggests that the observed $\mathcal{D}_\ell^{\omega^{\beta}_2 \times \omega^{\beta}_2}$ with the {\tt d1} simulations originates from the temperature variations and not those in the spectral index. This observation shows that it is less convenient to use the $\beta$-2 fitting scheme than the $\beta$-$T$ one in order to correctly recover the moment-expansion parameters and $\hat{r}$ when temperature varies spatially.

Moreover, we saw that $ \sigma_{\hat{r}}$ is lower when using the fitting scheme $r\beta$-$T$ instead of $r\beta$-2 for every simulation type, even if both have the same number of free parameters. This second observation additionally encourages an approach where the SED is expanded with respect to both $\beta$ and $T$. Nevertheless, the uncertainty on $\hat{r}$ we obtain in this case ($\sigma_{\hat{r}}=1.17\times10^{-3}$) is larger than the \lb{} requirements. 

\subsection{Increasing the accuracy on the tensor-to-scalar ratio}
\label{sec:Opt}

In Sect.~\ref{sec:d1} and Fig.~\ref{fig:chi2dust}, we see that the MBB and $\beta$-1 fitting schemes do not provide good fits for the {\tt d1} dust simulations, especially at low multipoles ($\ell \lesssim 100$). Conjointly, in Fig.~\ref{fig:moments3}, we can see that the $\beta$-$T$ moment parameters are significantly detected for $\ell \lesssim 100$ and compatible with zero above that threshold, suggesting that their corrections to the SED are predominantly required at large angular scales.

This implies that we can improve the pipeline presented in Sect.~\ref{sec:fit} to keep only the required parameters in order to recover $\hat{r}$ compatible with $r_{\rm sim}$ with a minimal $\sigma_{\hat{r}}$. It can be achieved by applying the $r\beta$-1 fitting sheme over the whole multipole range, while restricting the $r\beta$-$T$-specific ($\mathcal{D}_\ell^{\omega^\beta_1\times\omega^T_1}$ and $\mathcal{D}_\ell^{\omega^T_1\times\omega^T_1}$) moment-expansion parameters fit to the low multipoles range. We note that in order to correct the bias, it is still necessary to keep the $r\beta$-1 moment parameters even at high multipoles, because the MBB does not provide a good fit even for $\ell \in [100,200]$, as we can see in Fig.~\ref{fig:chi2dust}. We define $\ell_{\rm cut}$ as the multipole bin under which we keep all the $r\beta$-$T$ moment parameters and above which we use the $r\beta$-1 scheme.

The best-fit values and standard deviations of $\hat{r}$ for different values of $\ell_{\rm cut}$ are displayed in Table~\ref{tab:rellcut}. We can see that a trade-off has to be found: the smaller the $\ell_{\rm cut}$ , the bigger the shift from $r_{\rm sim}$, and the bigger the $\ell_{\rm cut}$, the higher the value of $\sigma_{\hat{r}}$. The trade-off point seems to be found for $\ell_{\rm cut} \sim 80$, allowing us to recover $\hat{r}$ without tension, with $\sigma_{\hat{r}} =8.8\times10^{-4}$. The error on $r$ is thus reduced by more than $\sim30\,\%$ with respect to the nonoptimized fit and meets the \lb{} requirements.

\subsection{\label{sec:fsky} Tests with smaller sky fractions}

\begin{table}[t!]
\centering\scalebox{1}{
    \centering
    \begin{tabular}{c|c}
        $\ell_{\rm cut}$ & $(\hat{r} \pm \sigma_{\hat{r}})\times 10^{4}$\\\hline
        50 &  \cellcolor{myred}$12.0 \pm 7.3$ \\
        60 &  \cellcolor{mygreen}$7.3 \pm 7.9$ \\
        70 &  \cellcolor{mygreen}$4.9 \pm 8.1$ \\
        80 & \cellcolor{mygreen}$-0.9 \pm 8.8$ \\
        90 &  \cellcolor{mygreen}$-2.1 \pm 9.9$ \\
    \end{tabular}}
\caption{\footnotesize Best-fit values of $\hat{r} \pm \sigma_{\hat{r}}$ in units of $10^{-4}$ for different values of $\ell_{\rm cut}$ for the {\tt d1c} simulations with $f_{\rm sky}=0.7$, when applying the $r\beta$-$T$ fitting scheme. The green values are compatible with $r_{\rm sim}=0$ at $1\,\sigma_{\hat{r}}$.}
\label{tab:rellcut}
\end{table}

\begin{table}[t!]
\centering\scalebox{1}{
    \centering
    \begin{tabular}{c | ccc}
        $(\hat{r} \pm \sigma_{\hat{r}})\times10^4$ & $r_{\rm sim}=0.01$ & $f_{\rm sky}=0.5$ & $f_{\rm sky}=0.6$\\
        \hline
        $r$MBB     & \cellcolor{myred} $204.8 \pm 7.7$ & \cellcolor{myred} $47.3 \pm 5.6$ & \cellcolor{myred}  $59.2 \pm 5.4$\\
        $r\beta$-1  & \cellcolor{myred} $129.0 \pm 8.3$ & \cellcolor{myyellow} $-8.4 \pm 6.7$ & \cellcolor{mygreen} $1.8 \pm 6.2$\\
        $r\beta$-$T$ & \cellcolor{mygreen} $94.6 \pm 15.1$ &\cellcolor{mygreen} $0.02 \pm 13.4$ & \cellcolor{mygreen} $-1.1 \pm 12.0$\\
        $r\beta$-2   & \cellcolor{myyellow} $62.5 \pm 25.0$ & \cellcolor{mygreen} $4.3 \pm 24.2$ & \cellcolor{mygreen} $-3.2 \pm 22.4$\\
         \hline
    \end{tabular}}
\caption{\footnotesize Best-fit values of $\hat{r}$ in units of $10^{-4}$ for an alternative {\tt d1c} simulation with $r_{\rm sim}=0.01$ on $f_{\rm sky}=0.7$, and with $r_{\rm sim}=0$ but on $f_{\rm sky}=0.5$ and $f_{\rm sky}=0.6$. The green values are compatible with $r_{\rm sim}$ at $1\,\sigma_{\hat{r}}$, the yellow values are compatible with $r_{\rm sim}$ at $2\,\sigma_{\hat{r}}$ , and the red values are incompatible with $r_{\rm sim}$ at more than $2\,\sigma_{\hat{r}}$.}
\label{tab:rvalues2}
\end{table}

In all the results presented in Sect.~\ref{sec:results}, we were considering a sky fraction of $f_{\rm sky}=0.7$. This sky mask keeps a considerable fraction of the brightest Galactic dust emission. To quantify the impact of the sky fraction on our analysis, we ran the pipeline as in Sect.~\ref{sec:d1c} with the different masks introduced in Sect.~\ref{sect:mask} ($f_{\rm sky}=0.5$ and $f_{\rm sky}=0.6$). This was done with the {\tt d1c} simulation type. 

\begin{figure}[t!]
    \centering
    \includegraphics[width=\columnwidth]{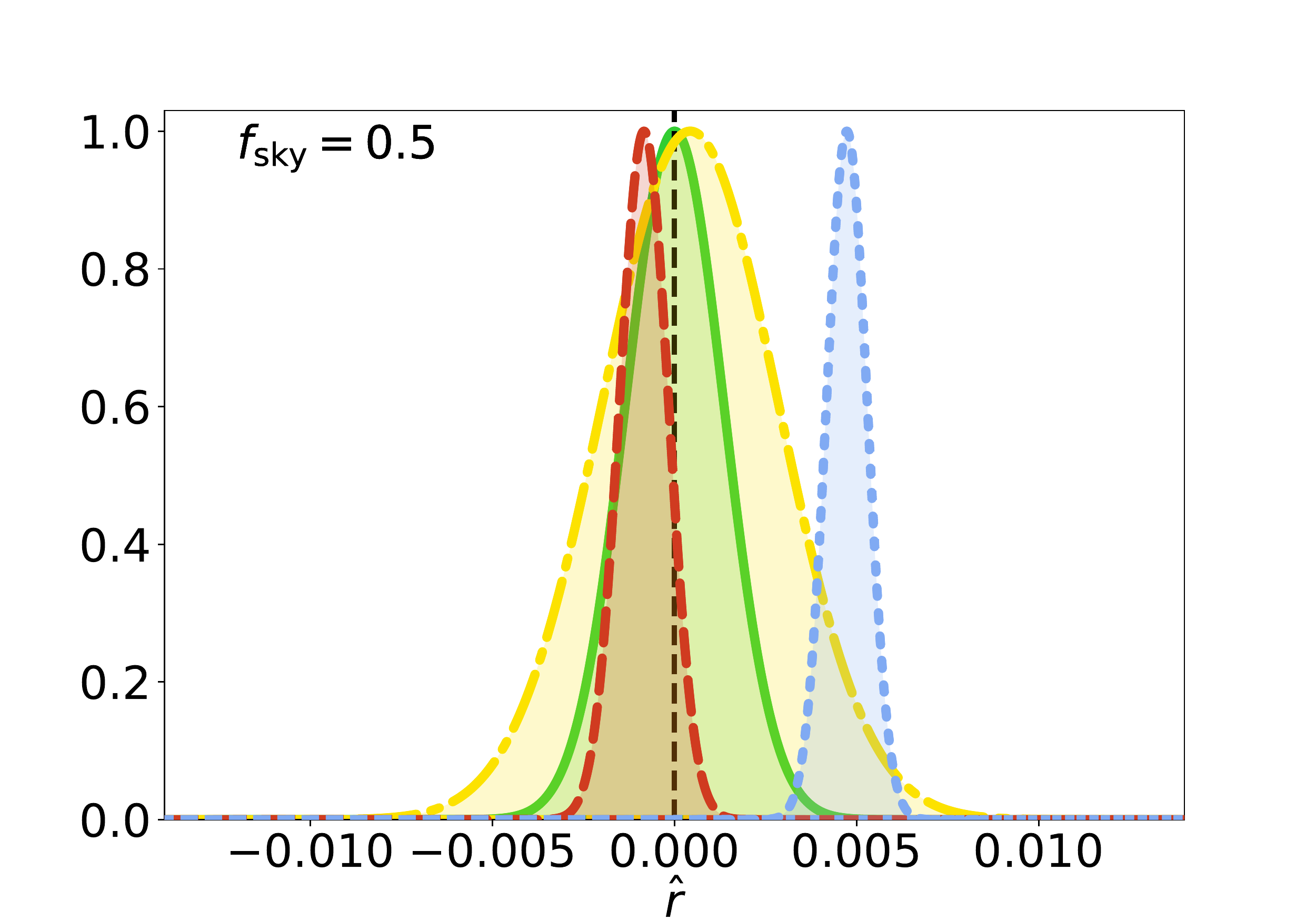}
    \includegraphics[width=\columnwidth]{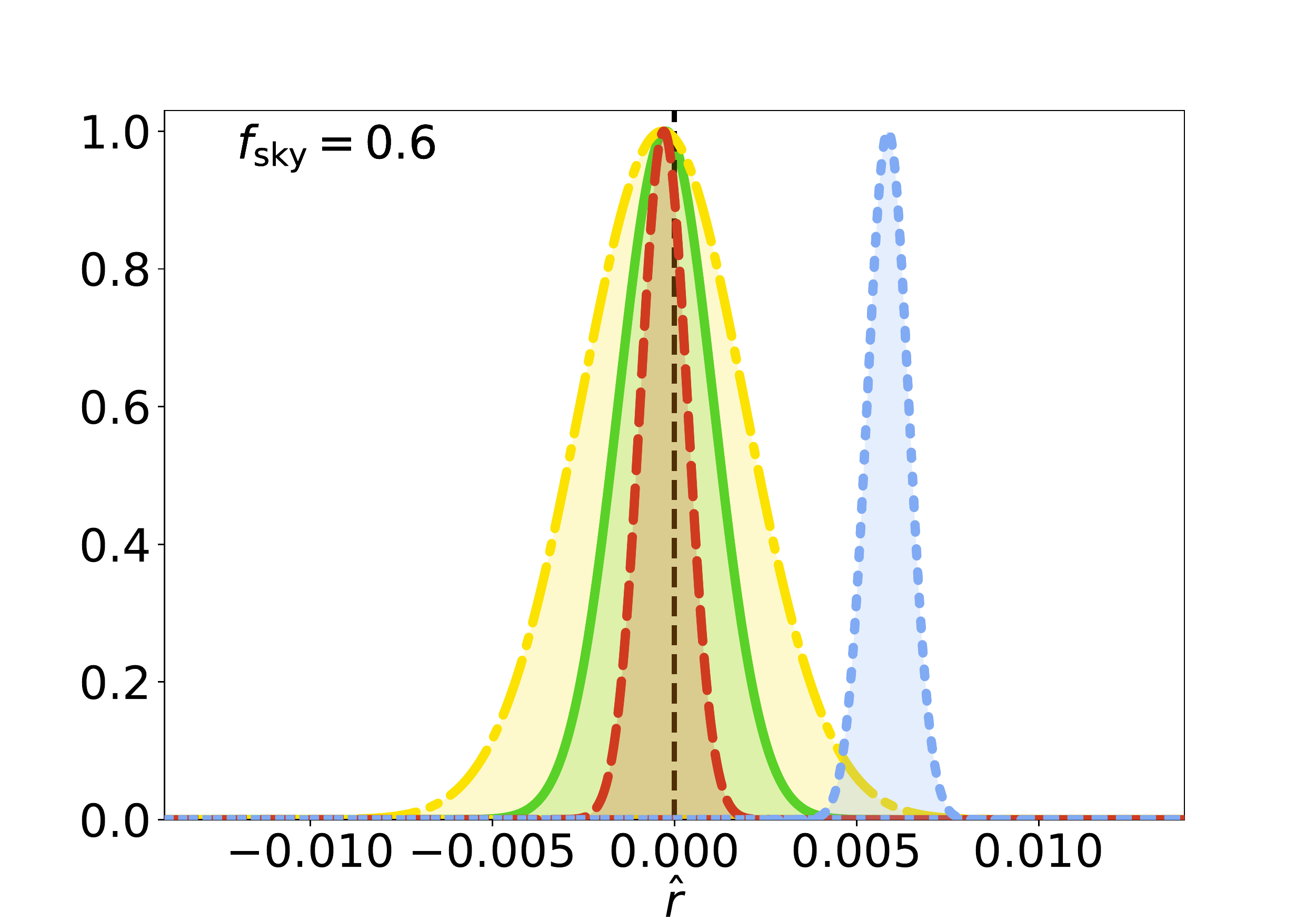}
    \caption{\footnotesize \emph{(Top panel)}: Posterior on $\hat{r}$ in the {\tt d1c} simulation type on $f_{\rm sky}=0.5$ for the different fitting schemes: $r$MBB (blue, dotted line), $r\beta$-1 (red, dashed line), $r\beta$-$T$ (green, solid line), and $r\beta$-2 (yellow, dash-dotted line). The vertical black dashed line marks the value of $r_{\rm sim}=0$.\emph{(Bottom panel)}: Same, in the case of the {\tt d1c} simulation type on $f_{\rm sky}=0.6$.}
    \label{fig:rgauss2}
\end{figure}

The posteriors on $\hat{r}$ for the different fitting schemes are displayed in Fig.~\ref{fig:rgauss2} and Table~\ref{tab:rvalues2}. We can see that, while the $r$MBB fiting scheme always leads to biased estimates, the $r\beta$-1 case allows us to recover $\hat{r}$ at $1.25\,\sigma_{\hat{r}}$ for $f_{\rm sky}=0.5$ and within $1\,\sigma_{\hat{r}}$ for $f_{\rm sky}=0.6$. In the two situations, the results using the $r\beta$-$T$ and $\beta$-2 fitting schemes are both unbiased with estimates of $\hat{r}$ compatible with $r_{\rm sim}$ within $1\,\sigma_{\hat{r}}$. The $\sigma_{\hat{r}}$ hierarchy between the $r$MBB, $r\beta$-1, $r\beta$-$T,$ and $r\beta$-2 fitting schemes is the same as for $f_{\rm sky}=0.7$ (see Sect.~\ref{sec:d1c}). Nevertheless, we observe that $\sigma_{\hat{r}}$ increases as the sky fraction decreases, as does the statistical error (cosmic variance of the lensing and noise). The bias, on the other hand, decreases for all the fitting schemes with the sky fraction, which is expected because less dust emission contributes to the angular power spectra. The negative tension observed on the $\hat{r}$ posterior in Sect.~\ref{sec:d1c} for the $r\beta$-2 case is not present when using smaller sky fractions. In Fig.~\ref{fig:fskydust}, the $r\beta$-2 moment parameters are displayed. We can see that they are not significantly detected for the $f_{\rm sky}=0.5$ and 0.6, unlike for $f_{\rm sky}=0.7$. As we have seen that some of the moments in the $\beta$-2 fitting scheme failed to model SED distortions coming from temperature, we can suppose that, in our simulations, the temperature variations play a less significant role in the dust SED on the $f_{\rm sky}=0.5$ and 0.6 masks than they play in the $f_{\rm sky}=0.7$ one. As a consequence, they have a smaller impact on $r$ when not properly taken into account.

\subsection{Tests with nonzero input tensor modes}
\label{sec:rnot0}

We show in Sect.~\ref{sec:d1c} that the $r\beta$-$T$ fitting scheme allows us to retrieve $\hat{r}$ compatible with zero when $r_{\rm sim}=0$. We now want to assess the potential leakage of $\hat{r}$ in the moment expansion parameters if $r_{\rm sim} \neq 0$. In this case, primordial tensor signals would be incorrectly interpreted as dust complexity. We run the pipeline as described in Sect.~\ref{sec:d1c} with $r_{\rm sim}=0.01$, in the ${\tt d1c}$ simulation type. This value of $r_{\rm sim}=0.01$ is larger than the value targeted by \lb,{} but given the order of magnitude of the error on $\hat{r}$ observed in the previous sections, a potential leakage could be left unnoticed using a smaller $r_{\rm sim}$. 

\begin{figure}[t!]
    \centering
    \includegraphics[width=\columnwidth]{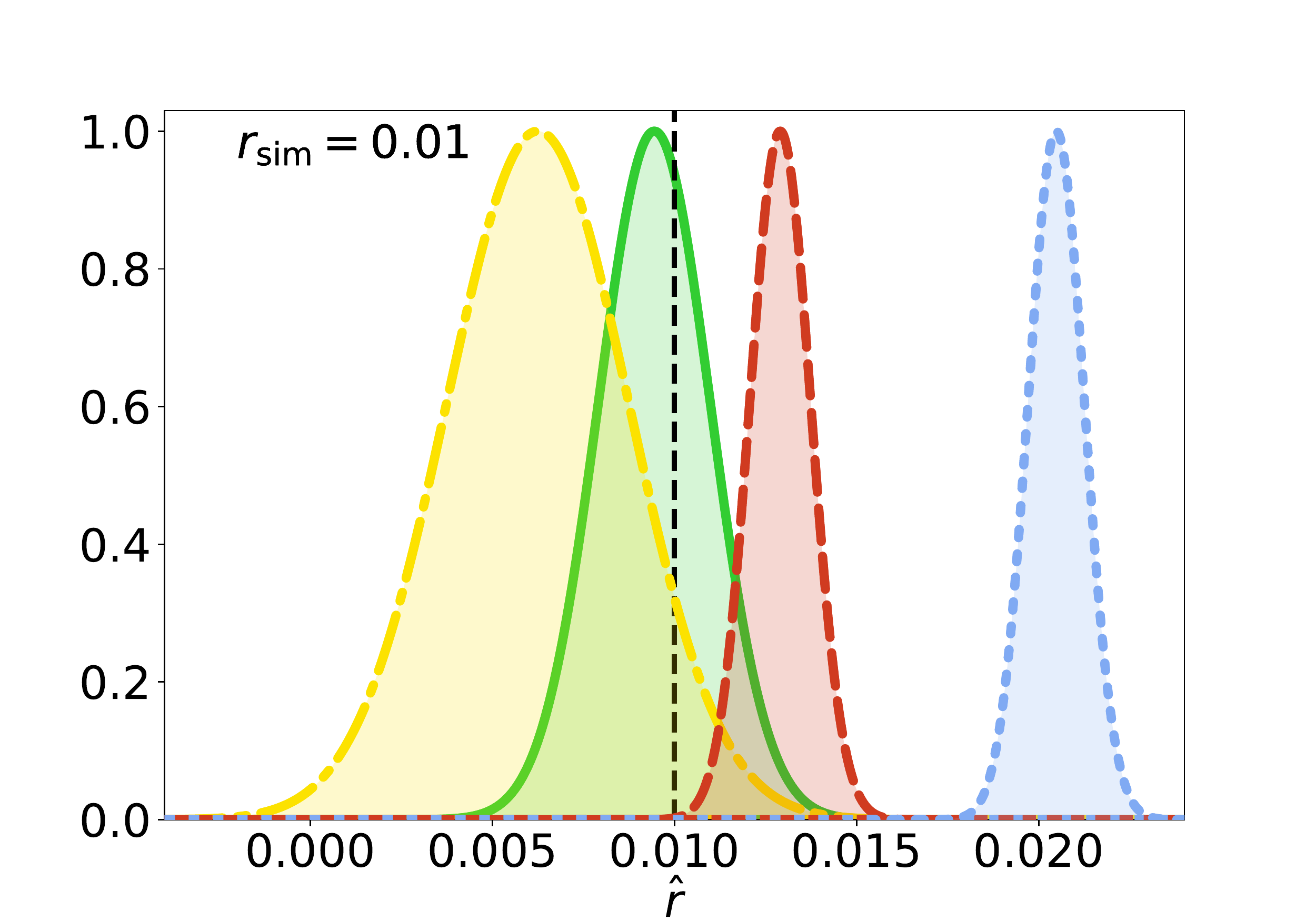}
    \caption{\footnotesize Posterior on $\hat{r}$ in the {\tt d1c} simulation type with $r_{\rm sim}=0.01$ and $f_{\rm sky}=0.7$ for the different fitting schemes: $r$MBB (blue, dotted line), $r\beta$-1 (red, dashed line), $r\beta$-$T$ (green, solid line), and $r\beta$-2 (yellow, dash-dotted line). The vertical black dashed line marks the value of $r_{\rm sim}$.}
    \label{fig:rgauss3}
\end{figure}

Looking at the final posterior on $\hat{r}$ (Fig.~\ref{fig:rgauss3} and Table~\ref{tab:rvalues2}), we can see that the results are comparable with the $r_{\rm sim}=0$ case, but centered on the new input value $r_{\rm sim}=0.01$. The $r$MBB fitting scheme gives a highly biased posterior of $\hat{r}=(2.048\pm0.077)\times10^{-2}$ ; the bias is reduced but still significant when using the $r\beta$-1 scheme ($\hat{r}=129.0\pm 8.3\times10^{-4}$) ; in the $\beta$-$T$ case we get an estimate of $\hat{r}=94.6\pm 15.1\times10^{-4}$ compatible with the input value of $r_{\rm sim}=100\times10^{-4}$ ; and  finally, the $\beta$-2 fitting scheme leads to a negative $2\,\sigma_{\hat{r}}$ tension ($\hat{r}=62.5\pm25.0\times10^{-4}$). This demonstrates the robustness of our method and its potential application to component separation. We note that the negative bias at second order is still present in the $r_{\rm sim }=0.01$ case, illustrating that setting a positive prior on $\hat{r}$ would not have been a satisfying solution when $r_{\rm sim}=0$.

\subsection{\label{sec:bias} Exploring the correlations between the parameters}

We now examine the substantial increase in the dispersion on the $\hat{r}$ posteriors between the $r\beta$-1 fitting scheme on the one hand and the $r\beta$-$T$ and $r\beta$-2 ones on the other. Indeed, in Sect.~\ref{sec:d1c}, we show that $\sigma_{\hat{r}}$ is about two times greater when using the $r\beta$-$T$ scheme than the $r\beta$-1 one, and about four times larger in the case of $r\beta$-2, while the $r\beta$-$T$ and $r\beta$-2 schemes share the same number of free parameters. Some other points to clarify are the shift on $\hat{r}$ appearing for $r\beta$-2 in the {\tt d1c} scenario, discussed in Sect.~\ref{sec:d1c}, and the inability to correctly recover $\mathcal{D}_\ell^{\omega_2^\beta \times \omega_2^\beta}$ when $\hat{r}$ is added to the fit illustrated in Fig.~\ref{fig:momentsd1d1c}. 

The 2D-SED shapes of the parameters $\mathcal{D}_\ell^{\mathcal{N} \times \mathcal{M}}(\nu_i \times \nu_j)$ in the $(\nu_i, \nu_j)$ space\footnote{For example, $\mathcal{S}(\nu_i,\nu_j) = \frac{I_{\nu_i}(\beta_0,T_0)I_{\nu_j}(\beta_0,T_0)}{I_{\nu_0}(\beta_0,T_0)^2}\cdot \left[\ln\left(\frac{\nu_i}{\nu_0}\right)\ln\left(\frac{\nu_j}{\nu_0}\right) \right]$ is associated to the $\mathcal{D}^{\omega_1 \times \omega_1}_\ell$ parameter (see Eq.~\ref{eq:moments}).} are displayed in Fig.~\ref{fig:momentshapes}. We used the nine frequencies of \lb{} presented in Sect.~\ref{sec:Instrsim} and fixed $\beta_0=1.54$ and $T_0 = 20$\,K. We also introduce the CMB 2D-SED shape with the black body function:

\begin{equation}
    B_{\rm CMB}(\nu_i \times \nu_j) = \frac{B_{\nu_i}(T_{\rm CMB})B_{\nu_j}(T_{\rm CMB})}{B_{\nu_0}(T_{\rm CMB})^2},
\end{equation}

\noindent where $T_{\rm CMB} = 2.726$\,K. 

\begin{figure}[t]
    \centering
    \includegraphics[width=\columnwidth]{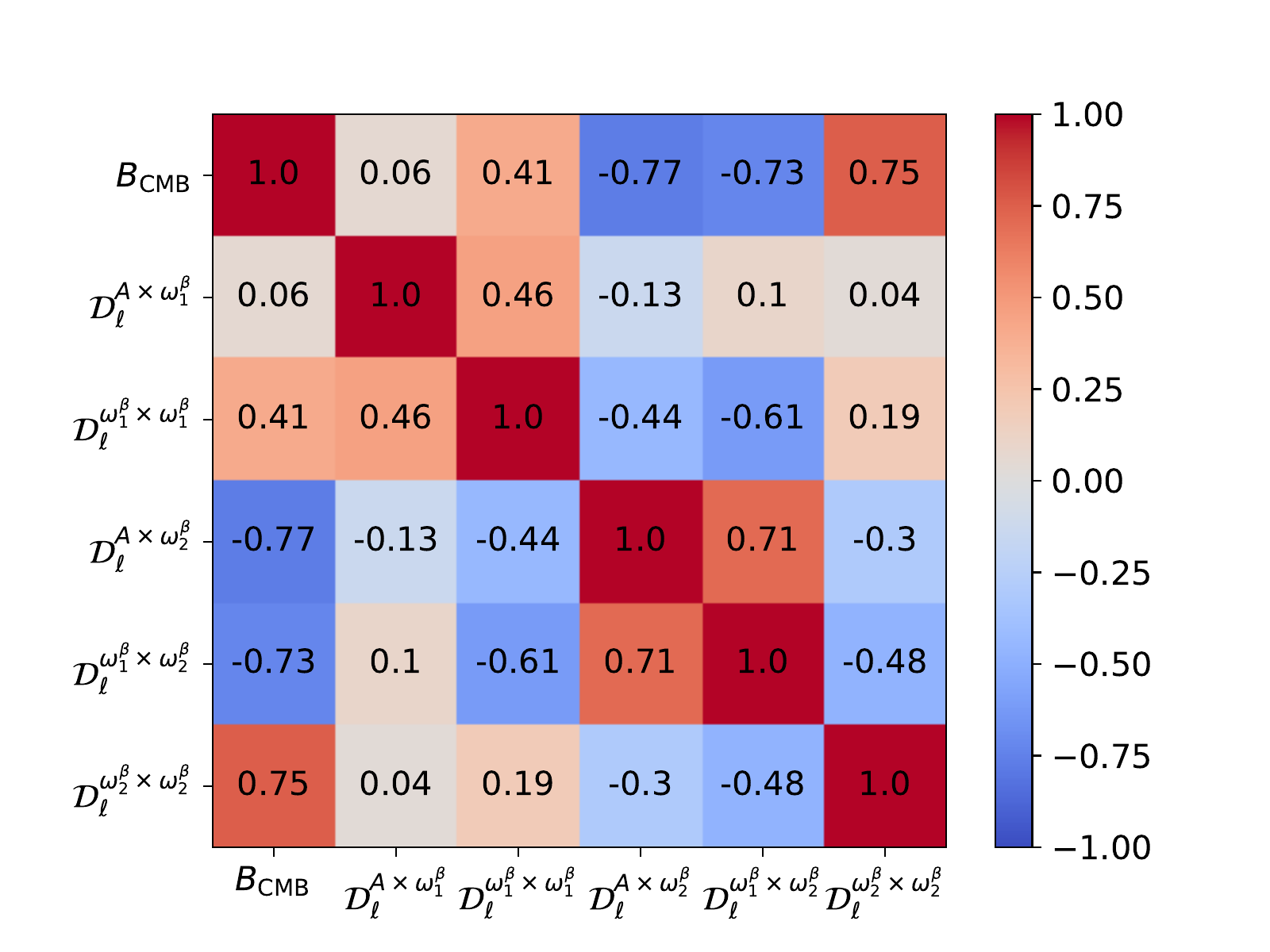}
    \includegraphics[width=\columnwidth]{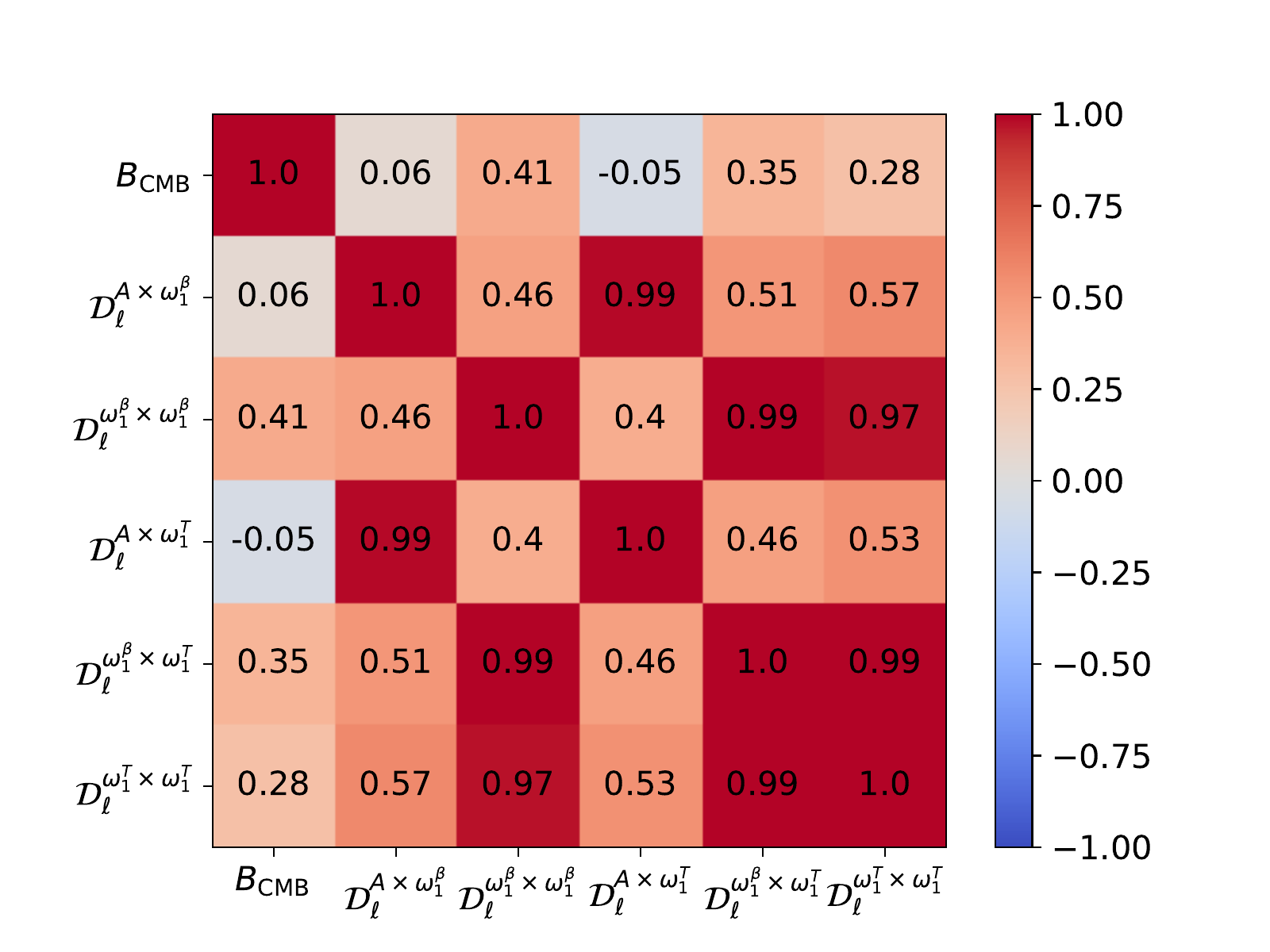}
    \caption{\footnotesize Correlation matrices of the 2D-SED shapes of the CMB ($B_{\rm CMB}(\nu_i\times\nu_j)$ and dust moments $\mathcal{D}_\ell^{\mathcal{N} \times \mathcal{M}}(\nu_i \times \nu_j)$ in the $(\nu_i, \nu_j)$ space). Each element represents the Pearson correlation coefficient between any 2 of these 2D-SED shapes. The correlation matrices are displayed in the case of the $\beta$-2 fitting scheme (top panel) and the $\beta$-$T$ one (bottom panel).}
    \label{fig:corr3D}
\end{figure}

The 2D correlation coefficients between these 2D-SED shapes are displayed in Fig.~\ref{fig:corr3D}. We present the correlations between the shapes of the parameters in the case of the $r\beta$-$T$ and $r\beta$-2 fitting schemes. We can see that all the moment parameters in $\omega^{\beta}_2$ are strongly correlated with the CMB SED signal, while the ones in $\omega^{T}_1$ are not. 

We showed that, when fitting $\beta$-2 on {\tt d1c}, the SED distortions due to spatial variations of $T$ are incorrectly detected by the second-order moment parameters with respect to the spectral index $\beta$. Due to the correlations highlighted above, those spurious moment parameters could then leak into $\hat{r}$ when adding it to the fit in $r\beta$-2. This explains both the negative shift on the $\hat{r}$ posterior using $\beta$-2 in the {\tt d1c} simulation type with $f_{\rm sky}=0.7$ presented in Sect.~\ref{sec:d1c} and \ref{sec:rnot0}, and the inability to  correctly recover the  $\omega^{\beta}_2 \times \omega^{\beta}_2$ dust moment parameter presented in Fig.~\ref{fig:momentsd1d1c}. In addition, it gives a natural reason for the surge of $\sigma_{\hat{r}}$ when the second-order moments in $\beta$ are added to the fit. 

On the other hand, the moment parameters in $\omega^{T}_1$ are strongly correlated with the moments in $\omega^{\beta}_1$. This behavior is expected due to the strong correlation between $\beta$ and $T$  \citep[see e.g.,][]{betatcorr}. However those moment parameters are less correlated with the CMB signal than the second-order parameters of $\beta$-2. This points out that the factor of $\sim 2$ on $\sigma_{\hat{r}}$ between $\beta$-$T$ and $\beta$-2 is due to this correlation of the 2D-SED shapes.
As the parameters in $\omega^{T}_1$ are highly correlated with one another, we expect them to be highly redundant in the fit. However, repeating the process described in Sect.~\ref{sec:d1c} using only $\mathcal{D}^{A \times \omega^{T}_1}_\ell$ for $\beta$-$T$ ---which is equivalent to applying the $\beta$-1 fitting scheme with an iterative correction to the temperature $T_0(\ell)$--- gives a $\hat{r}$ posterior similar to the one obtained for $\beta$-1 alone. Taking the other $\omega^{T}_1$ terms  into account appears to be necessary in order to recover an unbiased distribution of $\hat{r}$.

\subsection{Adding synchrotron to the simulations }
\label{sec:synchrotron}

Thermal dust is not the only source of polarized foreground that must be considered for CMB studies. Although subdominant at high frequencies ($\geq 100\,$GHz), the synchrotron emission due to accelerated electrons in the interstellar medium is still expected to represent a significant fraction of the total polarized signal. 

In order to take one more step towards realistic forecasts for the \lb{} instrument, we add a synchrotron contribution to the {\tt d1c} simulations presented in \ref{sec:simulations} using the {\tt s1} template included in the {\sc PySM}. In this scenario, the synchrotron SED for each line of sight is given by a power law of the form  (in antenna temperature units)

\begin{equation}
    S_\nu^{\tt s1} = A_{\tt s1}(\vec{n}) \left(\frac{\nu}{\nu^{\tt s1}_0}\right)^{\beta_{\tt s1}(\vec{n})},
\label{eq:s1powerlaw}
\end{equation}

\noindent where the amplitude $A_{\tt s1}(\vec{n})$ and the spectral index $\beta_{\tt s1}(\vec{n})$ maps are derived from the combination of the \wmap{} mission 23 GHz map \cite{WMAPfg} and Haslam 408 GHz map \cite{Haslam1}. $\nu^{\tt s1}_0$ is defined as 23 GHz. The simulations containing synchrotron are referred to as {\tt d1s1c} below.

If not treated in the fit, the presence of synchrotron is expected to induce a bias on the $\hat{r}$ posterior distribution. Regarding the dust MBB discussed in Sect.~\ref{sect:limitsmbb}, the synchrotron SED is expected to have distortions. However, as the synchrotron polarized emission is significantly lower than that of dust, in the frequency range considered in the present work, we expect the distortions to be small compared to the ones induced by dust and we leave their modeling to a further study. 

In order to minimize the number of free parameters used for fitting the synchrotron emission, we model its power spectrum as a power law of the multipole $\ell$ \citep{krachmalnicoff_s-pass_2018}. Therefore, combining with the synchrotron SED in Eq.~\ref{eq:s1powerlaw}, the synchrotron component of the cross-angular power spectra reads 

\begin{equation}
    \mathcal{D}^{\rm sync}_\ell (\nu_i \times \nu_j) = A_{\rm s}  \left(\frac{\nu_i \nu_j}{\nu_0}\right)^{\beta_{\rm s}} \ell^{\alpha_{\rm s}},
\label{eq:syncromoment}
\end{equation}

where the amplitude coefficient $A_{\rm s}$ is treated as a free parameter while we fix $\beta_{\rm s}=-3$ (median value of the ${\tt s1}$ $\beta_{\rm s}$ map on our $f_{\rm sky}=0.7$ mask) and $\alpha_{\rm s}=-1$ \citep{krachmalnicoff_s-pass_2018}. 

\begin{figure}[t]
    \centering
    \includegraphics[width=\columnwidth]{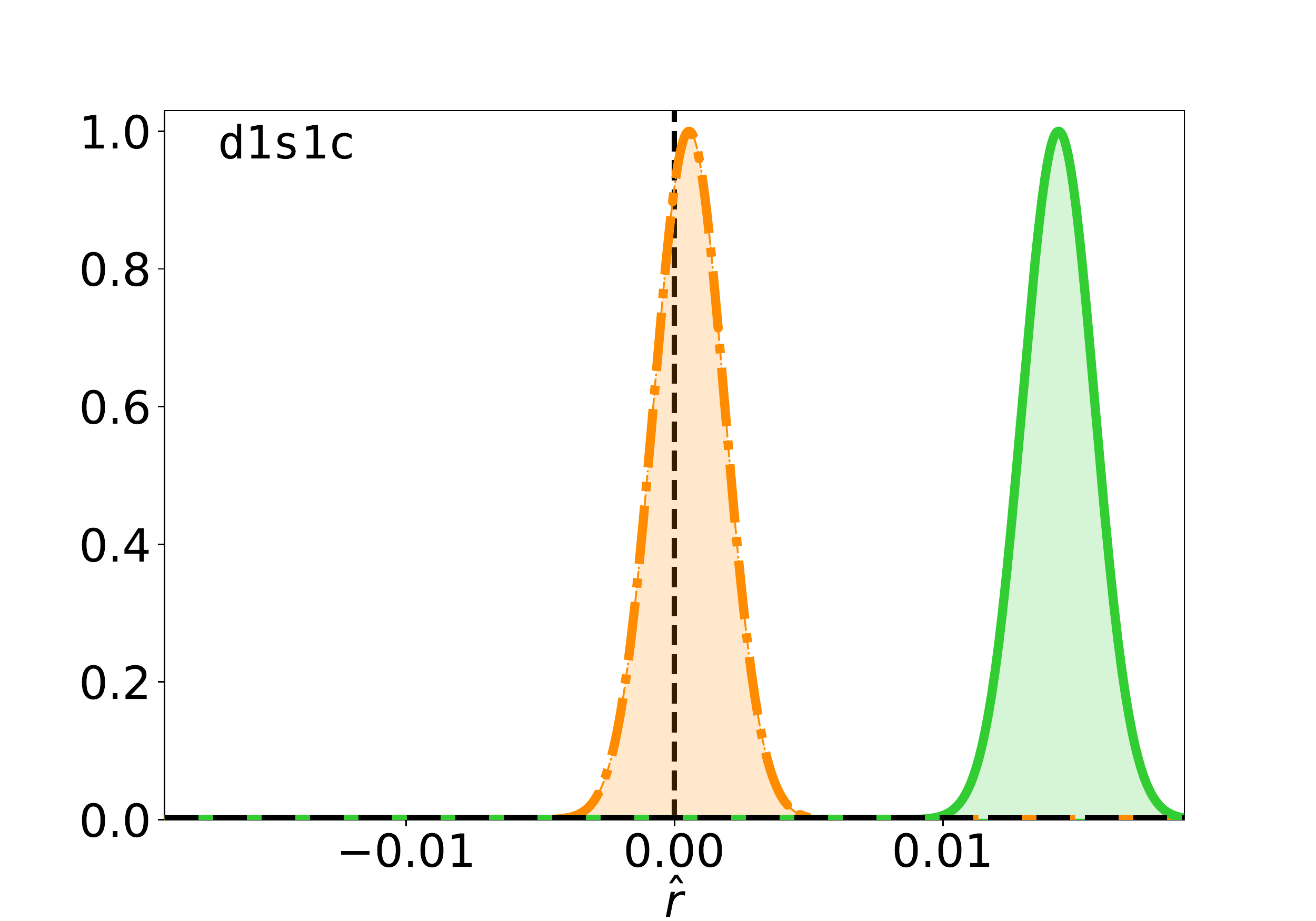}
    \caption{\footnotesize Posterior on $\hat{r}$ in the {\tt d1s1c} simulation type with $r_{\rm sim}=0$ and $f_{\rm sky}=0.7$ for the different fitting schemes: $r\beta$-$T$ (green, solid line) and {\tt s}$r\beta$-$T$ (orange, dash-dotted line). The vertical black dashed line marks the value $r_{\rm sim}=0$.}
    \label{fig:synchrotron}
\end{figure}

When fitting the {\tt d1s1c} simulations, we either use the $r\beta$-$T$ fitting scheme, neglecting the synchrotron component, or we add the synchrotron component in Eq.~\ref{eq:syncromoment} to the model in Eq.~\ref{eq:model}. We refer to this latter case as the {\tt s}$r\beta$-$T$ fitting scheme.
In Fig.~\ref{fig:synchrotron}, the $\hat{r}$ posteriors derived from the {\tt d1s1c} simulations are displayed with $r_{\rm sim}=0$ and $f_{\rm sky}=0.7$. 

Using the $r\beta$-$T$ fitting scheme, we find $\hat{r} = (143.1 \pm 13.5)\times 10^{-4}$. As expected, even at high frequencies, modeling the synchrotron component is critical and cannot be neglected in order to recover an unbiased value of $\hat{r}$.
On the other hand, using {\tt s}$r\beta$-$T$ fitting scheme, we recover $\hat{r} = (-5.4 \pm 13.2)\times 10^{-4}$. This result is comparable with the one obtained for the {\tt d1c} simulations in Sect.~\ref{sec:d1c}, with a minor increase in $\sigma_{\hat{r}}$. We can therefore conclude that a model as simple as that of Eq.~\ref{eq:syncromoment} is sufficient to take into account the {\tt s1} component at $\nu > 100$\,GHz and the corresponding SED distortions can be neglected in order to recover an unbiased value of $\hat{r}$. In principle, as we know that the dust-synchrotron spatial correlation is significant at large scales \citep{PlanckDust2},  Eq.~\ref{eq:model} should include a dust-synchrotron term \citep[see e.g.,][]{SOgalactic}. In our study, where we consider cross-spectra from 100 to 402\,GHz, this dust-synchrotron term is subdominant, but it could be significant when considering cross-spectra between LiteBIRD's extreme frequency bands (e.g., the 40$\times$402 cross-spectrum). The moment expansion might be more complicated as well in this case, as we could expect some correlation between the dust and synchrotron moment-terms.

This result shows that a full polarized foreground content can be treated at high frequencies when using a power law SED for the synchrotron coupled with the moment expansion of the MBB up to first order in both $\beta$ and $T$ for the dust SED. A full study remains to be done in that direction using all the frequency bands of the \lb{} instrument. Eventually, Eq.~\ref{eq:syncromoment} will also have to be expanded in moments with respect to its parameters. Doing so, one can expect to recover an unbiased value of $\hat{r}$ associated with a decrease in $\sigma_{\hat{r}}$ down to a value compatible with the full success criterion of the mission.

\subsection{Limitations of this work and caveats}

{As discussed in Sect.~\ref{sec:formalismspectra}, we neglected polarization effects through this work by treating the $BB$ signal as an intensity signal. This is not problematic in the present work, because no variations along the lines of sight were present in the simulations. However, this point has to be addressed using complex simulations or real sky data.}

The choice of reference frequency $\nu_0$ used for the normalization of the MBB in Eq.~\ref{eq:MBB}, which is not discussed in this study, can potentially have a significant impact on the moment expansion and, in turn, on the measurement of $\hat{r}$. Indeed, $\nu_0$ is the pivot frequency of the moment expansion (moments are equal to zero at $\nu_0$) and will determine the shape of the SED distortion around it. A poor choice for this reference frequency can have disastrous consequences for the moment fit: for example, if it is chosen far away from the observed bands, all the moments will become degenerated. In our case, the reference frequency (353\,GHz) is within the observed frequency range (100 to 402\,GHz), but we have not tried to optimize its position. In addition, the $\nu_0$ pivot of our moment expansion coincides with the one used to extrapolate the dust template map in the {\sc PySM} and we have not quantified how much this impacts our results. 

Finally, as pointed out several times in this work, the quantitative results depend strongly on the sky model of our simulations. Moreover, we lack dedicated sky models where we can control the complexity of the dust SED, either by directly including moments or by averaging the emission from the 3D structure of the Galaxy. However, both methods are beyond the scope of the present work.

\section{Conclusion}

Being able to precisely characterize the complexity of the Galactic thermal dust polarized SED has become critical for the measurement of the faint primordial $B$-mode signal of the CMB, especially at the sensitivity targeted by future CMB experiments such as the \lb{} satellite mission.

In this work, we applied the moment expansion formalism to the dust emission SED as a component-separation tool to recover the tensor-to-scalar ratio parameter $r$ in \lb-simulated data. This formalism, proposed by \citet{Chluba} and implemented in harmonic space by \citet{Mangilli}, allows us to deal with the spectral complexity of the Galactic dust signal by modeling its deviations from the canonical MBB model at the cross-angular power spectrum level. In the case of the data-driven realistic dust emission model ---we explore ({\sc PySM} {\tt d1}) 
here---, suitably taking into account the dust SED distortions prevents the spurious detection of the primordial $B$-mode signal. 

We show that the dust spectral index $\beta$ and dust temperature $T$ spatial variations significantly distort the dust cross-power spectrum SED. The MBB is not a good model to describe the data in that case and the estimation of $r$ is dramatically affected. In the case where no primordial signal is included in the simulated data sets, not taking into account the dust SED complexity leads to a highly significant spurious detection of $r$ with \lb{} (from $\hat{r}\simeq5\times10^{-3}$ to $1.25\times10^{-2}$, with a 8.4 to 21.2\,$\sigma$ significance, from 50 to 70\,\% of the sky, respectively). 

To overcome this obstacle, we applied the moment expansion formalism in order to model these SED distortions. We demonstrate that, at \lb{} sensitivity, the previously studied moment expansion with respect to the dust spectral index $\beta$  \citep{Mangilli,Azzoni} does not give satisfactory results. Indeed, expanding in $\beta$ to first order (following the angular power spectrum definition of the order) leads to a significant bias on 70\,\% of the sky ($\hat{r}=(3.29\pm0.65)\times10^{-3}$ when $r_{\rm sim}=0$ and $\hat{r}=(1.29\pm0.08)\times10^{-2}$, when $r_{\rm sim}=10^{-2}$). At second order in $\beta$, we observe a $\sim$2\,$\sigma$ negative tension ($\hat{r}=(-3.7\pm1.9)\times10^{-3}$ when $r_{\rm sim}=0$ and $\hat{r}=(6.25\pm2.50)\times10^{-3}$, when $r_{\rm sim}=10^{-2}$). 

We introduce for the first time in this work the expansion of the dust angular cross-power spectra with respect to both $\beta$ and $T$. We show that by using this expansion up to first order, we correctly model the dust SED distortions due to spatial variations of both $\beta$ and $T$ at the map level. This allows us to recover $r$ parameter without bias, with $\hat{r}=(-3.3\pm11.7)\times10^{-4}$ if $r_{\rm sim}=0$ and $\hat{r}=(0.95\pm0.15)\times10^{-2}$ if $r_{\rm sim}=10^{-2}$. Thus, despite the known degeneracy between the dust spectral index and its temperature in the Rayleigh-Jeans domain, it is important to correctly model the latter in order to accurately retrieve the tensor-to-scalar ratio $r$ at the unprecedented precision reached by experiments such as \lb{}. 

Adding parameters to tackle the dust SED complexity means an increase in the error budget. Given the \lb{} bands and sensitivities we consider in this work (frequency bands above 100\,GHz), the ideal sensitivity on $r$ without delensing is $\sigma_{\hat{r}}=3.4\times10^{-4}$. In the ideal case, where the dust $\beta$ and $T$ are constant over the sky ({\sc PySM} {\tt d0}), separating the CMB from dust leads to $\sigma_{\hat{r}}=3.9\times10^{-4}$ on 70\,\% of the sky. Adding the expansion to first order in $\beta$ does not significantly increase the error ($\sigma_{\hat{r}}=4.5\times10^{-4}$), but expanding to first order in both $\beta$ and $T$ multiplies it by a factor of $\sim2$ ($\sigma_{\hat{r}}=9.5\times10^{-4}$) and to second order in $\beta$ by a factor of $\sim4$ ($\sigma_{\hat{r}}=16.4\times10^{-3}$). We show that the surge of $\sigma_{\hat{r}}$ between the two latter cases, sharing the same number of free parameters, is due to strong correlations between the SED of the second-order moments in $\beta$ and the CMB. This is an important point, as it could lead to some intrinsic limitation for component-separation algorithms based exclusively on the modeling of the SED. Furthermore, when dealing with real data, if the dust SED is complex enough to have significant second-order distortions with respect to $\beta$, CMB experiments might reach a dilemma: either include the second order in the modeling at the cost of losing sensitivity on $r,$ or neglect it at the cost of a potential spurious detection. Coupling the SED-based separation with methods exploiting the diversity of spatial distribution between components \citep[e.g.,][]{Wavelets} seems a natural way to overcome this issue.

Nevertheless, moment expansion at the cross-angular power spectrum level provides a powerful and agnostic tool, allowing us to analytically recover the actual dust complexity without making any further assumptions. We additionally show that this method is robust, in the sense that it can effectively distinguish the primordial tensor signal from dust when $r_{\rm sim} \neq 0$, as in the case of \lb{} simulations. The dust moments in $\beta$ and $T$ at first order are needed in order to retrieve a reliable measure of $r$; they are significantly detected for $\ell\lesssim100$. We can therefore define a cut in $\ell$ above which we do not fit for the whole complexity of the dust (we fit only the expansion up to first order in $\beta$ and not in $\beta$ and $T$). Doing so, we can reduce the error on $\hat{r}$ while keeping the bias negligible ($\hat{r}=(-0.9\pm8.8)\times10^{-4}$). We could imagine other ways to reduce the number of free parameters in our model \cite[e.g., assuming a power-law of $\ell$ behavior for the moments, as in][]{Azzoni} and hence reduce the error on $r$. However, this optimization really depends on the simulated sky complexity and has not been comprehensively explored in the present work.

The {\sc PySM} {\tt d1} sky simulations, being data-driven, are widely used by the CMB community as they contain some of the real sky complexity. Nevertheless, at high-Galactic latitudes, the dust spectral index and temperature templates from \planck{} are dominated by systematic errors (uncertainty on the assumed zero-level of the \planck{} intensity maps, residual cosmic infrared background (CIB), anisotropies, instrumental noise, etc.). Therefore, some of the complexity we observe far from the Galactic plane in this sky model is not real. On the other hand, the modeled SED of the dust is {exactly} a MBB in each pixel, and line-of-sight averages or more complex dust models are ignored. As a consequence, our method and CMB $B$-mode component-separation algorithms in general need to be confronted with more complex models in order to really assess their performances in a  quantitative manner. 

Finally, although we demonstrate that the synchrotron component can be tackled at frequencies above 100\,GHz with a minimal model under our assumptions, a study over the full \lb{} frequency bands, including synchrotron and the potential moment expansion of its SED, will be considered as a natural next step for a further application.

\section*{Acknowledgments}

This work is supported in Japan by ISAS/JAXA for Pre-Phase A2 studies, by the acceleration program of JAXA research and development directorate, by the World Premier International Research Center Initiative (WPI) of MEXT, by the JSPS Core-to-Core Program of A. Advanced Research Networks, and by JSPS KAKENHI Grant Numbers JP15H05891, JP17H01115, and JP17H01125. The Italian \lb{} phase A contribution is supported by the Italian Space Agency (ASI Grants No. 2020-9-HH.0 and 2016-24-H.1-2018), the National Institute for Nuclear Physics (INFN) and the National Institute for Astrophysics (INAF). The French \lb{} phase A contribution is supported by the Centre National d’Etudes Spatiale (CNES), by the Centre National de la Recherche Scientifique (CNRS), and by the Commissariat à l’Energie Atomique (CEA). The Canadian contribution is supported by the Canadian Space Agency. The US contribution is supported by NASA grant no. 80NSSC18K0132. 
Norwegian participation in \lb{} is supported by the Research Council of Norway (Grant No. 263011). The Spanish \lb{} phase A contribution is supported by the Spanish Agencia Estatal de Investigación (AEI), project refs. PID2019-110610RB-C21 and AYA2017-84185-P. Funds that support the Swedish contributions come from the Swedish National Space Agency (SNSA/Rymdstyrelsen) and the Swedish Research Council (Reg. no. 2019-03959). The German participation in \lb{} is supported in part by the Excellence Cluster ORIGINS, which is funded by the Deutsche Forschungsgemeinschaft (DFG, German Research Foundation) under Germany’s Excellence Strategy (Grant No. EXC-2094 - 390783311). This research used resources of the Central Computing System owned and operated by the Computing Research Center at KEK, as well as resources of the National Energy Research Scientific Computing Center, a DOE Office of Science User Facility supported by the Office of Science of the U.S. Department of Energy.

MR acknowledges funding support from the ERC Consolidator Grant CMBSPEC (No. 725456) under the European Union's Horizon 2020 research and innovation program.

The authors would like to thank David Alonso, Josquin Errard, Nicoletta Krachmalnicoff and Giuseppe Puglisi, for useful discussions as well as Jens Chluba for insightful comments on earlier version of this work.

\bibliographystyle{aa}
\bibliography{bi}

\begin{appendix}

\section{Complementary figures}

\begin{figure}[h!]
    \centering
    \includegraphics[width=\columnwidth]{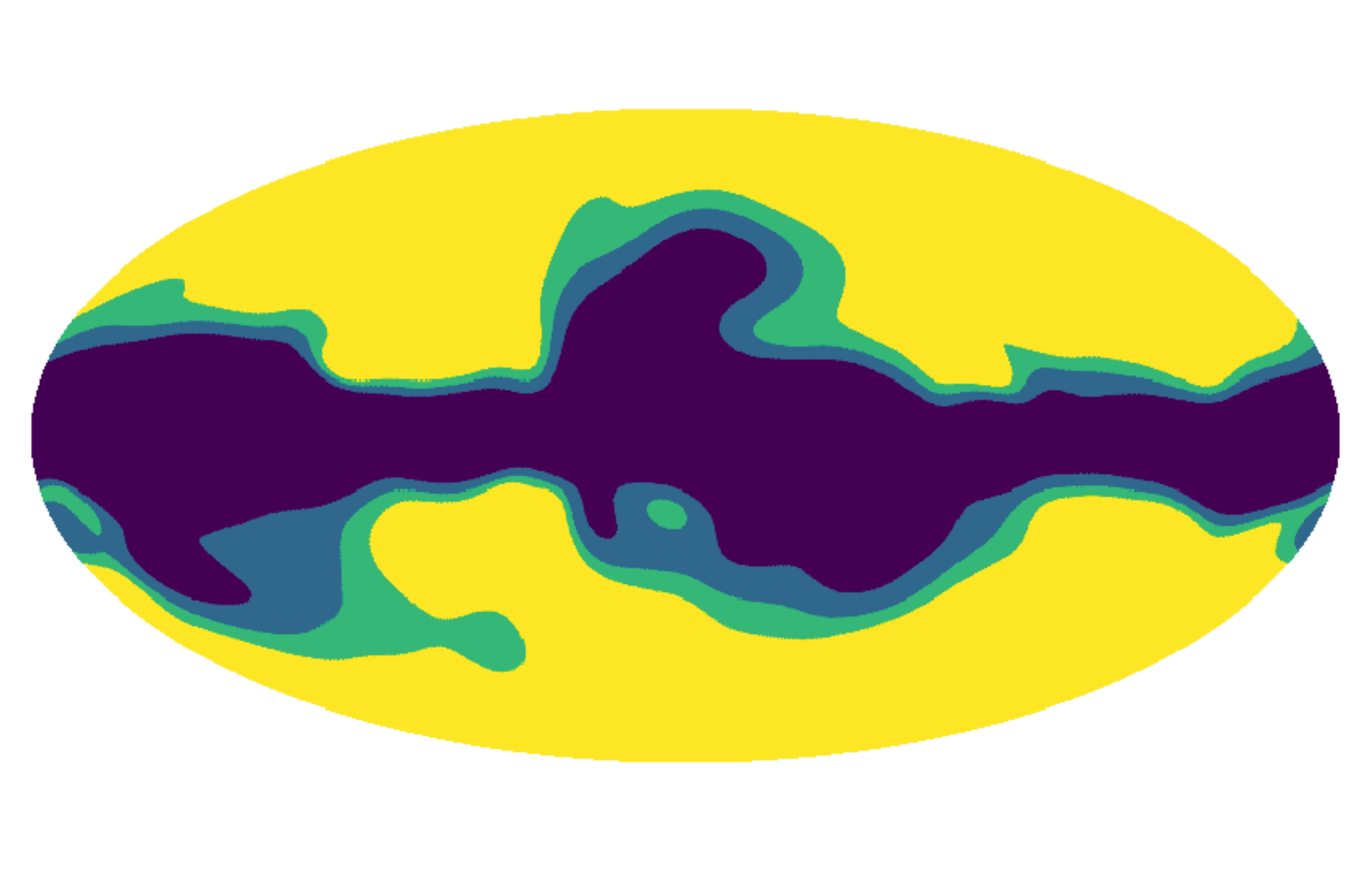}
    \caption{\footnotesize Raw masks used in the simulations:  $f_{\rm sky}=0.7$ (dark blue), $f_{\rm sky}=0.6$ (light blue) and $f_{\rm sky}=0.5$ (green).}
    \label{fig:mask}
\end{figure}

\begin{figure*}[t!]
    \centering
    \includegraphics[width=\columnwidth]{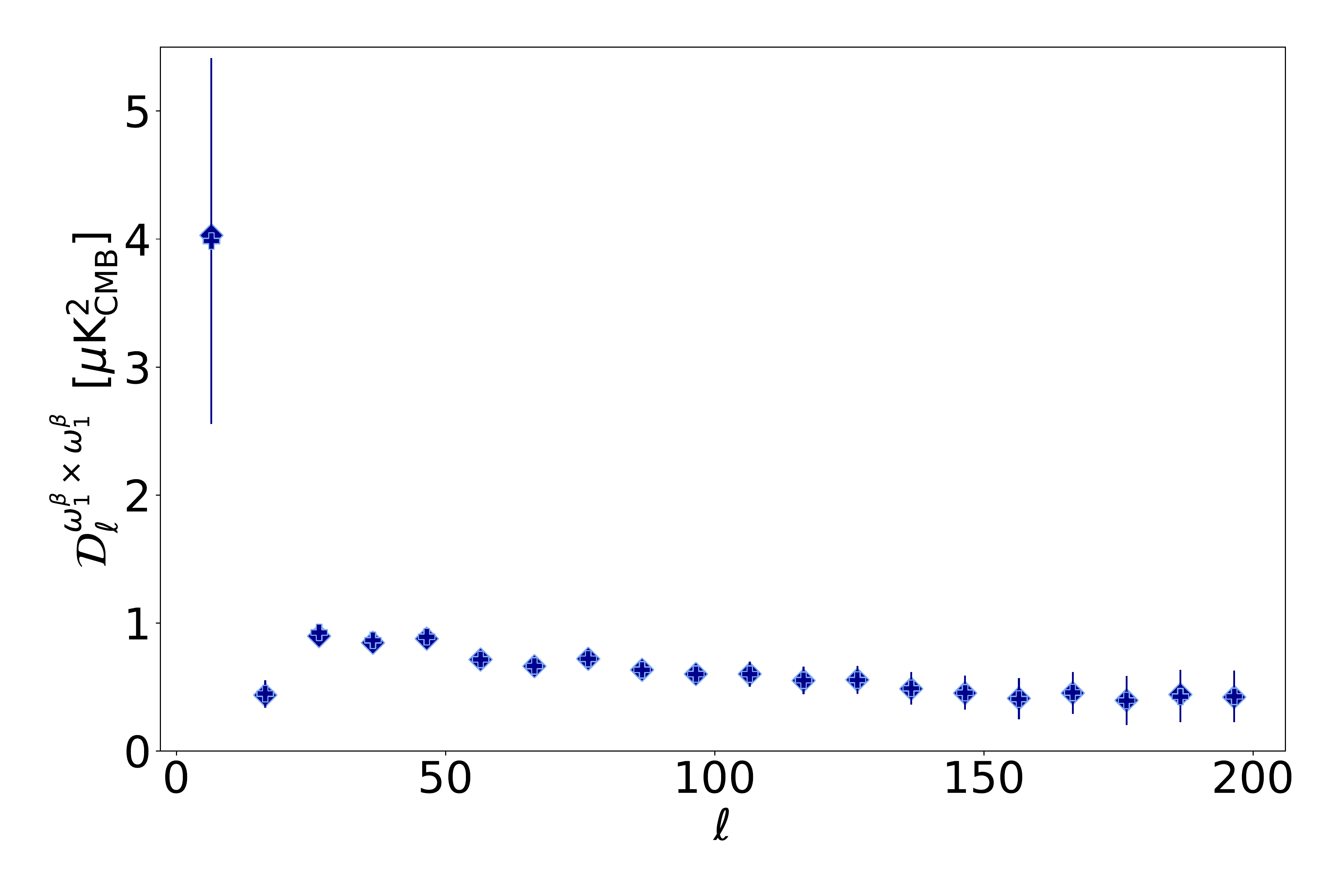}
    \includegraphics[width=\columnwidth]{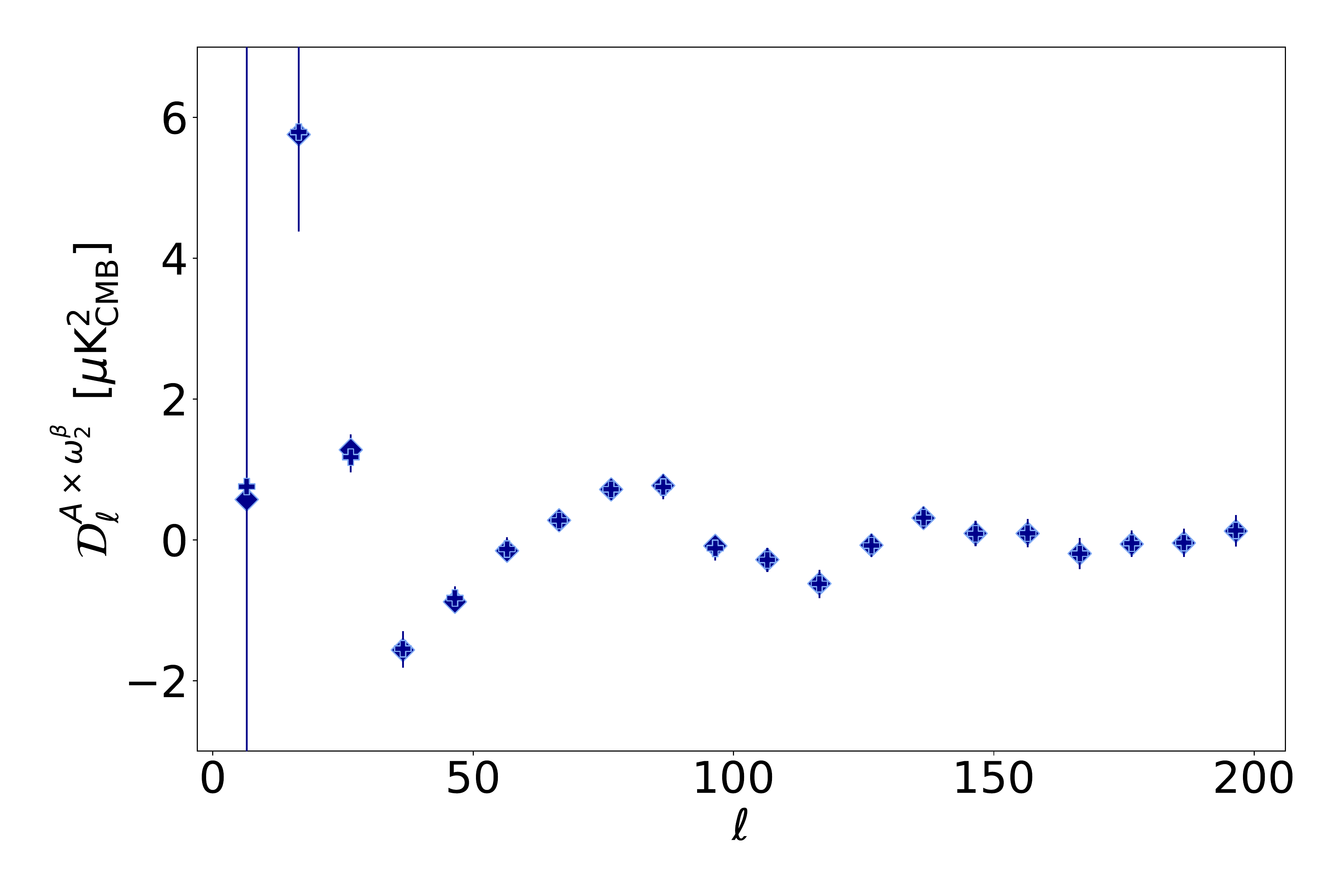}
    \includegraphics[width=\columnwidth]{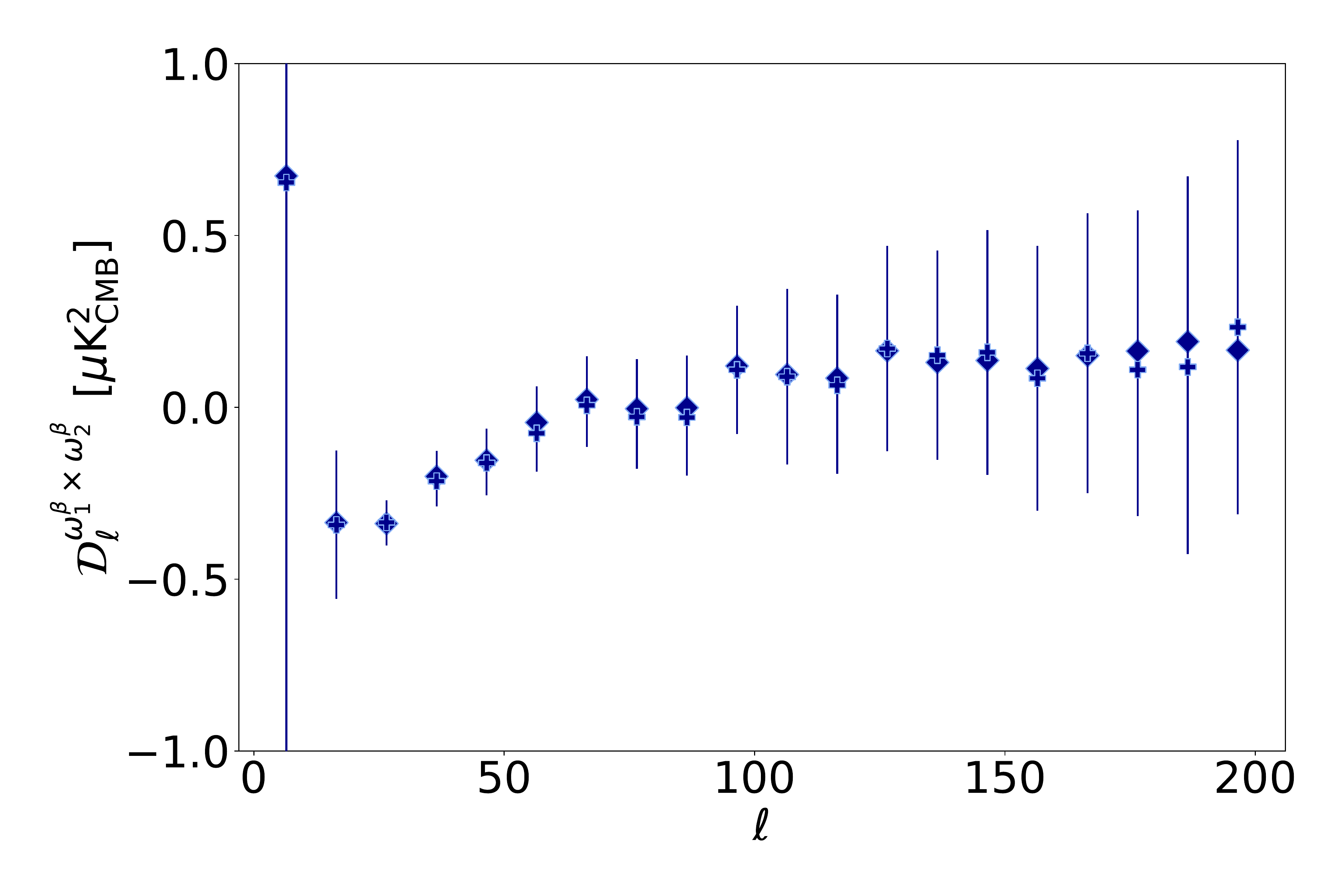}
    \includegraphics[width=\columnwidth]{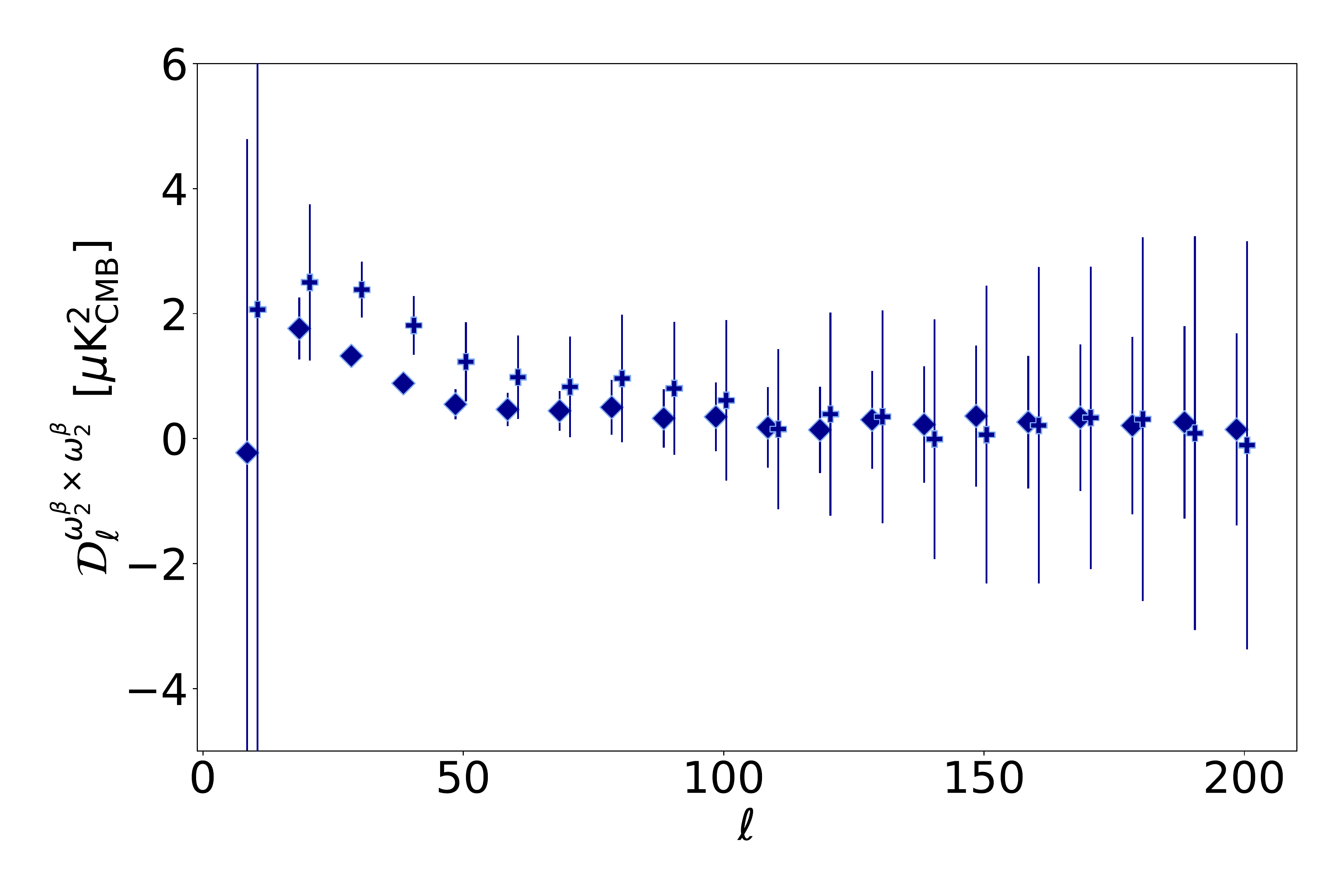}
    \caption{\footnotesize Best-fit values of the $\mathcal{D}_\ell^{\omega^{\beta}_1\times\omega^{\beta}_1}$ (top left),  $\mathcal{D}_\ell^{A\times\omega^{\beta}_2}$ (top right), $\mathcal{D}_\ell^{\omega^{\beta}_1\times\omega^{\beta}_2}$ (bottom left) and $\mathcal{D}_\ell^{\omega^{\beta}_2\times\omega^{\beta}_2}$ (bottom right) moment parameters for the $\beta$-2 fitting scheme applied on the {\tt d1c} simulation type (diamonds) and $r\beta$-2 on {\tt d1c} (plus sign).}
    \label{fig:momentsd1d1c}
\end{figure*}

\begin{figure*}[t!]
    \centering
    \includegraphics[width=\columnwidth]{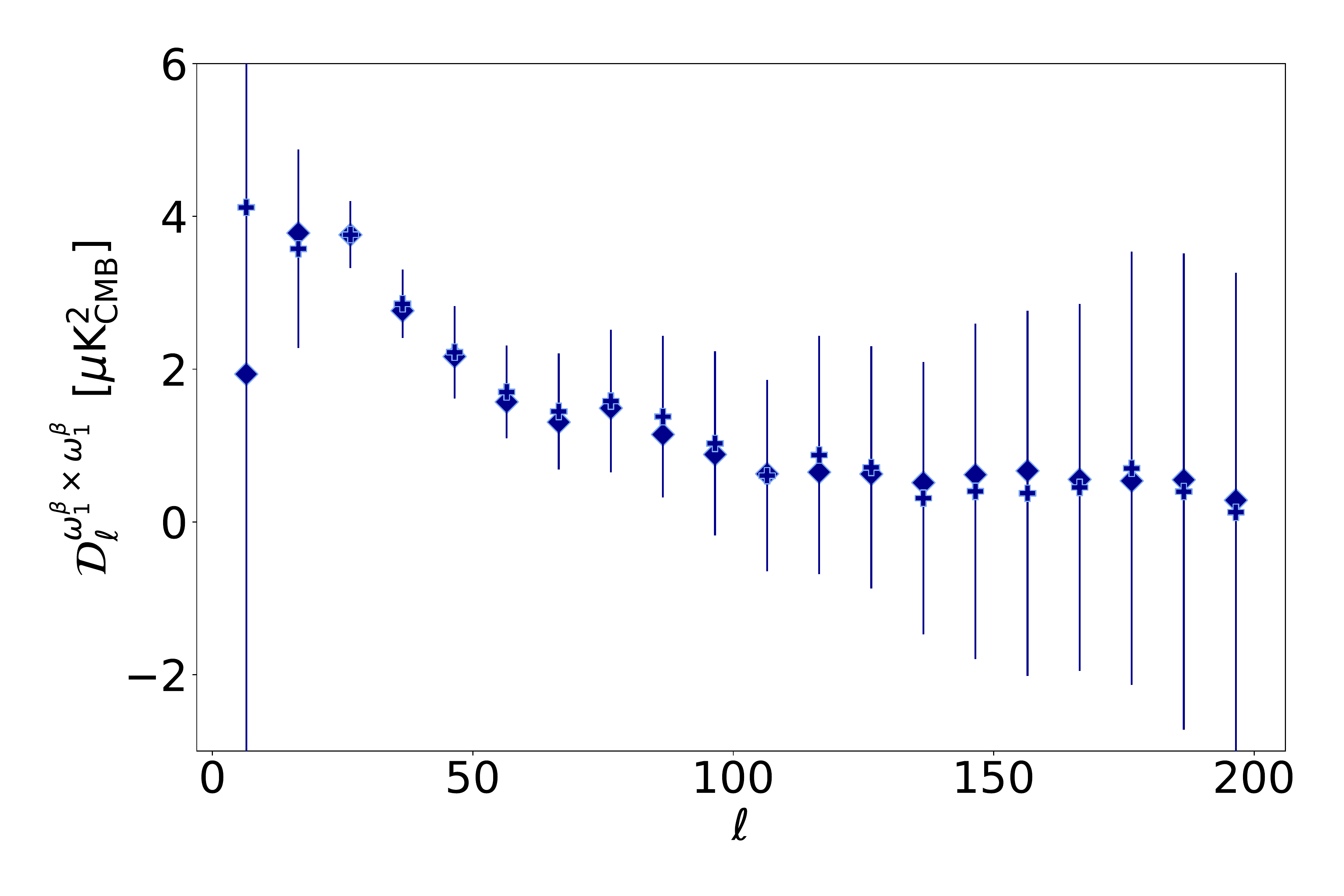}
    \includegraphics[width=\columnwidth]{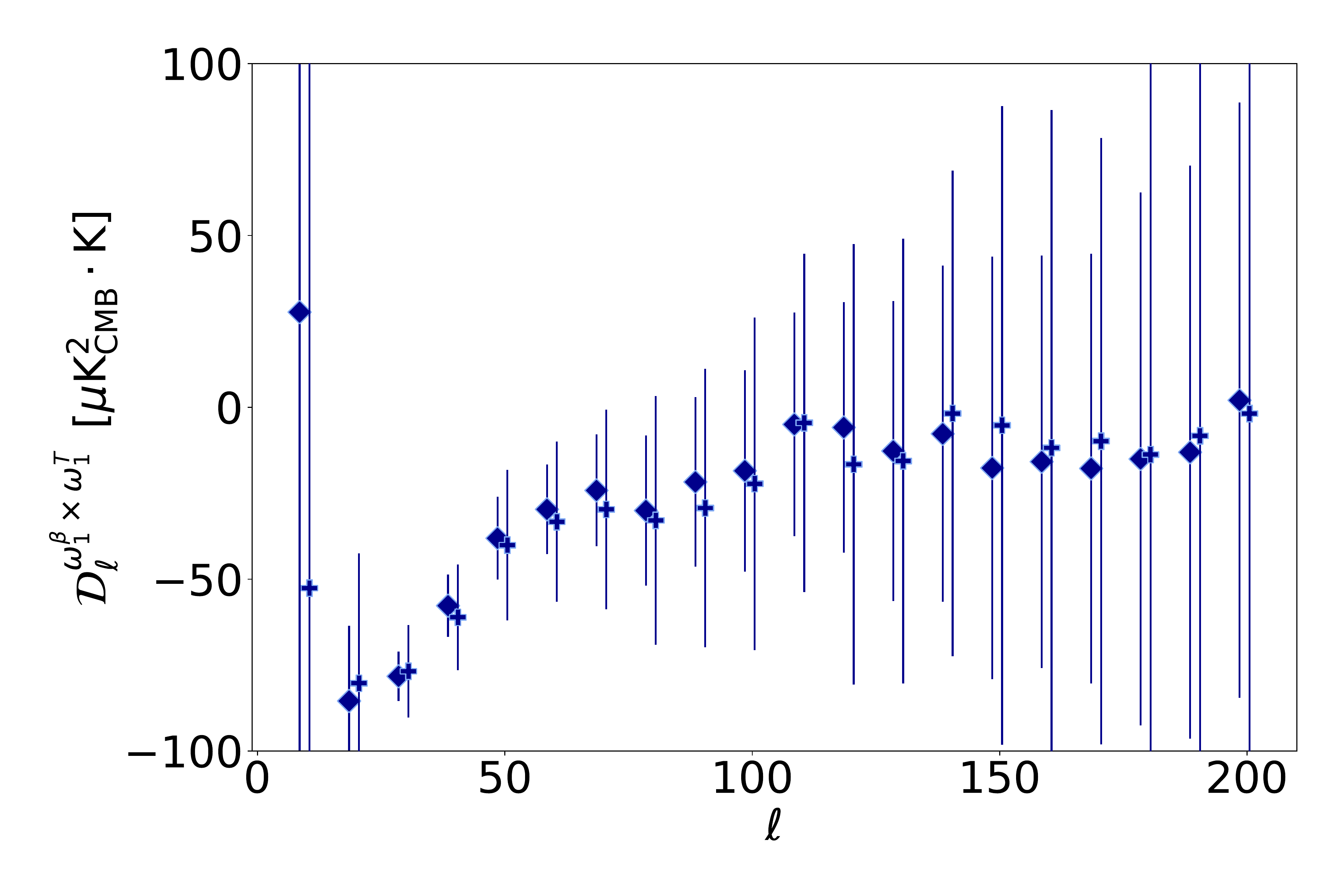}
    \includegraphics[width=\columnwidth]{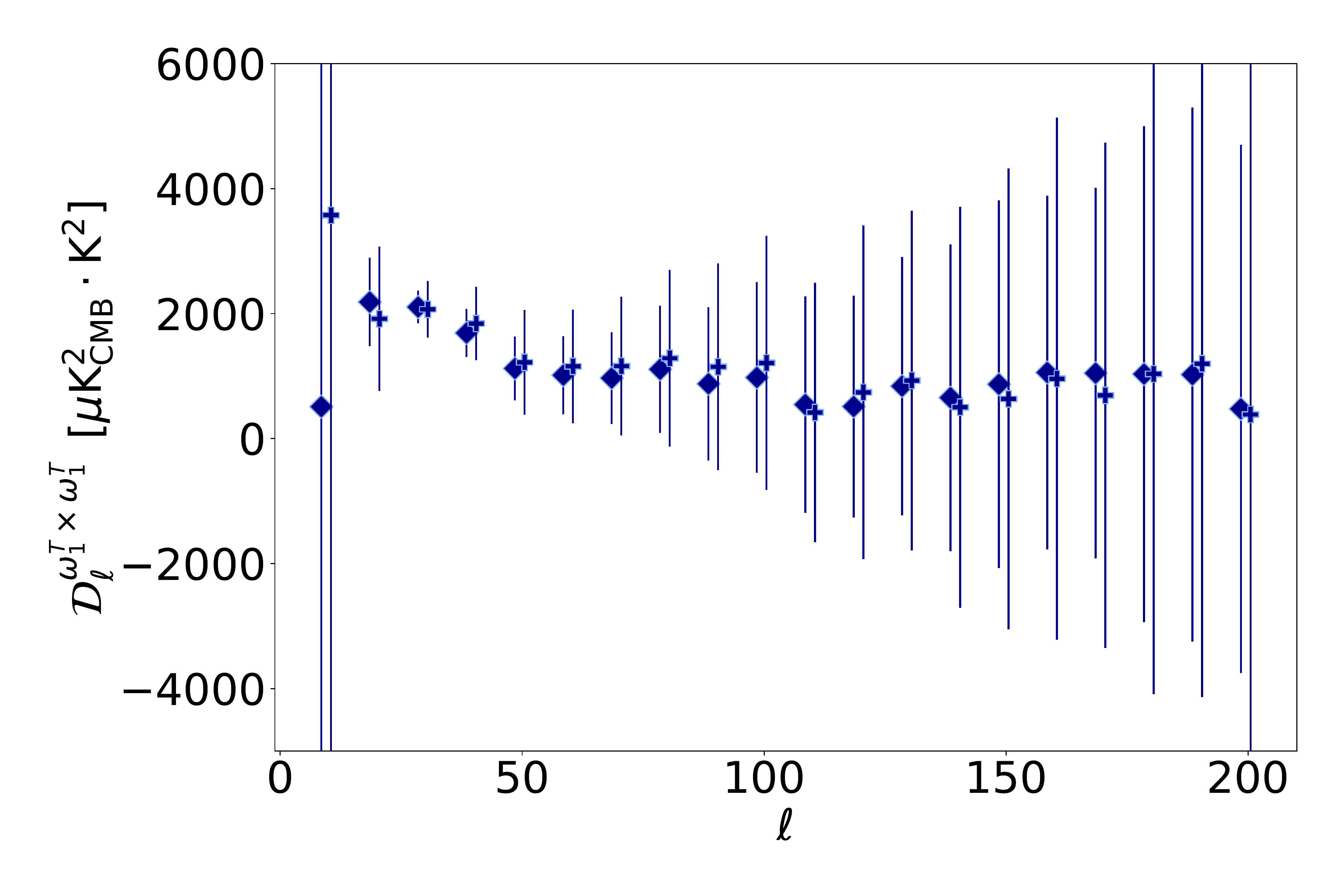}
    \caption{\footnotesize {Best-fit values of the $\mathcal{D}_\ell^{\omega^{\beta}_1\times\omega^{\beta}_1}$ (top left), $\mathcal{D}_\ell^{\omega^{\beta}_1\times\omega^{T}_1}$ (top right) and $\mathcal{D}_\ell^{\omega^{T}_1\times\omega^{T}_1}$ (bottom) moment parameters for the $\beta$-$T$ fitting scheme applied on the {\tt d1c} simulation type (diamonds) and $r\beta$-$T$ on {\tt d1c} (plus sign).}}
    \label{fig:moments2d1d1c}
\end{figure*}

\begin{figure*}[t!]
    \centering
    \includegraphics[width=2\columnwidth]{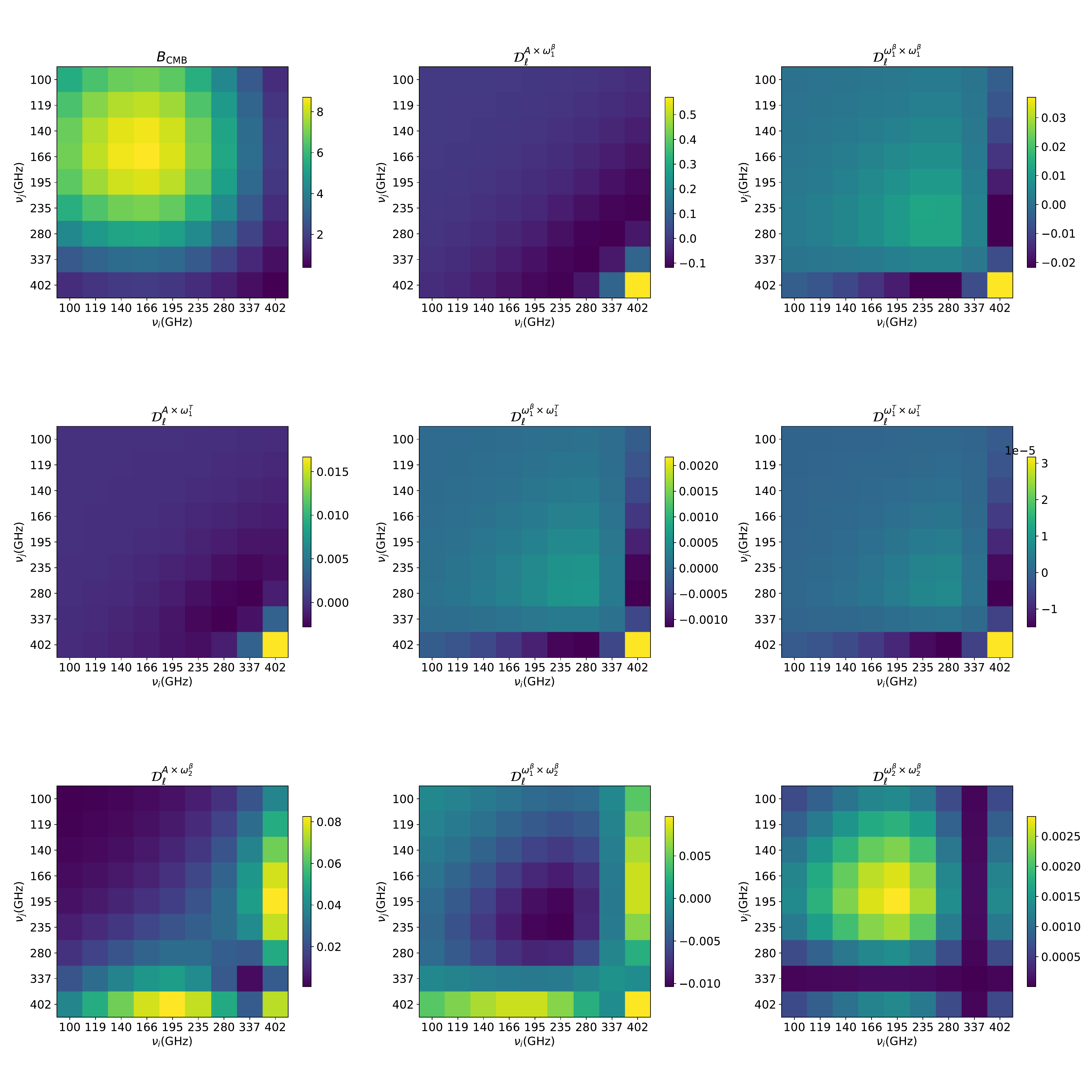}
    \caption{\footnotesize Two-dimensional SED shapes of the moment expansion parameters and the CMB in the $(\nu_i, \nu_j)$ space for the $nine$ \lb{} frequencies used throughout this work. {The intensities are all expressed in MJy$^2$ normalized by the squared SED at $\nu_0=353$ GHz.}}
    \label{fig:momentshapes}
\end{figure*}

\begin{figure}[t!]
    \centering
    \includegraphics[width=\columnwidth]{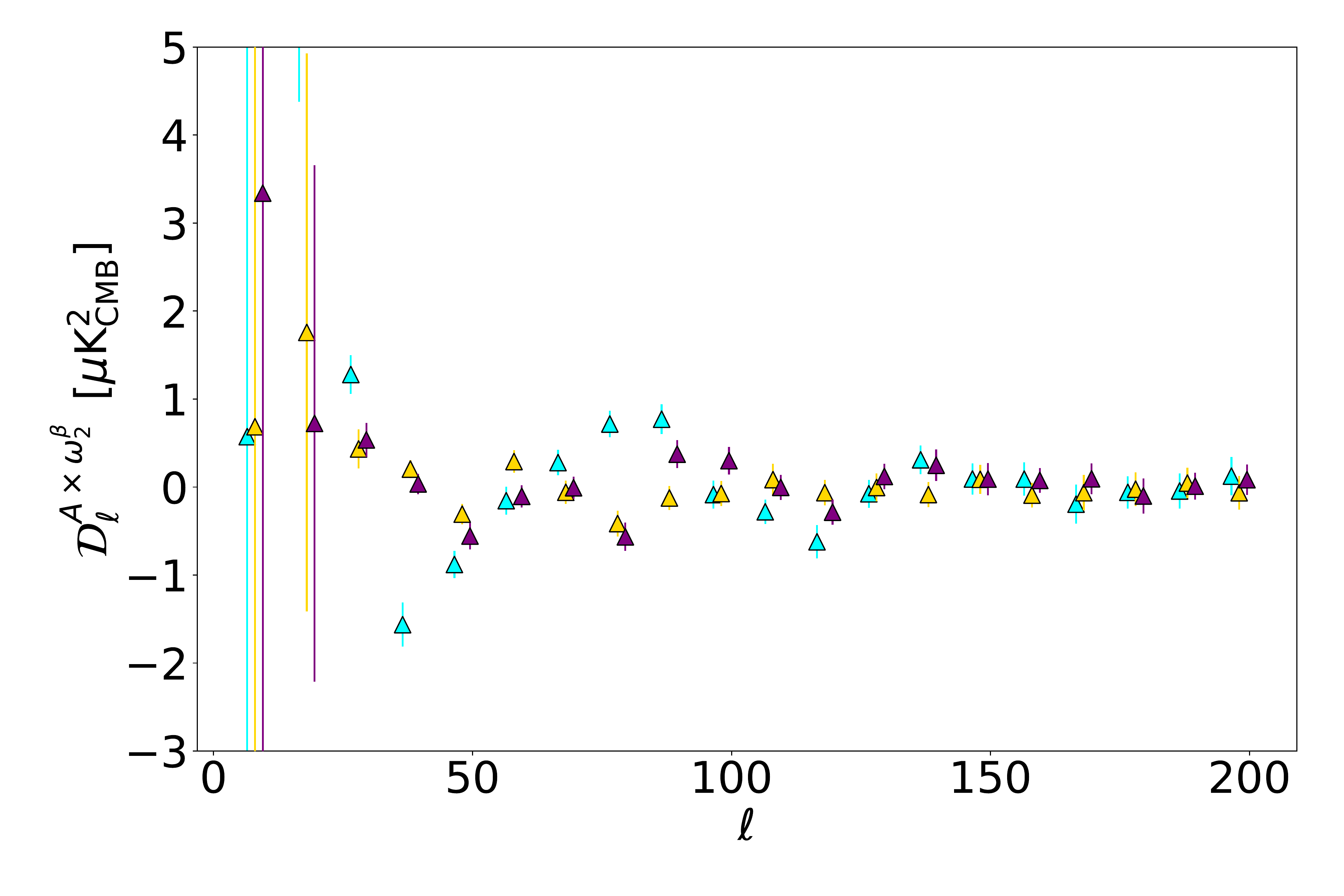}
    \includegraphics[width=\columnwidth]{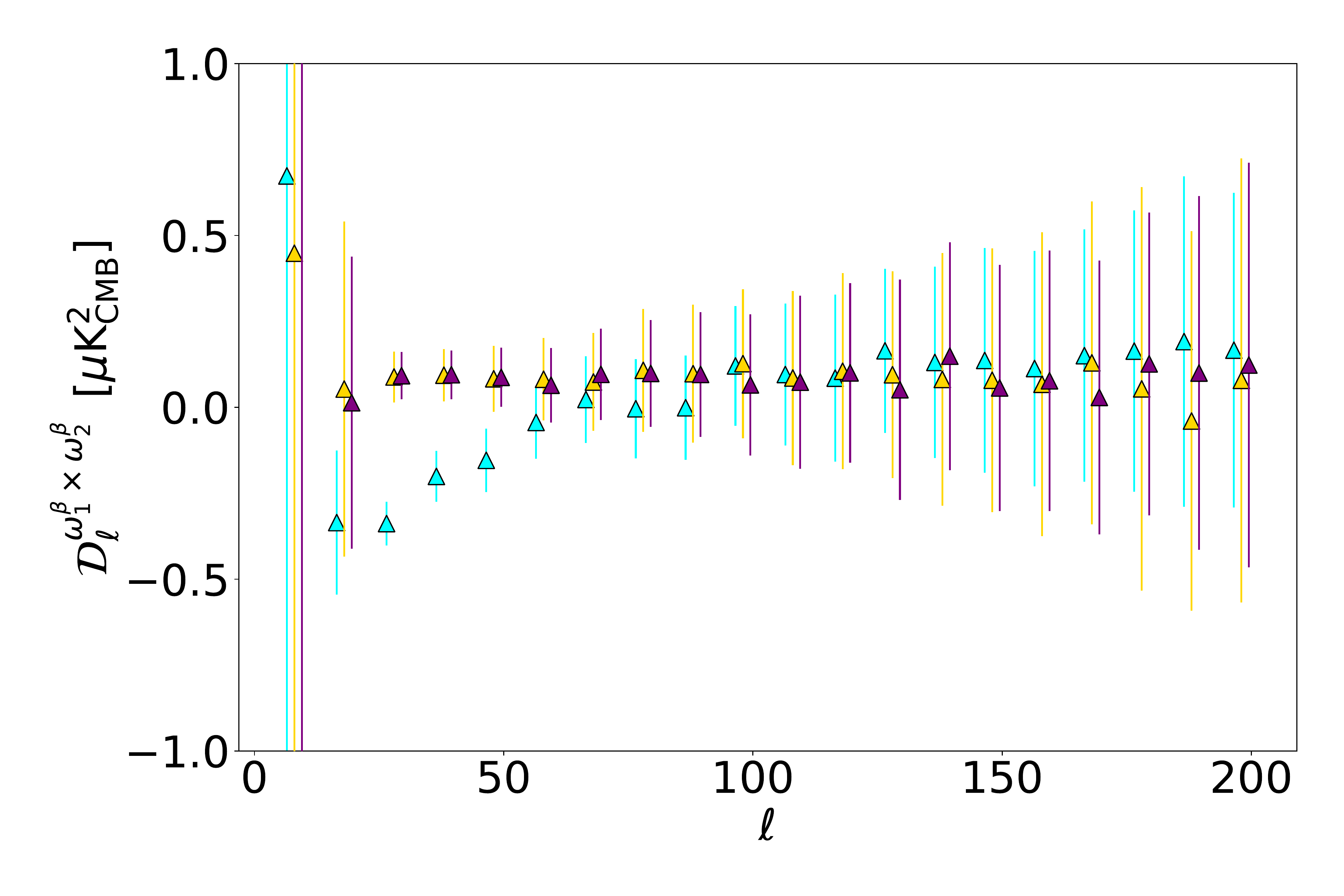}
    \includegraphics[width=\columnwidth]{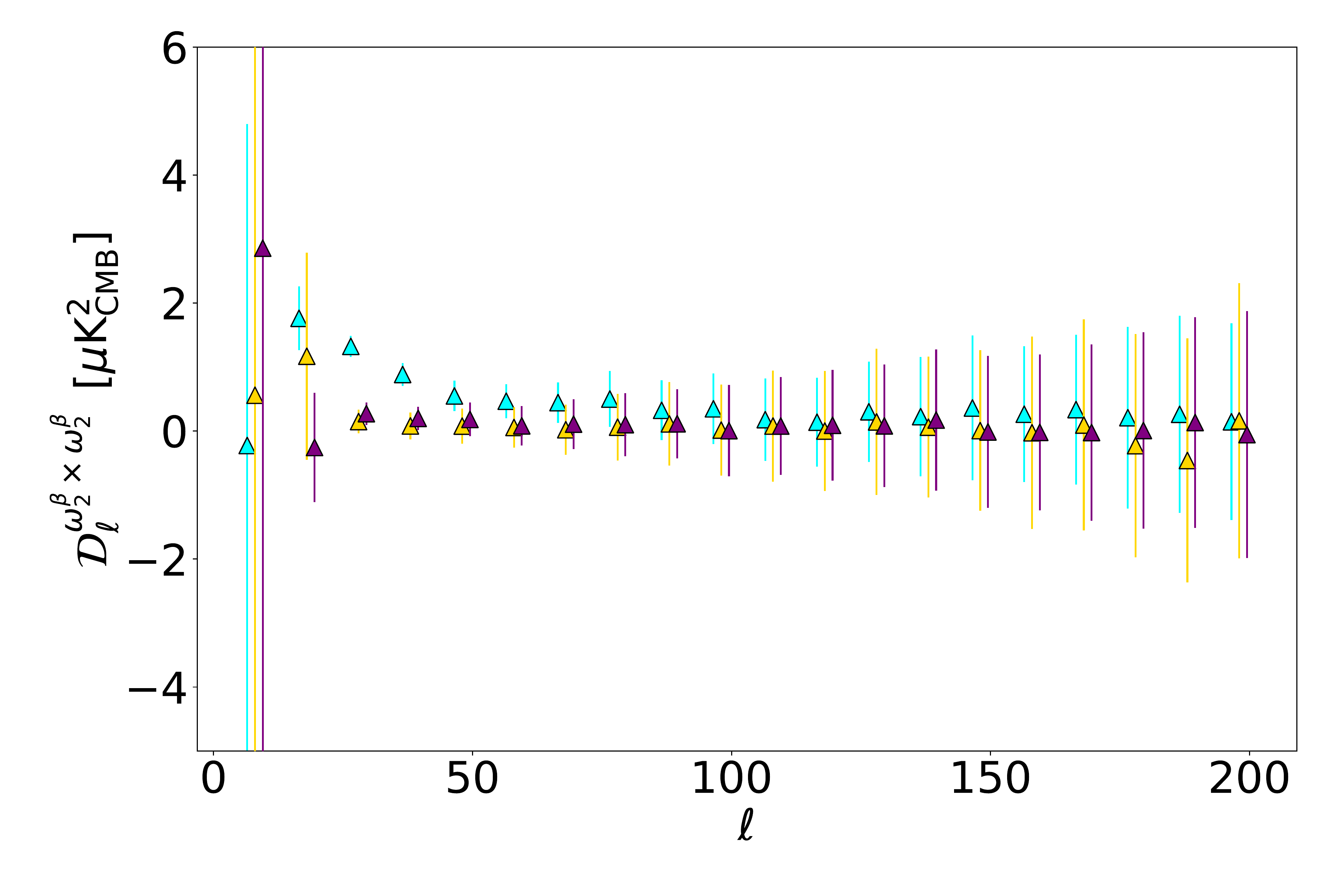}
    \caption{\footnotesize $\beta-2$ moment coefficients for different $f_{\rm sky}$ :  $f_{\rm sky}=0.7$ (cyan), $f_{\rm sky}=0.6$ (purple) and $f_{\rm sky}=0.5$ (gold).}
    \label{fig:fskydust}
\end{figure}

\end{appendix}
\end{document}